\documentclass{raa}

\usepackage{graphicx,times}
\usepackage{natbib}
\usepackage{amssymb,amsmath}
\usepackage{subcaption}
\usepackage[T1]{fontenc}
\usepackage{longtable}
\usepackage{pdflscape}
\usepackage{url}
\graphicspath{{./}{fig/}}
\bibpunct{(}{)}{;}{a}{}{,}

\usepackage[pagebackref=true]{hyperref}

\begin{document}

   \title{The Evolution of Thermal and Non-thermal Emission Components in GRB 250920C}

\volnopage{ {\bf 20XX} Vol.\ {\bf X} No. {\bf XX}, 000--000}
	\setcounter{page}{1}

	\author{Jia-Ming Chen
   		\inst{1}
   	\and Ke-Rui Zhu
   		\inst{1}
   	\and Shan Chang
   		\inst{1}
	\and Dan Zhu
	    \inst{2}
	\and Yu-Lu Gong
	    \inst{1}
   	\and Zhao-Yang Peng
   		\inst{3}
   	\and Yong-Gang Zheng
   		\inst{4}
   	\and Li Zhang
   		\inst{1}
   }

	\institute{Department of Astronomy, School of Physics and Astronomy,
    Key Laboratory of Astroparticle Physics of Yunnan Province, Yunnan University, Kunming 650091, People's Republic of China; 
    \and
	South-Western Institute for Astronomy Research, Yunnan University, Kunming, Yunnan 650504, People's Republic of China;
	\and
    College of Physics and Electronics information, Yunnan Normal University, Kunming 650500, People's Republic of China; 
    \and
    Department of Physics, Yunnan Normal University, Kunming 650500, People's Republic of China
} 

\vs \no
{\small Received 20XX Month Day; accepted 20XX Month Day}

\abstract{
	We present a temporal and time-resolved spectral analysis of GRB 250920C using \textit{Fermi}/GBM and \textit{Swift}/BAT data. The prompt emission consists of two distinct episodes, EI and EII, separated by a significant quiescent interval. We perform Bayesian spectral fitting with empirical, thermal, composite, and physical synchrotron models. In EI, the spectra show clear evidence for an additional thermal component. The BB+PL model is preferred in most bright time bins, and joint \textit{Fermi}/GBM+\textit{Swift}/BAT fits further support the presence of this component. The blackbody temperature generally decreases with time, the blackbody flux follows the pulse profile, and the thermal flux fraction remains high. Fireball-parameter estimates give a photospheric Lorentz factor of a few hundred and an initial radius of $r_{0}\sim10^{9}$--$10^{10}$ cm, with $1+\sigma_{0}$ of order unity and $\eta\gg1$, supporting a thermally dominated baryonic outflow. In contrast, EII is dominated by non-thermal emission. Its low-energy photon indices do not significantly exceed the synchrotron line of death, and the spectra are well described by fast-cooling synchrotron radiation in a decaying magnetic field. The magnetization constraint gives $\sigma_{\min}\sim1.1$--$4.3$. Since these values exceed unity, they suggest that EII may be Poynting-flux dominated. These results therefore suggest a possible transition in GRB~250920C from a fireball-dominated EI phase to a Poynting-flux-dominated EII phase.
}

   \authorrunning{J-M. Chen, et al. }            
   \titlerunning{GRB 250920C: Thermal and Non-thermal Emission Components}  
   \maketitle

%
\section{Introduction}           
\label{sec:intro}
The radiative mechanism of Gamma-Ray Burst (GRB) prompt emission remains one of the central unresolved issues in high-energy astrophysics \citep{2011CRPhy..12..206Z,2015PhR...561....1K,2018pgrb.book.....Z}. Although most prompt emission spectra can be described by the empirical Band function \citep{1993ApJ...413..281B,2014ApJS..211...12G,2021ApJ...913...60P}, these empirical models do not directly correspond to a unique physical process \citep{2015AdAst2015E..22P}. Current mainstream interpretations primarily focus on two perspectives \citep{2014IJMPD..2330002Z}: one is non-thermal radiation from optically thin regions, such as synchrotron radiation and its variants \citep{2000MNRAS.313L...1G,2011A&A...526A.110D,2020NatAs...4..174B}; the other is thermal emission released near the photosphere of the jet \citep{2000ApJ...530..292M,2009ApJ...700L..47L}, as well as the broadened quasi-thermal spectra formed through sub-photospheric dissipation and Comptonization \citep{2010MNRAS.407.1033B,2010ApJ...725.1137L,2012MNRAS.420..468P,2013MNRAS.436L..54A}. Recent time-resolved spectral studies extending to the soft X-ray band have also revealed an additional low-energy break below $E_{\text{peak}}$ in many long GRBs, with spectral indices consistent with synchrotron radiation expectations \citep{2017ApJ...846..137O,2018A&A...616A.138O,2019A&A...625A..60R}. This further suggests that prompt emission may not be dominated by a single mechanism, but more likely involves joint contributions from both thermal and non-thermal components \citep{2014ApJ...784L..43B}.

In this context, identifying the thermal component in prompt emission has become a crucial breakthrough for understanding the composition, dynamics, and energy dissipation sites of GRB jets. Early work by \citet{2005ApJ...625L..95R} proposed that the time-resolved spectra of many GRBs could be described by a hybrid "thermal + non-thermal" model, noting that the thermal component might be the key factor determining spectral evolution. Subsequently, systematic analysis of bright long GRB samples by \citet{2009ApJ...702.1211R} demonstrated that the thermal component typically exhibits regular temporal behavior during the prompt phase, with the temperature remaining approximately stable in the early pulse phase followed by a power-law decay. High-quality data from the \textit{Fermi} era have further consolidated this picture: for instance, the prompt emission spectrum of GRB 090902B can be interpreted as photospheric quasi-thermal radiation superimposed on a separate non-thermal power-law tail \citep{2009ApJ...706L.138A,2011MNRAS.417.1584B,2012MNRAS.420..468P}, while GRB 100724B also shows a statistically significant thermal component \citep{2011ApJ...727L..33G,2015MNRAS.454L..31A}. Recently, more bursts have expanded the sample of thermal or quasi-thermal events. For example, GRB 220426A shows narrow spectra and a strong photospheric component, similar to GRB 090902B \citep{2022ApJ...940..142W, 2022ApJ...934L..22D, 2022MNRAS.517.2088S}. Similar thermal features are also found in GRB 220304A and GRB 231129C \citep{2024ApJ...964...45C, 2024ApJ...972..132C}. Together, these studies suggest that thermal emission is not a rare exception but an essential constituent of the GRB prompt phase that warrants systematic investigation.

However, the physical interpretation of the thermal component is non-trivial. On one hand, a pure blackbody shape in real GRB spectra is often significantly broadened by geometric effects, jet structure, and sub-photospheric dissipation \citep{2008ApJ...682..463P,2013MNRAS.428.2430L,2013A&A...551A.124H,2019MNRAS.485..474A}. Consequently, observations more commonly reveal quasi-thermal peaks, multicolor blackbodies, or composite spectra that can be approximated by empirical models such as BB+PL, mBB+PL, or 2BB+PL \citep{2010ApJ...709L.172R,2014MNRAS.442..419B,2018ApJ...866...13H}. On the other hand, blackbody components in time-integrated spectra can sometimes be mimicked by rapid spectral evolution, making time-resolved spectral analysis vital for verifying the authenticity of thermal components \citep{2015MNRAS.447.3087B,2015AdAst2015E..22P}. Meanwhile, recent research indicates that the inclusion of a thermal component systematically alters non-thermal spectral parameters, while sub-photospheric dissipation models can produce a broad non-thermal appearance through the Comptonization of thermal photons without explicitly adding an independent synchrotron component \citep{2015MNRAS.454L..31A,2019MNRAS.485..474A}. Therefore, establishing robust links between empirical models, physical models, and competing interpretations has become a critical step in GRB prompt spectral analysis \citep{2015AdAst2015E..22P,2017IJMPD..2630018P}.

Rather than discussing thermal or non-thermal components in isolation, a more physically meaningful direction in recent years is to examine whether a coupled evolution exists between the two. Analysis of several bright \textit{Fermi} GRBs by \citet{2014ApJ...784L..43B} showed a significant correlation between the characteristic energies of the photospheric thermal and non-thermal components, suggesting they are likely modulated by the same set of jet physical parameters rather than being independently superimposed. Similarly, \citet{2012MNRAS.420..468P} discussed the connection between the thermal peak and the high-energy non-thermal tail in GRB 090902B, pointing out that thermal photons can either directly constitute the quasi-thermal main peak or serve as seed photons for subsequent inverse Compton scattering. \citet{2012MNRAS.420..468P,2010ApJ...709L.172R} found systematic differences between models such as BB+PL, mBB+PL, and 2BB+PL in time-resolved analyses of single-pulse GRBs, noting that the power-law component often appears with a delay and exhibits a late-time tail relative to the thermal component. These results indicate that the temporal evolution of thermal--non-thermal coupling serves as a sensitive probe for constraining photospheric radiation, sub-photospheric dissipation, and particle acceleration in optically thin regions \citep{2015AdAst2015E..22P,2017IJMPD..2630018P}.

GRB 250920C provides a suitable sample for investigating this issue \citep{2025GCN.41917....1M}. We perform a time-resolved spectral analysis of the prompt emission of GRB 250920C focusing on testing whether its spectra can be described by a composite model of thermal and non-thermal components. Furthermore, we investigate the temporal evolution and correlations of key parameters, including blackbody temperature, thermal flux, and the non-thermal power-law index. The structure of this paper is as follows. In Section \ref{sec2}, we describe the \textit{Fermi}/GBM and \textit{Swift}/BAT observations and the corresponding data reduction procedures. In Section \ref{sec3}, we present the temporal properties of GRB 250920C, including the minimum variability timescale and spectral lag. In Section \ref{sec4}, we perform time-resolved spectral analysis and investigate the temporal evolution and correlations of the key spectral parameters. In Section \ref{sec5}, we derive physical constraints on the fireball and magnetized outflow properties. Finally, the discussion and summary are given in Section \ref{sec6}. Throughout this paper, we adopt a flat $\Lambda$CDM cosmology with $H_{0}=67.4~{\rm km~s^{-1}~Mpc^{-1}}$, $\Omega_{\rm M}=0.315$, and $\Omega_{\Lambda}=0.685$ \citep{2020A&A...641A...6P}. Unless otherwise stated, all quoted uncertainties correspond to the $1\sigma$ confidence level.

\section{Observations and Data Reduction}
\label{sec2}
\subsection{\textit{Fermi}/GBM observations}

GRB 250920C triggered \textit{Fermi}/GBM at 15:25:17.06 UT on 2025 September 20 \citep{2025GCN.41917....1M}. The GBM light curve shows multiple emission episodes, with a duration of $T_{90}\approx36.6$ s in the 50--300 keV band. The time-averaged spectrum over the interval from $T_{0}+0.003$ s to $T_{0}+37.377$ s is well fitted by a Band function with $E_{\rm p}=117\pm4$ keV, $\alpha=-0.76\pm0.04$, and $\beta=-2.34\pm0.05$ \citep{2025GCN.41917....1M}. The burst was also independently detected by \textit{Konus}-Wind, which reported a multi-peaked prompt emission episode with a total duration of $\sim41$ s and emission extending up to $\sim2$ MeV \citep{2025GCN.41957....1P}.

\begin{figure}[htbp]
\centering
\begin{minipage}[t]{0.9\textwidth}
\centering
\includegraphics[width=\textwidth]{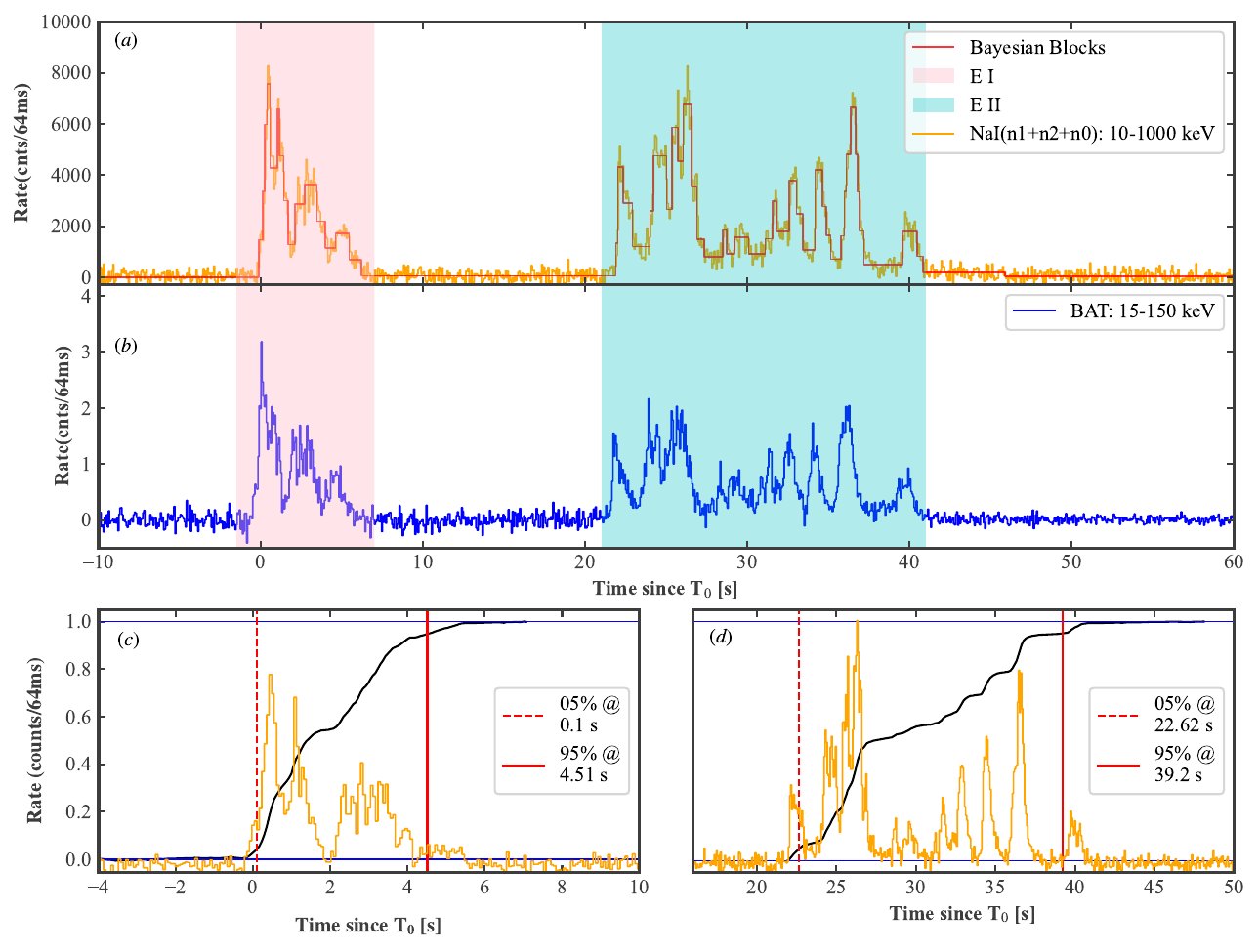}
\end{minipage}
\caption{Composite light-curve analysis of GRB 250920C. Panel (a) shows the \textit{Fermi}/GBM NaI light curve in the 10--1000~keV band, with Bayesian-block intervals overlaid. Panel (b) shows the \textit{Swift}/BAT light curve in the 15--150~keV band. Panels (c) and (d) show the cumulative count distributions for EI and EII, respectively. The vertical dashed and solid red lines mark the 5\% and 95\% cumulative-count levels used to estimate $T_{90}$. The yellow and cyan shaded regions denote EI and EII, respectively. \label{fig1}}
\end{figure}

Figure~\ref{fig1} (a) shows the prompt light curve of GRB 250920C in the 10--1000 keV band. The burst exhibits a two-episode emission structure. The first episode (EI), spanning from $T_{0}+0$ s to $T_{0}+6.5$ s, consists of multiple pulses and has a duration of $T_{90}=4.41$ s in the 50--300 keV band. The second episode (EII), occurring from $T_{0}+22$ s to $T_{0}+41$ s, has $T_{90}=16.58$ s. Applying the Bayesian blocks algorithm with a false-alarm probability of $p=0.01$, we identify a significant quiescent interval between the two episodes, extending from $T_{0}+6.5$ s to $T_{0}+22$ s, with a duration of $\Delta t_{\rm QP}=15.5$ s.

\subsection{\textit{Swift}/BAT observations}
\label{BAT observations}

The burst was also detected by \textit{Swift}/BAT. The refined BAT analysis gave $T_{90}=38.02\pm1.27$ s in the 15--350 keV band, and the time-averaged spectrum is well fitted by a cutoff power-law model with $E_{\rm p}=133.4\pm43.9$ keV \citep{2025GCN.41924....1G}. The \textit{Swift} data of GRB 250920C were retrieved from the UK \textit{Swift} Science Data Centre \footnote{\url{https://www.swift.ac.uk/index.php}}. We processed the BAT data using \textsc{HEASoft} v6.36, \textsc{FTOOLS}, and the standard procedures described in the \textit{Swift}/BAT Software Guide. A BAT light curve with a time resolution of 64 ms was generated. We first applied gain correction using \texttt{bateconvert}. After creating detector plane images (DPIs), identifying problematic detectors and removing hot pixels, mask weighting and background subtraction were performed using \texttt{batdetmask}, \texttt{bathotpix}, \texttt{batmaskwtevt}, and \texttt{batbinevt}. The background-subtracted BAT light curve was then extracted with \texttt{batbinevt}. The coded-mask background subtraction improves the signal-to-noise ratio and enables an accurate reconstruction of the burst light curve, as shown in Figure~\ref{fig1}(b).

For spectral extraction, we followed the same basic processing steps as those used for the light-curve generation. In addition, \texttt{batphasyserr} and \texttt{batupdatephakw} were applied to account for residual response features and to ensure the correct burst position in instrument coordinates. The detector response matrix (DRM) was generated using \texttt{batdrmgen}. These \textit{Swift}/BAT data products allow a detailed joint temporal and spectral analysis of the prompt emission of GRB 250920C in combination with the \textit{Fermi} data.

For completeness, follow-up \textit{Swift}/XRT observations revealed a fading X-ray afterglow from 110 s to 131 ks after the trigger \citep{2025GCN.41935....1D}. Subsequent optical spectroscopic observations established a redshift of $z=1.40$, later confirmed to be consistent with $z=1.399$ \citep{2025GCN.41928....1S,2025GCN.41955....1I}.

\section{Temporal Properties}
 \label{sec3}
\subsection{Minimum Variability Timescale}

The minimum variability timescale (MVT) is an important diagnostic of the central-engine activity and the characteristic size of the emitting region. In this work, we applied a wavelet analysis to the light curves of EI and EII in the \textit{Fermi}/GBM 10--1000 keV band with a time resolution of 8 ms in order to determine the MVT in each episode. Similar techniques have been widely used in previous studies of GRB variability. We find that the MVT is $76.88 \pm 12.13$ ms for EI and $52.14 \pm 20.67$ ms for EII.

Based on the measured MVTs, we further estimated the physical properties of the emitting region in GRB 250920C. Following \citet{2015ApJ...811...93G}, the lower limit on the bulk Lorentz factor and the characteristic radius of the emission region, $R_{\rm c}$, can be estimated as
\begin{equation}
\Gamma_{\rm min} \gtrsim 110
\left(
\frac{L_{\gamma,\mathrm{iso}}}{10^{51}\ \mathrm{erg\ s^{-1}}}
\cdot
\frac{1+z}{t_{\rm MV}/0.1\ \mathrm{s}}
\right)^{1/5},
\end{equation}
and
\begin{equation}
R_{\rm c} \approx 7.3\times10^{13}
\left(
\frac{L_{\gamma,\mathrm{iso}}}{10^{51}\ \mathrm{erg\ s^{-1}}}
\right)^{2/5}
\left(
\frac{t_{\rm MV}/0.1\ \mathrm{s}}{1+z}
\right)^{3/5}
\ \mathrm{cm}.
\end{equation}
Using these relations, we obtain $\Gamma_{\rm min} \gtrsim 234.2$ and $R_{\rm c} \approx 1.04\times10^{14}$ cm for EI, and $\Gamma_{\rm min} \gtrsim 254.8$ and $R_{\rm c} \approx 8.34\times10^{13}$ cm for EII.

\subsection{Spectral Lag}

\begin{figure}[htbp]
\centering
\begin{minipage}[t]{0.6\textwidth}
\centering
\includegraphics[width=\textwidth]{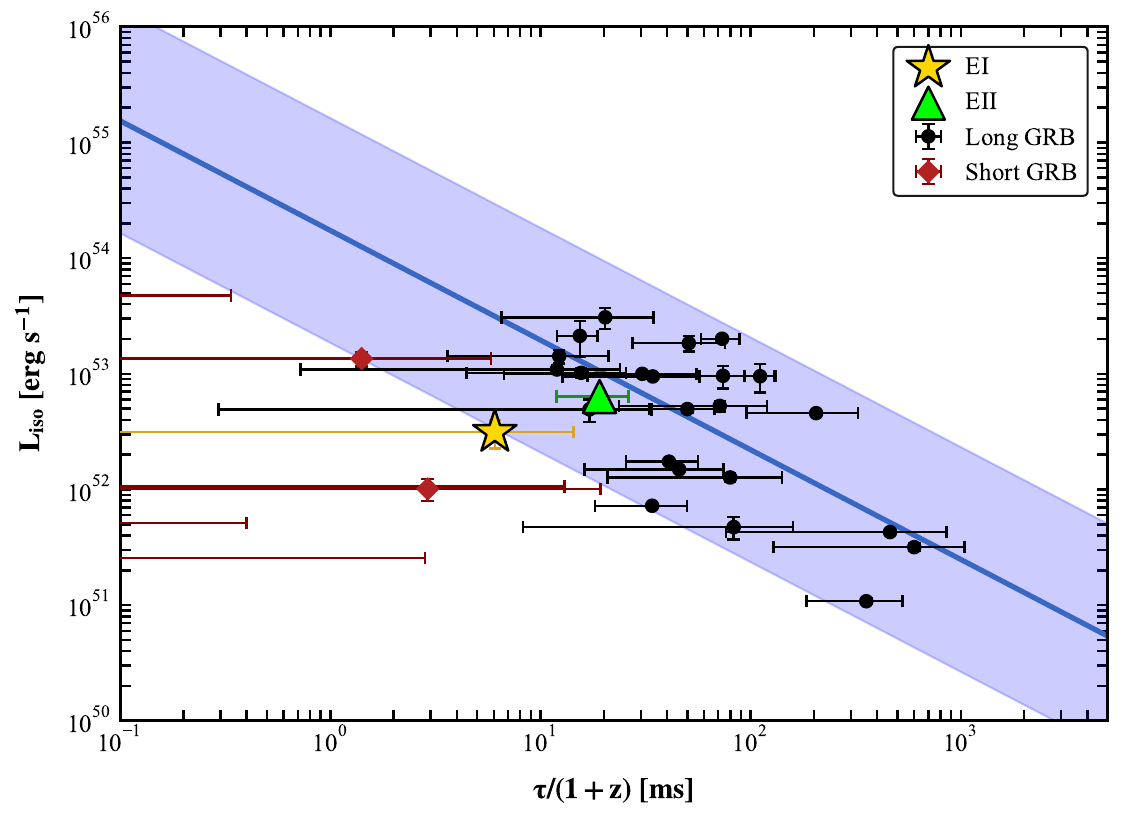}
\end{minipage}
\caption{Placement of the EI and EII phases of GRB 250920C in the peak-luminosity versus rest-frame spectral-lag plane. The comparison sample is taken from \citet{2015MNRAS.446.1129B}. Orange and green symbols represent EI and EII, respectively. EI and EII both lie along the track occupied by long GRBs.} \label{fig2}
\end{figure}

Another important temporal property of GRBs is the spectral lag, namely the systematic delay of low-energy photons relative to high-energy photons in the prompt light curve \citep{2000ApJ...534..248N,2006Natur.444.1044G}. Following \citet{2015MNRAS.446.1129B}, we measure the spectral lag with the cross-correlation function using the same fixed rest-frame energy bands of 100--150~keV and 200--250~keV. We obtain observer-frame lags of $0.0145 \pm 0.0199$~s for EI and $0.045819 \pm 0.017281$~s for EII. After correcting for cosmological time dilation, we compare GRB~250920C with the samples of \citet{2015MNRAS.446.1129B} and related studies. As shown in Figure~\ref{fig2}, EI and EII both fall within the region occupied by long GRBs in the $L_{\rm iso}$--$\tau_{\rm RF}$ plane.

\section{Spectral Analysis}
 \label{sec4}
The \textit{Fermi} Gamma-ray Burst Monitor (GBM; \citealt{2009ApJ...702..791M}) consists of 12 sodium iodide (NaI) detectors ($8 \rm{\,keV}$ to $1 \rm{\,MeV}$) and two bismuth germanate (BGO) detectors ($200 \rm{\,keV}$ to $40 \rm{\,MeV}$). Based on the viewing angles and observed count rates, we selected the three brightest NaI detectors (n0, n1, and n2) and one BGO detector (b0) for our analysis. We retrieved the Time-Tagged Event (TTE) data, which offer a $2 \rm{\,\mu s}$ temporal resolution across 128 energy channels, from the Fermi Science Support Center (FSSC) \footnote{\url{https://fermi.gsfc.nasa.gov/ssc/}}. Time-resolved spectral analysis was conducted in this work using the Bayesian analysis tool \texttt{threeML} \citep{2015arXiv150708343V}.

For background estimation, we used the count data from the brightest NaI detector and selected source-free intervals before and after the burst. The background was fitted with polynomials of order 0--4, and the optimal polynomial order was determined through a likelihood ratio test. The resulting model was then used to subtract the background in all 128 energy channels.

For the time-resolved spectral analysis, the light curve must be divided into sufficiently fine temporal bins to capture the spectral evolution of the prompt emission. To this end, we reconstructed the TTE light curve of the brightest NaI detector using the Bayesian blocks algorithm with a false-alarm probability of $p=0.01$ \citep{2013ApJ...764..167S}. This procedure resulted in 14 time bins for EI (0--7.0 s) and 30 time bins for EII (22.0--43.0 s).

In addition, we performed joint time-resolved spectral analysis using both \textit{Fermi} and \textit{Swift}/BAT data. For EI, we defined three time intervals: $[0.0,1.40]$ s, $[1.40,3.80]$ s, and $[3.80,7.0]$ s. For EII, we defined eight intervals: $[21.5,23.20]$ s, $[23.20,24.80]$ s, $[24.80,27.00]$ s, $[27.0,32.0]$ s, $[32.0,33.5]$ s, $[33.5,35.0]$ s, $[35.0,37.0]$ s, and $[37.0,41.0]$ s. Because the BAT band lies within the broader GBM coverage, including BAT does not extend the total fitted energy range. Instead, the joint fits add an independent data set and instrument response in the overlapping energy range, providing a cross-instrument test and additional constraints on the spectral parameters. The BAT spectral extraction procedure is described in Section \ref{BAT observations}.

The prompt emission spectra of GRBs are commonly described by the empirical Band function \citep{1993ApJ...413..281B}. The Band function is defined as:
\begin{equation}
N(E)_{\rm{BAND}} = A_{\rm{BAND}}
\begin{cases}
\left( \frac{E}{100 \rm{\,keV}} \right)^\alpha \exp \left[ -\frac{E(2+\alpha)}{E_p} \right], & E \le \frac{\alpha-\beta}{2+\alpha}E_p \\
\left( \frac{(\alpha-\beta)E_p}{(2+\alpha)100 \rm{\,keV}} \right)^{\alpha-\beta} \exp(\beta-\alpha) \left( \frac{E}{100 \rm{\,keV}} \right)^\beta, & E \ge \frac{\alpha-\beta}{2+\alpha}E_p
\end{cases}
\end{equation}
where $\alpha$ and $\beta$ are the low- and high-energy photon indices, respectively. $E_p$ is the peak energy in the $\nu F_\nu$ spectrum. Furthermore, if the count rate of high-energy photons is relatively low, the high-energy spectral index $\beta$ may remain unconstrained. In such cases, a Cutoff Power-Law (CPL) function can be employed:
\begin{equation}
N(E)_{\rm{CPL}} = A_{\rm{CPL}} \left( \frac{E}{100 \rm{\,keV}} \right)^\alpha \exp(-E/E_c),
\end{equation}
where $\alpha$ is the photon index and $E_c$ is the cutoff energy in units of $\rm{keV}$. The peak energy reported in the tables is related to the cutoff energy by
$E_{\rm p}=(2+\alpha)E_{\rm c}$. For all the aforementioned models, $A$ represents the amplitude in units of $\rm{ph \, cm^{-2} \, keV^{-1} \, s^{-1}}$. 

When considering thermal emission components, the photon spectrum of blackbody (BB) radiation is typically written as:
\begin{equation}
N(E)_{\rm{BB}} = A_{\rm{BB}} \frac{E^2}{\exp(E/kT_{\rm{BB}}) - 1},
\end{equation}
where $k$ is the Boltzmann constant, and the parameter $kT_{\rm{BB}}$ is the usual fitted output. Compared to a single Planck function, a multi-color blackbody (mBB) model can often describe the photospheric emission component more accurately. By superimposing Planck functions of different temperatures, an empirical mBB model can be constructed, which has been widely applied in many GRBs exhibiting thermal spectra \citep{2010ApJ...709L.172R}. The mBB formalism adopted in this paper is a modified version from \citet{2018ApJ...866...13H}, and its expression is given by:
\begin{equation}
N(E)_{\rm{mBB}} = \frac{8.0525(m+1)A_{\rm{mBB}}}{\left[ \left( \frac{T_{\rm{max}}}{T_{\rm{min}}} \right)^{m+1} - 1 \right]} \left( \frac{kT_{\rm{min}}}{\rm{keV}} \right)^{-2} I(E),
\end{equation}
with
\begin{equation}
I(E) = \left( \frac{E}{kT_{\rm{min}}} \right)^{m-1} \int_{E/kT_{\rm{max}}}^{E/kT_{\rm{min}}} \frac{x^{2-m}}{e^x - 1} \, dx,
\end{equation}
where $x = E/kT$, and the temperature ranges from $kT_{\rm{min}}$ to $kT_{\rm{max}}$. The index $m$ controls the spectral shape. When $m = 2$, the mBB model serves as a good approximation to a pure blackbody spectrum.

We also include a power-law (PL) model added to the mBB and BB models to fit the spectra. The PL model is defined as:
\begin{equation}
N(E)_{\rm{PL}} = A \left( \frac{E}{100 \rm{\,keV}} \right)^{\alpha_{\rm {PL}}} .
\end{equation}

While the exact non-thermal emission mechanism of GRB prompt emission remains heavily debated, electron synchrotron radiation is widely considered the most viable physical candidate \citep{2016ApJ...816...72Z,2020NatAs...4..174B}. Fitting observations directly with physical synchrotron models, rather than traditional empirical functions, provides a more robust probe of the physical conditions within the jet. Here, we apply a fast-cooling synchrotron model with a decaying magnetic field \citep{2014ApJ...780...12Z,2014NatPh..10..351U} to the spectra of GRB 250920C. Because direct numerical integration of this physical model is computationally prohibitive for extensive Bayesian sampling, we utilize a Convolutional Neural Network (CNN) based spectral emulator \citep{2026ApJ..1005...76C}. This emulator reconstructs the numerical spectra with high precision and is seamlessly integrated into the \texttt{ThreeML} pipeline. The theoretical spectrum is parameterized as:
\begin{equation}
F_{\text{MDFSYN}} = F_{\nu,\rm{obs}} \left( E_{\rm{obs}}, \hat{t}, B_0, a, \Gamma, \gamma_{\rm{inj}}, p, R_0, Q_0, z \right).
\end{equation}
where $\hat{t}$ is the observer-frame time since the onset of an injection episode, $B_0$ is the magnetic-field strength at the emission radius $R_0$, $a$ is the magnetic-field decay index defined by $B(R)=B_0(R/R_0)^{-a}$, $\Gamma$ is the bulk Lorentz factor, $\gamma_{\rm inj}$ is the minimum injected electron Lorentz factor, $p$ is the electron injection index, $Q_0$ is the injection coefficient, and $z$ is the redshift. 

To determine the most appropriate model from a given set, we use the Bayesian Information Criterion (BIC), $\mathrm{BIC}=-2\ln L+k\ln n$, where $L$ is the maximized likelihood, $k$ is the number of free parameters, and $n$ is the number of data points \citep{Neath}.

Because no single statistic fully characterizes the fit quality of nonlinear count-spectrum models, we use several complementary diagnostics. We report the likelihood statistic at the sampled maximum-likelihood solution: the joint ${\rm PGSTAT}=-2\ln L_{\max}$ for GBM-only fits and the total $-2\ln L_{\max}$ for GBM+BAT fits, whose likelihood also contains the Gaussian BAT contribution. RSS is retained only to inspect residual structure; RSS/dof is neither treated as a reduced chi-square nor used for model selection or formal goodness-of-fit assessment \citep{Andrae2010}. Absolute fit quality is evaluated with a posterior predictive check (PPC), where replicated spectra are drawn from the posterior and $p=P(T_{\rm rep}\geq T_{\rm obs})$ for $T=-2\ln L$; values close to 0 or 1 indicate possible model--data tension. BIC is used only for relative model comparison. Our conclusions jointly consider the likelihood statistic, PPC, residual patterns, posterior identifiability, and BIC.

\subsection{Time-resolved spectral analysis results}
\label{sec4.1}

We performed a systematic time-resolved spectral analysis of GRB~250920C. The parameter analysis and discussion below focus primarily on intervals with a spectral significance of $S>10$, for which the spectral parameters can be constrained more reliably. Each \textit{Fermi}/GBM-only interval was fitted with the Band, CPL, mBB, and BB+PL models, and the same empirical models were applied to the broader joint \textit{Fermi}/GBM+\textit{Swift}/BAT intervals. Their parameters are reported in Tables~\ref{tab:fermi_empirical} and \ref{tab:joint_empirical}, respectively. We use the GBM-only fits as the primary sequence for tracing the finer temporal evolution, while the joint fits provide an independent cross-instrument robustness check. For EII, we also fitted the fast-cooling synchrotron model with a decaying magnetic field (MDFSYN); its GBM-only and joint-fit parameters are given in Tables~\ref{tab:eii_mdfsyn} and \ref{tab:joint_mdfsyn}.

We also examined the posterior identifiability of the more complex empirical models. The additional PL normalization in mBB+PL is frequently boundary-limited and strongly degenerate with the mBB temperature parameters, so its parameter estimates are not reported. The CPL+BB and Band+BB fits likewise yield broad or multimodal posteriors in which the BB normalization accumulates near its lower boundary and is strongly degenerate with $kT$, $E_{\rm p}$, and the continuum indices. Because all 11 joint intervals have $S>20$, this behavior cannot be attributed simply to the total photon statistics, but instead reflects the difficulty of separating strongly overlapping spectral components.

\begin{figure}[htbp]
\centering
\includegraphics[width=0.7\textwidth]{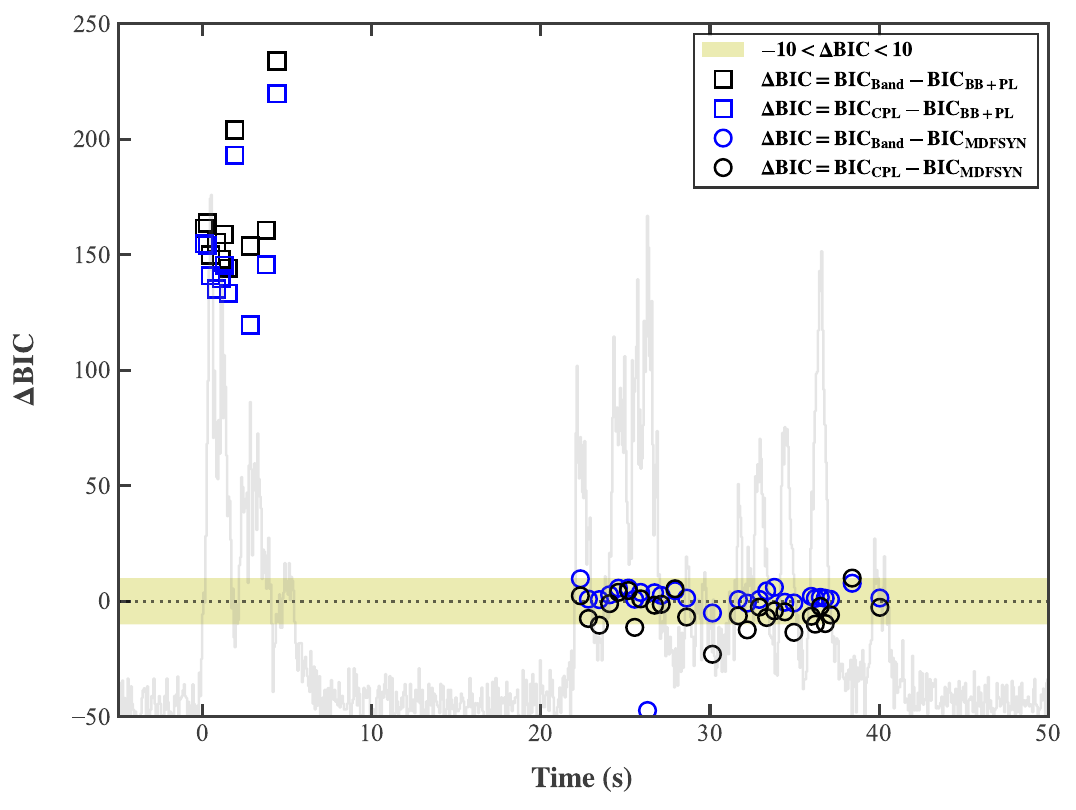}
\caption{Temporal evolution of the BIC differences for model comparison in GRB 250920C. In EI, the black and blue squares represent $\Delta{\rm BIC}={\rm BIC}_{\rm Band}-{\rm BIC}_{\rm BB+PL}$ and $\Delta{\rm BIC}={\rm BIC}_{\rm CPL}-{\rm BIC}_{\rm BB+PL}$, respectively. In EII, the blue and black circles represent $\Delta{\rm BIC}={\rm BIC}_{\rm Band}-{\rm BIC}_{\rm MDFSYN}$ and $\Delta{\rm BIC}={\rm BIC}_{\rm CPL}-{\rm BIC}_{\rm MDFSYN}$, respectively. Positive values indicate that BB+PL or MDFSYN is preferred over the corresponding empirical model. The shaded region marks $|\Delta{\rm BIC}|<10$, where the evidence is weak or inconclusive.}
\label{fig:delta_bic}
\end{figure}

For direct comparisons, we define $\Delta{\rm BIC}={\rm BIC}_{\rm ref}-{\rm BIC}_{\rm test}$, where the test model is BB+PL for EI and MDFSYN for EII. Positive $\Delta{\rm BIC}$ favors the test model. We regard $|\Delta{\rm BIC}|<10$ as weak or inconclusive evidence and $\Delta{\rm BIC}>10$ as strong evidence in favor of the test model. Figure~\ref{fig:delta_bic} shows the temporal evolution of these BIC differences.

\begin{figure}[htbp]
\centering
\includegraphics[width=\textwidth]{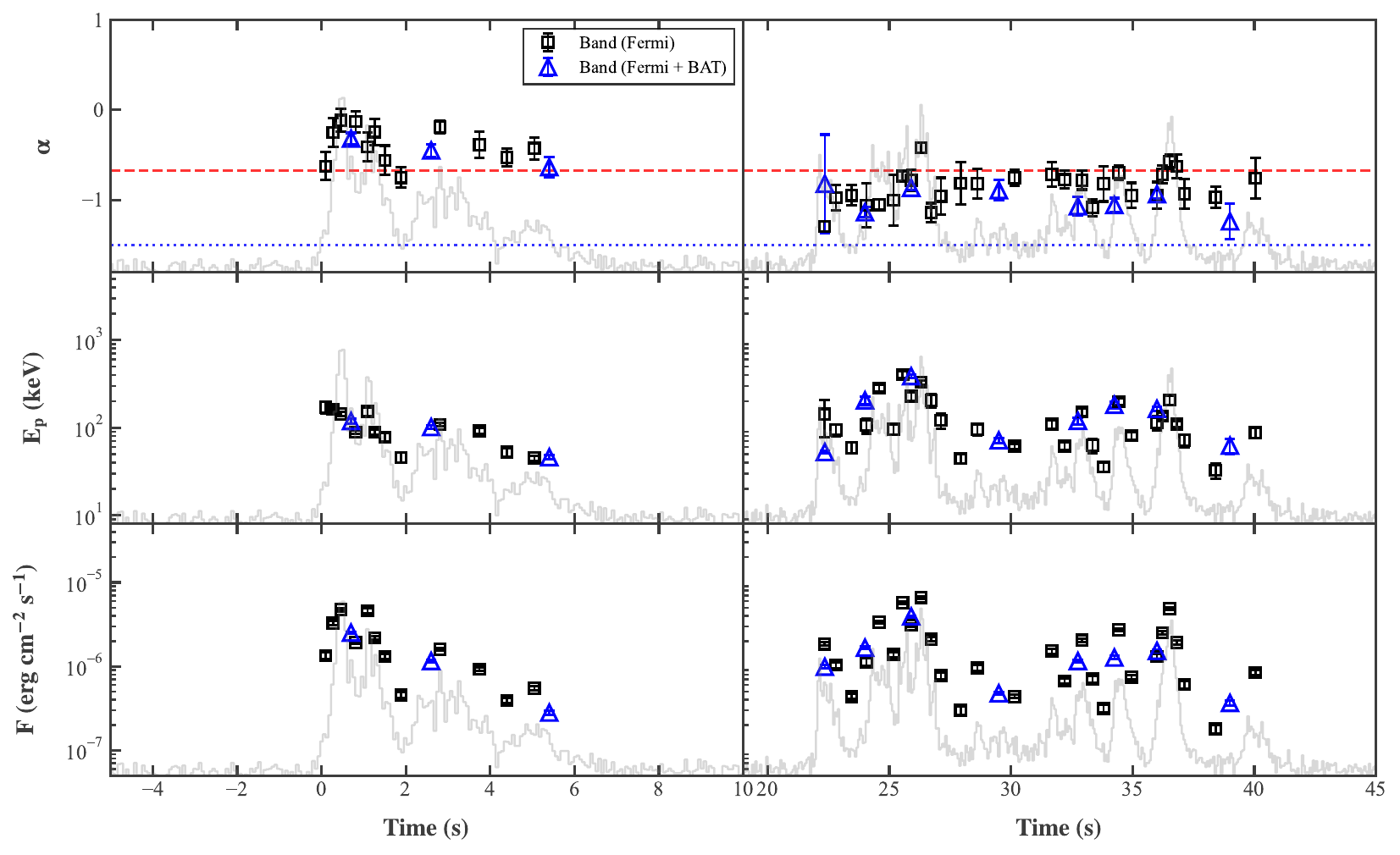}
\caption{Temporal evolution of the Band spectral parameters for GRB 250920C. Black squares show the \textit{Fermi}/GBM-only fits, and blue triangles show the joint \textit{Fermi}/GBM+\textit{Swift}/BAT fits. The left and right columns correspond to EI and EII, respectively.}
\label{fig:param_evol}
\end{figure}

Figure~\ref{fig:param_evol} compares the temporal evolution of the Band low-energy photon index $\alpha$, peak energy $E_{\rm p}$, and energy flux $F$ obtained from the GBM-only and joint fits. The two analyses show consistent overall trends despite their different temporal resolutions.

In EI, the Band and CPL models are strongly disfavored relative to BB+PL in the bright GBM-only intervals. For all 12 EI intervals with $S>10$, $\mathrm{BIC}_{\rm Band}-\mathrm{BIC}_{\rm BB+PL}$ ranges from 144.19 to 245.56, while $\mathrm{BIC}_{\rm CPL}-\mathrm{BIC}_{\rm BB+PL}$ ranges from 119.53 to 219.62. The single mBB model is also generally not preferred, indicating that the thermal-like component does not describe the full spectrum by itself but is accompanied by a non-thermal tail. In Figure~\ref{fig:param_evol}, $\alpha$ is harder than the synchrotron line of death, $\alpha=-2/3$, in most bright EI intervals, while $E_{\rm p}$ evolves predominantly from hard to soft. These properties are consistent with an additional thermal contribution in EI.

\begin{figure}[htbp]
\centering
\includegraphics[width=0.7\textwidth]{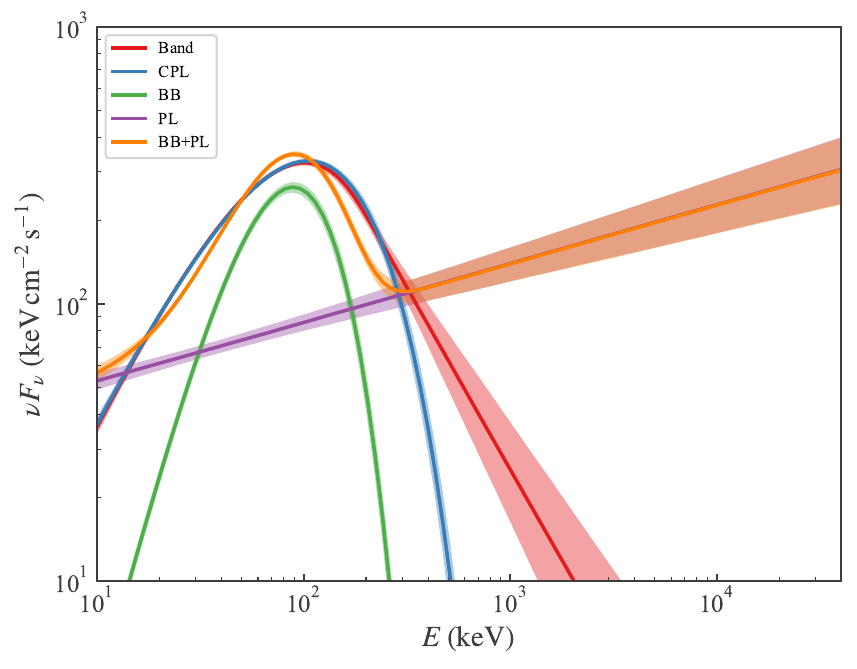}
\caption{Comparison of the Band, CPL, and BB+PL spectral-energy distributions for the second joint EI interval, $1.4$--$3.8$~s. The BB and PL components of the BB+PL model are plotted separately, together with their summed spectrum. The shaded regions show the propagated spectral uncertainties.}
\label{fig:ei_sed_comparison}
\end{figure}

The joint fits independently support this model preference. All three joint EI intervals favor BB+PL: its BIC improvements over Band are 105.84, 148.47, and 186.53, and those over CPL are 103.99, 121.99, and 170.31, respectively. Figure~\ref{fig:ei_sed_comparison} illustrates the spectral-shape differences among Band, CPL, and BB+PL for the representative $1.4$--$3.8$~s interval.

\begin{figure}[p]
\centering
\begin{minipage}[t]{0.48\textwidth}
\centering
\includegraphics[width=\textwidth]{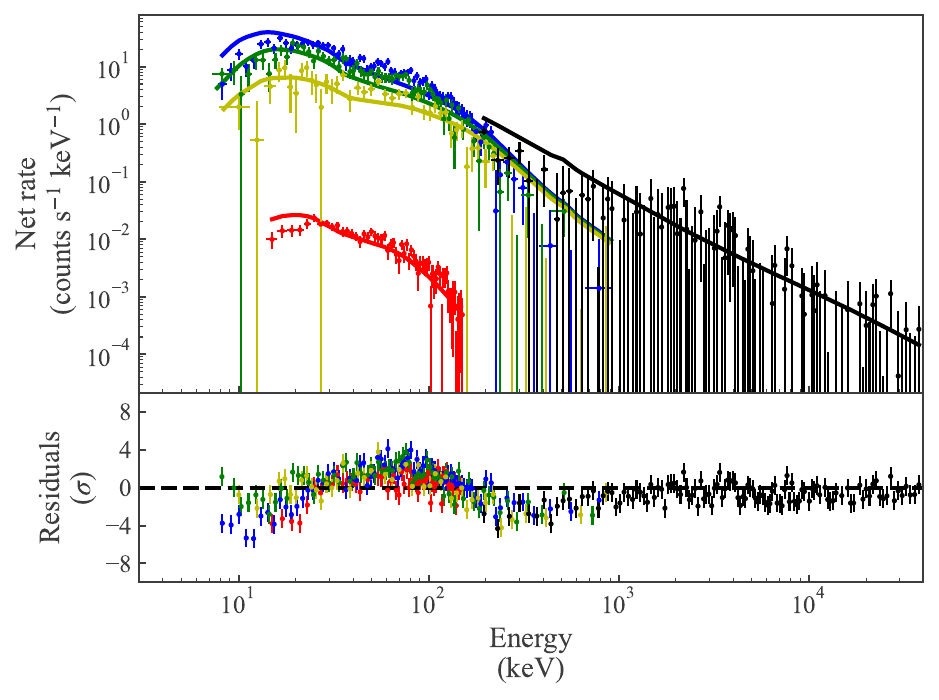}
\par\smallskip
\textbf{(a)}
\end{minipage}
\hfill
\begin{minipage}[t]{0.48\textwidth}
\centering
\includegraphics[width=\textwidth]{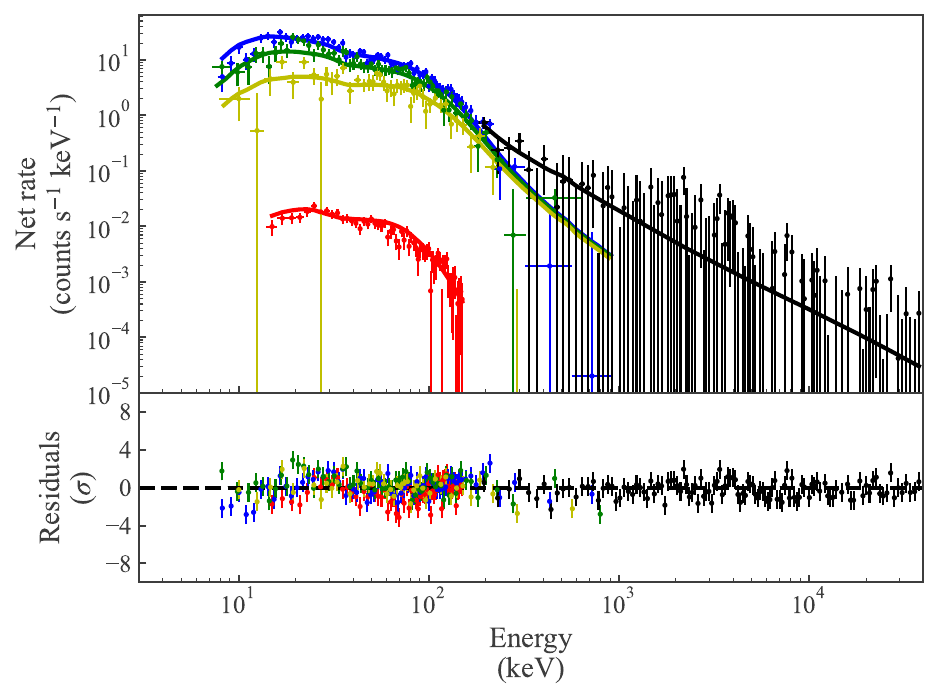}
\par\smallskip
\textbf{(b)}
\end{minipage}
\par\medskip
\begin{minipage}[t]{0.34\textwidth}
\centering
\includegraphics[width=\textwidth]{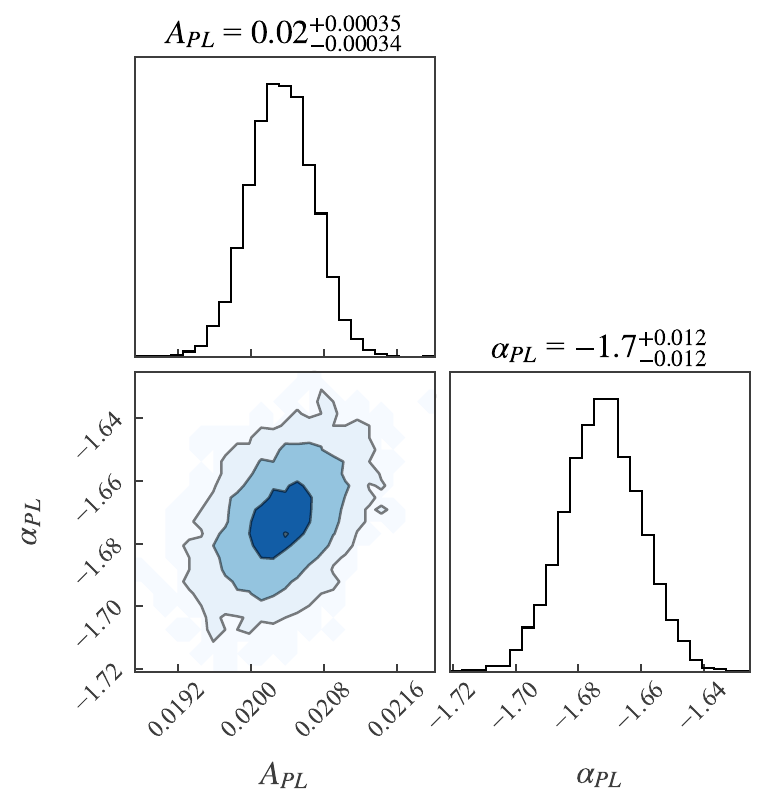}
\par\smallskip
\textbf{(c)}
\end{minipage}
\hfill
\begin{minipage}[t]{0.54\textwidth}
\centering
\includegraphics[width=\textwidth]{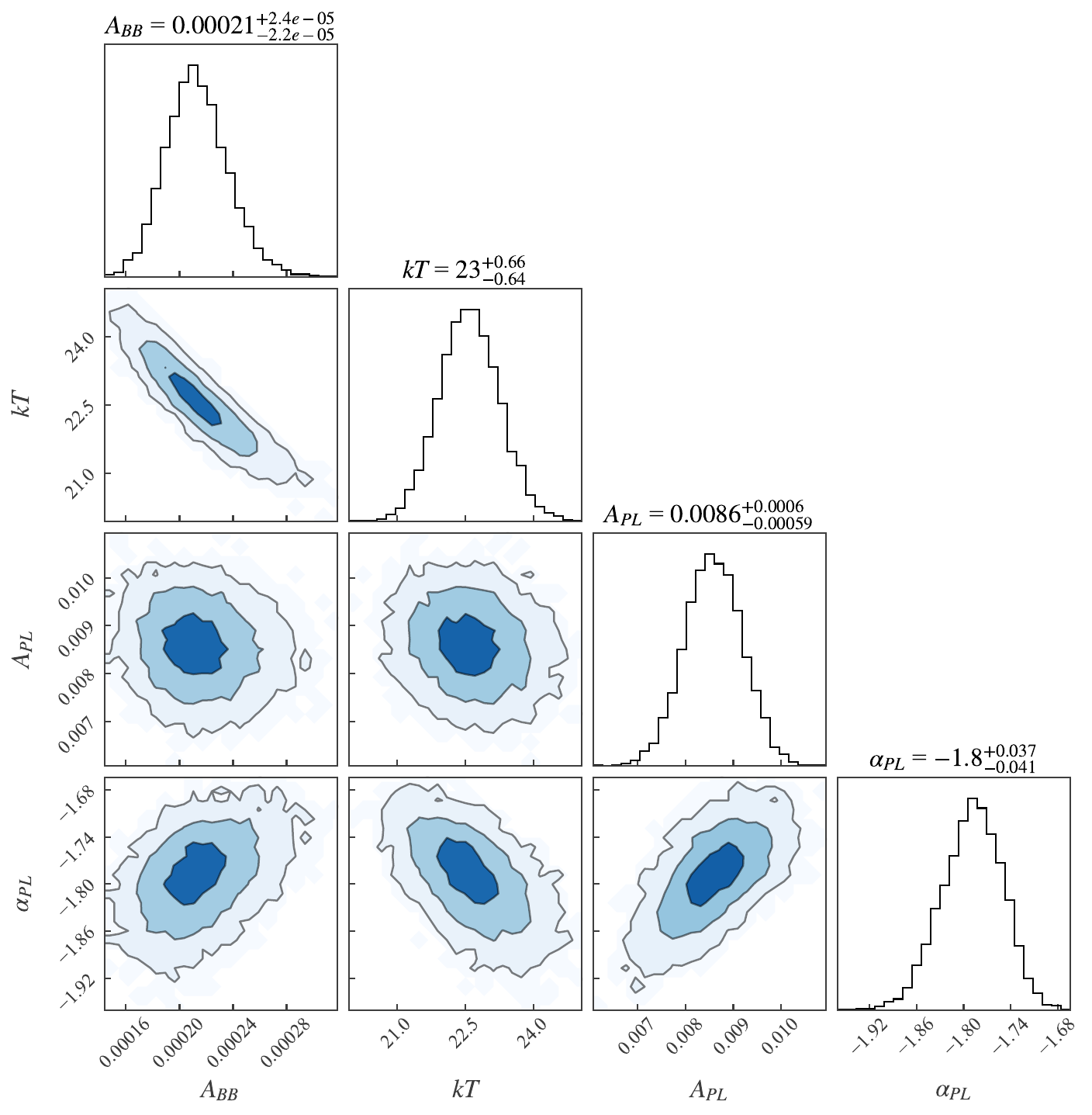}
\par\smallskip
\textbf{(d)}
\end{minipage}
\caption{Comparison of the independent PL-only and BB+PL fits for the second joint EI interval, $1.4$--$3.8$~s. Panels (a) and (b) show the observed count spectra, folded model predictions, and standardized residuals for PL-only and BB+PL, respectively. Panels (c) and (d) show their corresponding posterior distributions. The PL-only parameters are constrained, but its broad, coherent residual pattern demonstrates that a single power law cannot reproduce the spectral curvature. The BB+PL fit removes this structure and yields compact posterior distributions for the BB normalization $A_{\rm BB}$, temperature $kT$, PL normalization $A_{\rm PL}$, and photon index $\alpha_{PL}$.}
\label{fig:ei_bbpl_fit}
\end{figure}

For a more direct assessment of the additional thermal component, Figure~\ref{fig:ei_bbpl_fit} compares independent PL-only and BB+PL fits to the same interval. The PL-only parameters are constrained, but the fit leaves broad, coherent residuals, showing that a single power law cannot reproduce the spectral curvature. BB+PL removes this residual structure and yields compact posterior distributions for both components.

Although the spectra have sufficient signal for the simpler models to be constrained, CPL+BB and Band+BB do not yield identifiable additional blackbody components. Their posteriors are broad or multimodal, and the BB normalization is strongly degenerate with $kT$, $E_{\rm p}$, and the continuum indices. The inferred Band+BB blackbody fraction is only approximately 0.9--5.5\%. In contrast, the localized BB peak and scale-free PL component in BB+PL are more readily separated.

Motivated by the low-energy-break interpretation proposed by \citet{2017ApJ...846..137O,2018A&A...616A.138O}, we also fitted the three joint EI spectra with the double smoothly broken power-law model (2SBPL; \citealt{2018A&A...613A..16R}), which tests for an additional spectral break below $E_{\rm p}$. The 2SBPL fits do not yield an independently constrained additional break. The posterior distributions of $E_{\rm break}$ are broad and strongly asymmetric, with upper uncertainties extending over several hundred keV, and $E_{\rm break}$ is strongly degenerate with the two low-energy photon indices and $E_{\rm p}$. No statistically well-separated pair of $E_{\rm break}$ and $E_{\rm p}$ is recovered. Moreover, the 2SBPL BIC values exceed those of BB+PL by 107.57, 186.99, and 181.41 in the three joint EI intervals, respectively. Thus, the available spectra do not provide positive evidence for a distinct low-energy break within the observed energy range. This result does not exclude a break at a few keV: the breaks reported by \citet{2018A&A...616A.138O} were primarily revealed using \textit{Swift}/XRT data extending to approximately 0.3~keV, whereas our GBM and BAT spectra do not provide comparable coverage below 10~keV. Among the converged and statistically identifiable models tested here, BB+PL therefore remains the best-supported description of EI.

EII shows a different spectral behavior. Its BIC differences are more moderate and vary with time: MDFSYN is favored over Band or CPL in several intervals, whereas the preference is less decisive in others. Most EII intervals do not significantly exceed the synchrotron line of death, although a small number of short intervals show harder values. In addition, $E_{\rm p}$ rises near several sub-pulse peaks and decreases during the corresponding decays, producing an intensity-tracking pattern rather than the predominantly hard-to-soft evolution seen in EI.

\begin{figure}[p]
\centering
\includegraphics[width=0.95\textwidth]{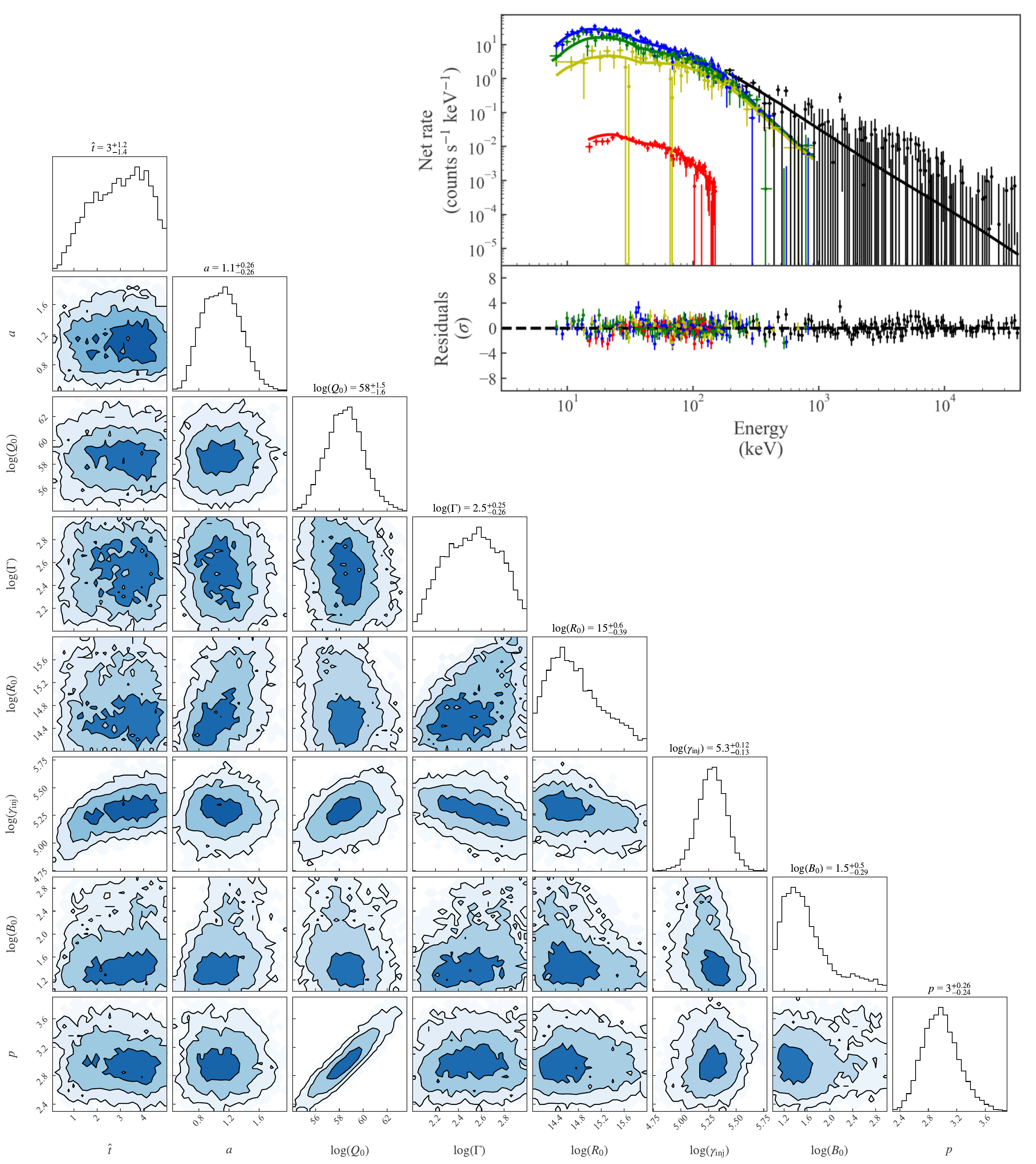}
\caption{Representative MDFSYN fit to the joint \textit{Fermi}/GBM+\textit{Swift}/BAT data for the second EII interval, $23.2$--$24.8$~s. The upper-right panel shows the observed count spectra, folded model predictions, and standardized residuals, while the lower-left corner plot shows the posterior distributions of the physical MDFSYN parameters.}
\label{fig:eii_mdfs_fit}
\end{figure}

Taken together, the low-energy slopes and the model comparisons indicate that MDFSYN provides a physically motivated and statistically competitive description of EII, but it is not statistically preferred in every interval. Figure~\ref{fig:eii_mdfs_fit} presents the representative joint GBM+BAT fit and its posterior correlations for the $23.2$--$24.8$~s interval.

The contrast between the best-supported BB+PL description of EI and the synchrotron-compatible EII spectra therefore suggests that the two episodes are characterized by different dominant emission components.

\clearpage

\subsection{Spectral components: thermal emission}

\begin{figure}[htbp]
\centering
\begin{minipage}[t]{\textwidth}
\centering
\includegraphics[width=0.7\textwidth]{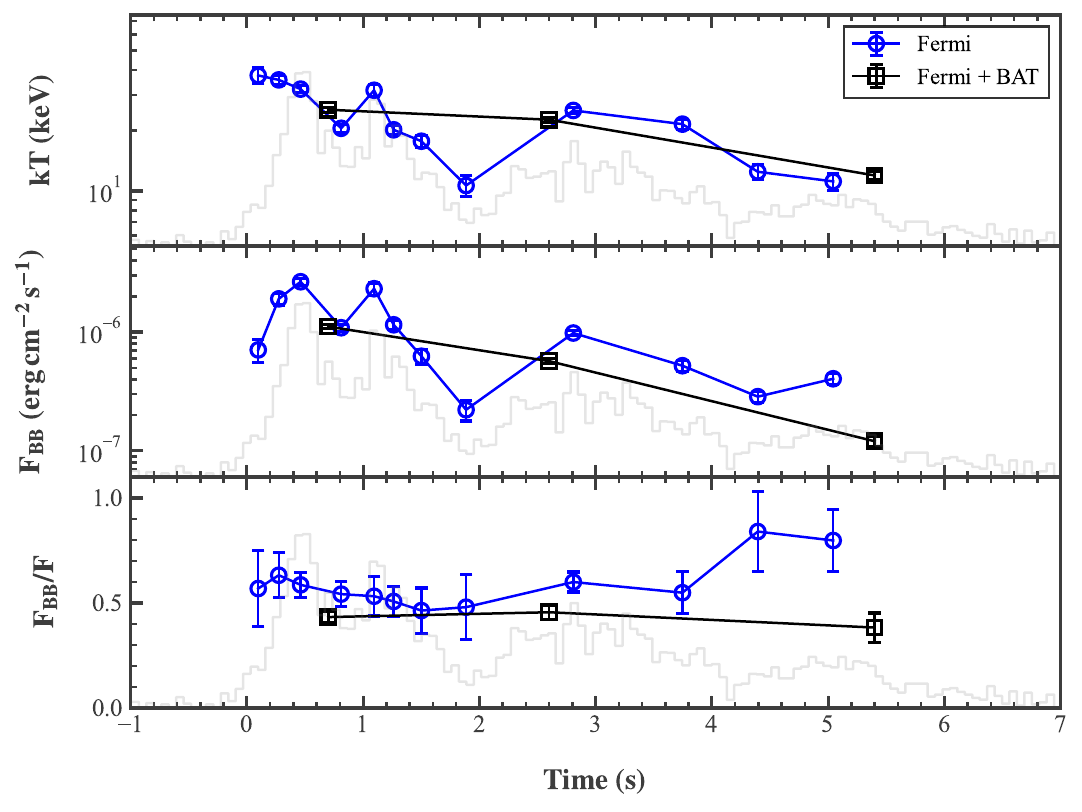}
\end{minipage}
\caption{Temporal evolution of the thermal-component parameters in EI. The top, middle, and bottom panels show the blackbody temperature $kT$, blackbody flux $F_{\rm BB}$, and thermal flux fraction $F_{\rm BB}/F$, respectively. Blue circles represent the \textit{Fermi}/GBM-only results, while black squares denote the joint \textit{Fermi}/GBM+\textit{Swift}/BAT fits. \label{fig:thermal_evol}}
\end{figure}

As shown in Section \ref{sec4.1}, the time-resolved spectra of EI reveal a clear additional thermal component. To investigate its temporal evolution, we calculated the blackbody temperature $kT$, the blackbody flux $F_{\rm BB}$, and the ratio of the blackbody flux to the total observed flux, $F_{\rm BB}/F$. The evolution of these parameters is shown in Figure~\ref{fig:thermal_evol}.

 For the \textit{Fermi}/GBM-only fits, the blackbody temperature $kT$ is initially several tens of keV and then generally decreases as the main pulse evolves, with a moderate increase around the later substructure. The blackbody flux $F_{\rm BB}$ broadly follows the light-curve evolution, increasing during the bright phase and decreasing during the decay phase. The thermal flux fraction $F_{\rm BB}/F$ remains high in most time bins, typically contributing about half or more of the total observed flux.

The joint \textit{Fermi}/GBM+\textit{Swift}/BAT fits are broadly consistent with the GBM-only results. For the three joint EI intervals, the BB+PL temperatures are $25.38_{-0.62}^{+0.75}$, $22.64_{-0.68}^{+0.63}$, and $11.95_{-0.92}^{+0.61}$~keV, and the corresponding thermal fractions are 0.433, 0.456, and 0.383. Thus, the joint-fit values of $kT$ and $F_{\rm BB}$ show an overall decreasing trend with time, while $F_{\rm BB}/F$ remains at approximately 0.4. Because the joint fits use broader time intervals, their parameter evolution is smoother than that from the GBM-only analysis, but the overall trend remains consistent.

\subsection{Correlations among spectral parameters}

\begin{figure}[htbp]
\centering

\begin{minipage}[t]{0.33\textwidth}
\centering
\includegraphics[width=\textwidth]{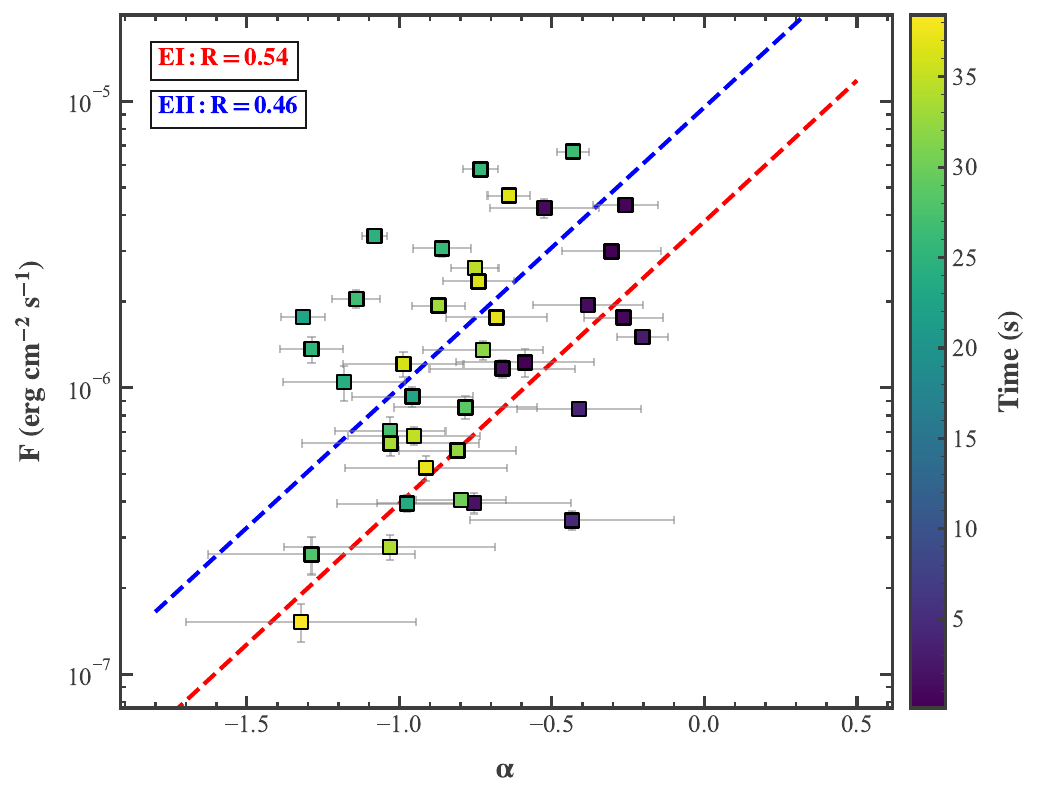}
\end{minipage}
\hfill
\begin{minipage}[t]{0.33\textwidth}
\centering
\includegraphics[width=\textwidth]{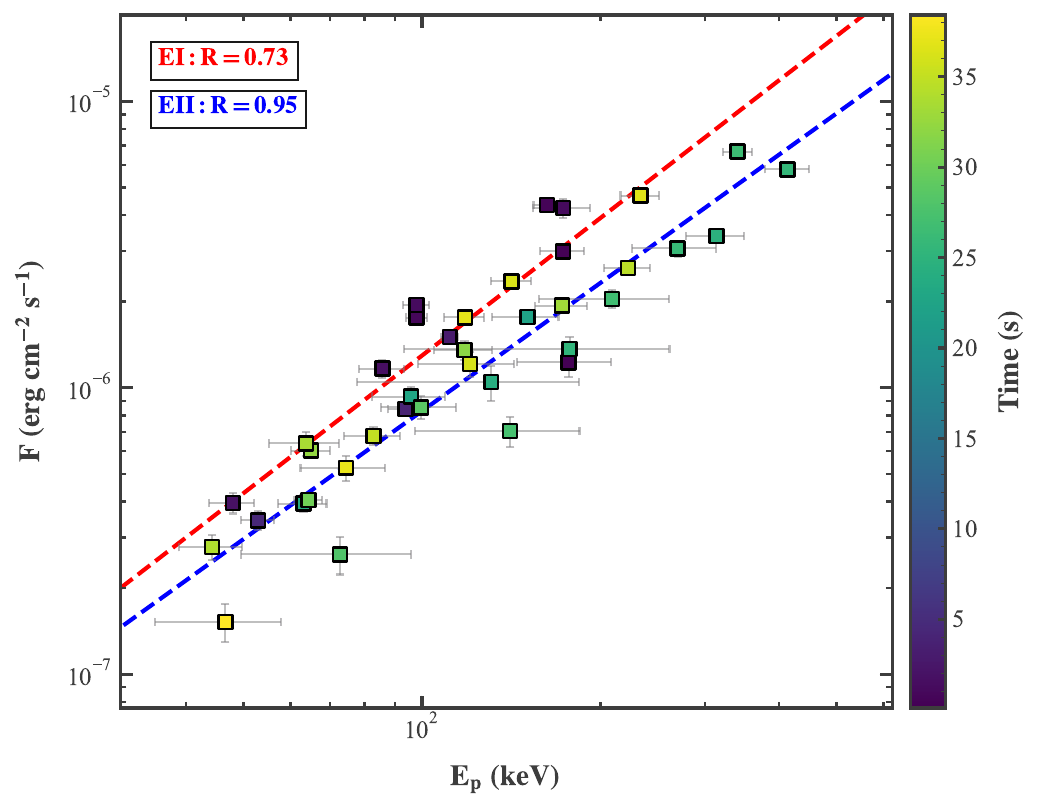}
\end{minipage}
\hfill
\begin{minipage}[t]{0.33\textwidth}
\centering
\includegraphics[width=\textwidth]{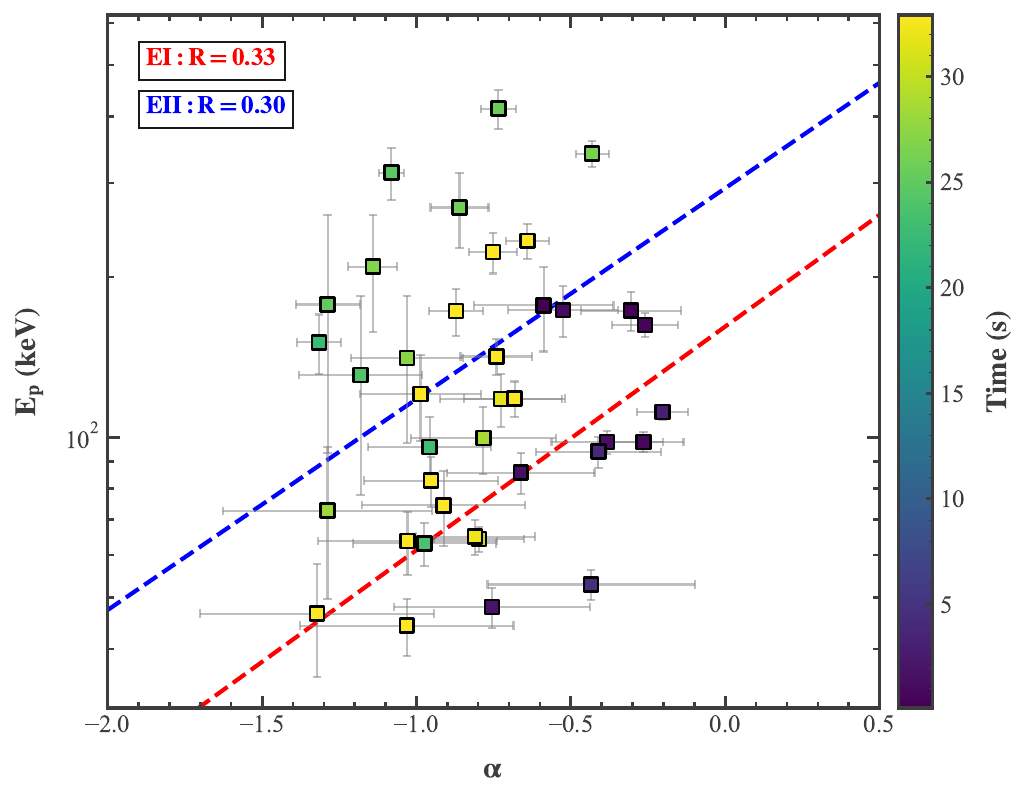}
\end{minipage}

\caption{Correlations among the spectral parameters of GRB 250920C. From left to right, the panels show the $F$--$\alpha$, $F$--$E_{\rm p}$, and $E_{\rm p}$--$\alpha$ relations, respectively. Red and blue dashed lines represent the best-fit relations for EI and EII, and the color scale indicates the time since the trigger.}
\label{fig:corr}
\end{figure}

To further investigate the relation between spectral evolution and radiation intensity, we examined the correlations of $F$--$\alpha$, $F$--$E_{\rm p}$, and $E_{\rm p}$--$\alpha$ for EI and EII, as shown in Figure~\ref{fig:corr}. The red and blue dashed lines represent the best-fit relations for EI and EII, respectively. The correlation between $F$ and $\alpha$ is moderate, with correlation coefficients of $R=0.54$ for EI and $R=0.46$ for EII. This indicates that the spectrum tends to become harder as the emission becomes brighter, although the evolution of the low-energy photon index is not fully controlled by the flux. In contrast, $F$ shows a much stronger positive correlation with $E_{\rm p}$, with $R=0.73$ for EI and $R=0.95$ for EII. In particular, the close coupling between $E_{\rm p}$ and the flux in EII indicates a clear intensity-tracking behavior.

On the other hand, the correlation between $E_{\rm p}$ and $\alpha$ is weak, with $R=0.33$ for EI and $R=0.30$ for EII, suggesting that the peak energy and the low-energy spectral slope do not evolve synchronously. Overall, the $F$--$E_{\rm p}$ relation is the most significant correlation, whereas correlations involving $\alpha$ are relatively weak. This further suggests that the spectral evolution in EII is mainly governed by the instantaneous radiation conditions, while the low-energy spectral shape in EI may be affected by an additional thermal component.

\section{Physical parameter constraints}
\label{sec5}

\subsection{Physical constraints in the fireball framework}

The identification of a significant quasi-thermal component in EI provides an opportunity to constrain the physical properties of the outflow within the fireball framework. Once a blackbody component is identified in the observed prompt spectrum, the photospheric parameters can be estimated from the observed blackbody flux and temperature. Following the method of \citet{2007ApJ...664L...1P}, we first define the effective transverse size parameter as
\begin{equation}
\mathcal{R} =
\left(
\frac{F_{\rm BB}^{\rm obs}}
{\sigma_{\rm SB} (T^{obs})^{4}}
\right)^{1/2},
\end{equation}
where $F_{\rm BB}^{\rm obs}$ is the observed blackbody flux, $T^{obs}$ is the observed blackbody temperature, and $\sigma_{\rm SB}$ is the Stefan--Boltzmann constant.

For a baryonic fireball whose photosphere lies in the coasting regime, the photospheric Lorentz factor and radius are estimated following \citet{2007ApJ...664L...1P} as
\begin{equation}
\Gamma_{\rm ph}=\left[\frac{1.06(1+z)^2d_LYF^{\rm obs}\sigma_{\rm T}}
{2m_{\rm p}c^3\mathcal{R}}\right]^{1/4},
\end{equation}
\textbf{and}
\begin{equation}
r_{\rm ph}=\frac{1.06d_L}{(1+z)^2}\mathcal{R}\Gamma_{\rm ph}.
\end{equation}
The initial radius of the outflow can be estimated as
\begin{equation}
r_{0} =
\frac{4^{3/2}}{(1.48)^6(1.06)^4}
\frac{d_{L}}{(1+z)^2}
\left(\frac{F_{\rm BB}^{\rm obs}}{YF^{\rm obs}}\right)^{3/2}
\mathcal{R}.
\end{equation}
In a standard baryonic fireball, the flow accelerates until its Lorentz factor approaches the dimensionless entropy $\eta=L/(\dot{M}c^2)$. The corresponding saturation radius is \citep{2007ApJ...664L...1P,2015AdAst2015E..22P}
\begin{equation}
r_{\rm s}=r_0\eta\simeq r_0\Gamma_{\rm ph},
\end{equation}
where the last approximation applies when the photosphere is above the saturation radius and the flow is coasting.
Here, $d_{L}$ is the luminosity distance, $z$ is the redshift, $F^{\rm obs}$ is the total observed flux, $\sigma_{\rm T}$ is the Thomson cross section, $m_{\rm p}$ is the proton mass, and $Y$ is the ratio between the total fireball energy and the gamma-ray radiated energy. We adopt $Y=1$ throughout this calculation.

\begin{figure}[htbp]
\centering
\begin{minipage}[t]{\textwidth}
\centering
\includegraphics[width=0.7\textwidth]{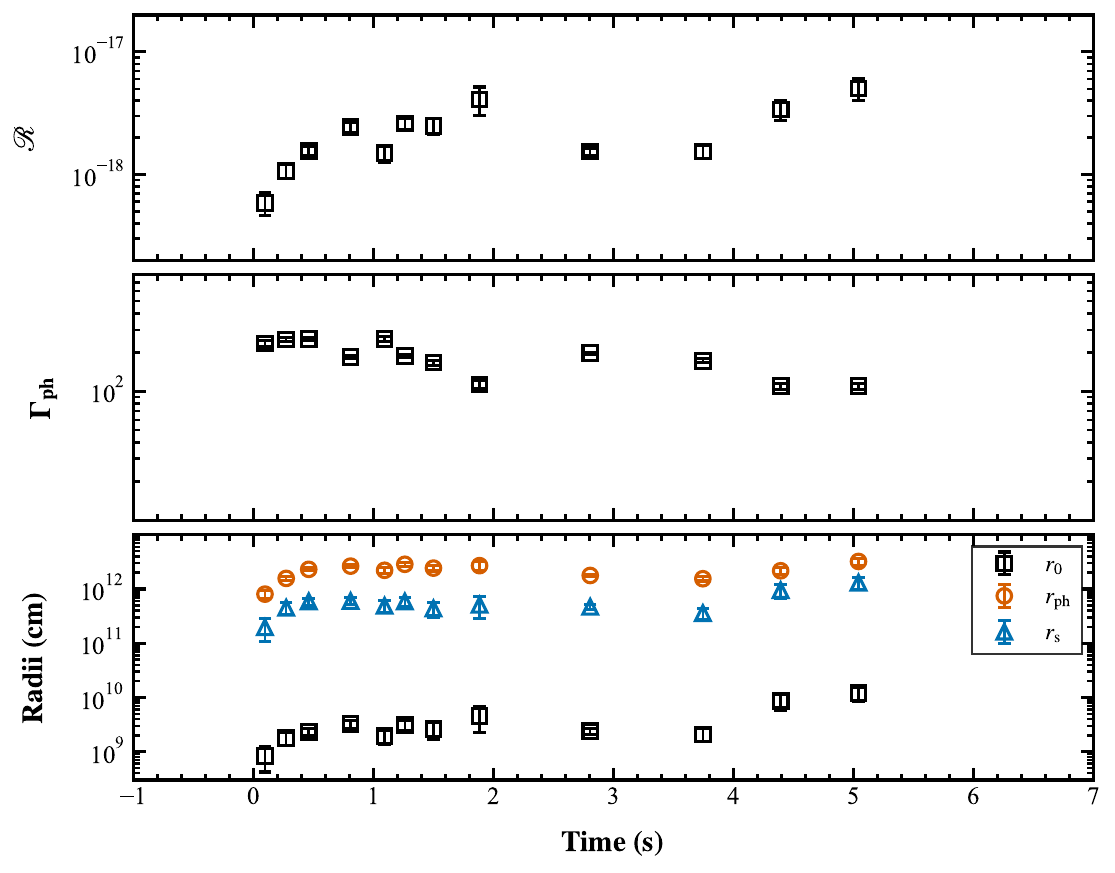}
\end{minipage}
\caption{Temporal evolution of the fireball parameters derived from the thermal component in EI. The top and middle panels show the effective transverse size parameter $\mathcal{R}$ and the photospheric Lorentz factor $\Gamma_{\rm ph}$, respectively. The bottom panel compares the initial radius $r_0$ (black squares), the photospheric radius $r_{\rm ph}$ (orange circles), and the saturation radius $r_{\rm s}=r_0\Gamma_{\rm ph}$ (blue triangles). \label{fig:fireball_param}}
\end{figure}

Using the time-resolved BB+PL fitting results, we calculated the photospheric parameters for the EI time bins with significant thermal emission. Figure~\ref{fig:fireball_param} shows the temporal evolution of $\mathcal{R}$ and $\Gamma_{\rm ph}$, together with a direct comparison of the three characteristic radii $r_0$, $r_{\rm ph}$, and $r_{\rm s}$. The effective transverse size parameter $\mathcal{R}$ remains at the level of $\sim10^{-18}$ and shows moderate fluctuations with time. The photospheric Lorentz factor $\Gamma_{\rm ph}$ is typically of the order of a few hundred, while the initial radius $r_{0}$ is mainly distributed around $\sim10^{9}$--$10^{10}$ cm. These values are consistent with a compact central engine and a relativistic photospheric outflow. For the 12 EI bins used in this calculation, we obtain $r_{\rm ph}=8.05\times10^{11}$--$3.22\times10^{12}$~cm and $r_{\rm s}=1.97\times10^{11}$--$1.31\times10^{12}$~cm. As shown directly in the bottom panel of Figure~\ref{fig:fireball_param}, the orange $r_{\rm ph}$ points lie above the blue $r_{\rm s}$ points in every time bin. The inferred radii are therefore consistent with the photosphere being above the saturation radius within the adopted $Y=1$ baryonic-fireball framework.

This ordering is consistent with previous studies of GRBs containing prominent thermal components. For GRB~061007, \citet{2011MNRAS.414.2642L} similarly verified that the photosphere was above the saturation radius before applying the coasting-regime fireball relations. For GRB~090902B, \citet{2010ApJ...709L.172R} inferred a characteristic photospheric radius of $(1.1\pm0.3)\times10^{12}$~cm, comparable to the range found here. In a sample of bright single-pulse \textit{Fermi} bursts, \citet{2016MNRAS.456.2157I} found photospheric radii clustered around $10^{11.8\pm0.4}$~cm and discussed their evolution under the same assumption that the photosphere lies above the coasting radius. In this regime, the outflow has reached its terminal Lorentz factor before becoming transparent, and the thermal radiation subsequently experiences adiabatic cooling \citep{2012MNRAS.420..468P,2015AdAst2015E..22P}. The coexistence of thermal and non-thermal components in EI is compatible with this picture. Nevertheless, the comparison remains a consistency check within the adopted fireball model rather than model-independent proof, because the inferred $r_0$ and $\Gamma_{\rm ph}$ themselves rely on the coasting-photosphere relations \citep{2007ApJ...664L...1P}.

We further applied the ``top-down'' diagnostic method proposed by \citet{2015ApJ...801..103G} to estimate the central-engine parameters. This method uses the observed blackbody temperature $kT^{\rm obs}$, blackbody flux $F_{\rm BB}$, total flux $F^{\rm obs}$, and initial radius $r_{0}$ to constrain the dimensionless entropy $\eta$ and the initial magnetization parameter $\sigma_{0}$. In different photospheric regimes, $\eta$ can be written as
\begin{equation}
\eta =
\begin{cases}
74.8(1+z)^{11/12}
\left(\frac{kT^{\rm obs}}{50~{\rm keV}}\right)^{11/12}
\left(\frac{F_{\rm BB}}{10^{-8}~{\rm erg~s^{-1}~cm^{-2}}}\right)^{1/48}
r_{0,9}^{5/24} d_{L,28}^{1/24},
& {\rm Regime~II}, \\

20.3(1+z)^{-5/6}
\left(\frac{kT^{\rm obs}}{30~{\rm keV}}\right)^{-5/6}
\left(\frac{F_{\rm BB}}{10^{-7}~{\rm erg~s^{-1}~cm^{-2}}}\right)^{11/24}
r_{0,9}^{-2/3} f_{{\rm th},-1}^{3/4} f_{\gamma}^{3/4}
d_{L,28}^{11/12},
& {\rm Regime~III~and~VI}, \\

105.01(1+z)^{7/6}
\left(\frac{kT^{\rm obs}}{10~{\rm keV}}\right)^{7/6}
\left(\frac{F_{\rm BB}}{10^{-9}~{\rm erg~s^{-1}~cm^{-2}}}\right)^{5/24}
r_{0,9}^{1/12} f_{{\rm th},-1}^{1/2} f_{\gamma}^{1/2}
d_{L,28}^{5/12},
& {\rm Regime~V}.
\end{cases}
\end{equation}
The corresponding initial magnetization parameter is given by
\begin{equation}
1+\sigma_{0} =
\begin{cases}
25.5(1+z)^{4/3}
\left(\frac{kT^{\rm obs}}{50~{\rm keV}}\right)^{4/3}
\left(\frac{F_{\rm BB}}{10^{-8}~{\rm erg~s^{-1}~cm^{-2}}}\right)^{-1/3}
r_{0,9}^{2/3} f_{{\rm th},-1}^{-1} f_{\gamma}^{-1}
d_{L,28}^{-2/3},
& {\rm Regime~II}, \\

5.99(1+z)^{4/3}
\left(\frac{kT^{\rm obs}}{30~{\rm keV}}\right)^{4/3}
\left(\frac{F_{\rm BB}}{10^{-7}~{\rm erg~s^{-1}~cm^{-2}}}\right)^{-1/3}
r_{0,9}^{2/3} f_{{\rm th},-1}^{-1} f_{\gamma}^{-1}
d_{L,28}^{-2/3},
& {\rm Regime~III~and~VI}, \\

6.43(1+z)^{4/3}
\left(\frac{kT^{\rm obs}}{10~{\rm keV}}\right)^{4/3}
\left(\frac{F_{\rm BB}}{10^{-9}~{\rm erg~s^{-1}~cm^{-2}}}\right)^{-1/3}
r_{0,9}^{2/3} f_{{\rm th},-1}^{-1} f_{\gamma}^{-1}
d_{L,28}^{-2/3},
& {\rm Regime~V}.
\end{cases}
\end{equation}
Here, $r_{0,9}=r_{0}/10^{9}~{\rm cm}$ and $d_{L,28}=d_{L}/10^{28}~{\rm cm}$. 
The parameter $f_{\rm th}$ denotes the observed thermal fraction, which is obtained from the time-resolved BB+PL fits as
$f_{\rm th}=F_{\rm BB}/F$, where $F_{\rm BB}$ is the blackbody flux and $F$ is the total observed flux in the same time bin. 
Thus, the normalized quantity used above is $f_{\rm th,-1}=f_{\rm th}/0.1$. 
We adopt a fiducial radiative efficiency of $f_{\gamma}=0.5$ \citep{2015ApJ...801..103G}. 
In the calculation, we used the time-resolved $r_{0}$ values derived above rather than assuming a fixed initial radius.

\begin{figure}[htbp]
\centering
\begin{minipage}[t]{\textwidth}
\centering
\includegraphics[width=0.6\textwidth]{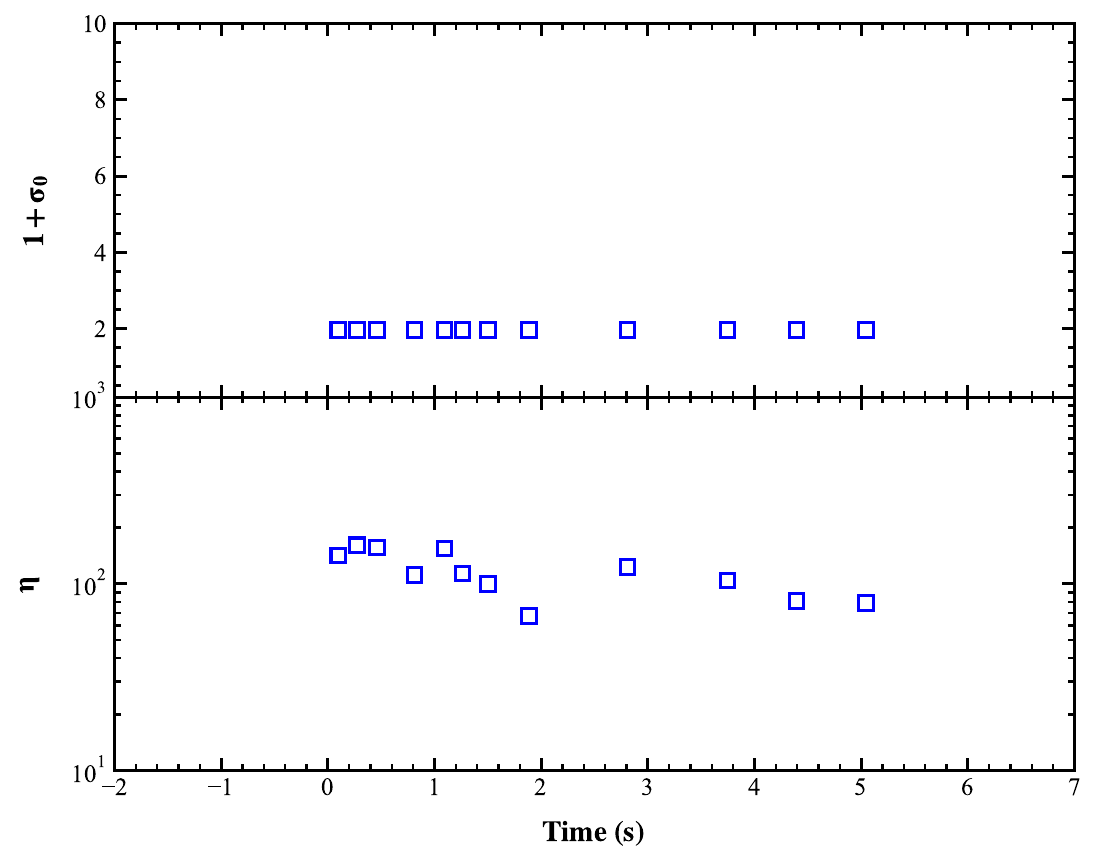}
\end{minipage}
\caption{Temporal evolution of the central-engine parameters inferred from the top-down method. The top and bottom panels show $1+\sigma_{0}$ and the dimensionless entropy $\eta$, respectively. The values of $1+\sigma_{0}$ remain of order unity, while $\eta$ is much larger than unity, indicating a thermally dominated outflow during EI. \label{fig:engine_param}}
\end{figure}

The resulting temporal evolution of $1+\sigma_{0}$ and $\eta$ is shown in Figure~\ref{fig:engine_param}. The inferred values of $1+\sigma_{0}$ remain of order unity throughout EI, indicating a low degree of magnetization. Meanwhile, $\eta$ is much larger than unity, with typical values of order $10^{2}$. These results suggest that the EI outflow is dominated by thermal energy rather than Poynting flux. This conclusion is consistent with the prominent photospheric-like component identified in the time-resolved spectral analysis and supports a fireball-dominated interpretation for the early prompt emission of GRB 250920C.

\subsection{Constraints on the photospheric magnetization parameter}

As shown by the time-resolved analysis in Section \ref{sec4.1}, no statistically significant thermal component is identified in EII, whose spectra are dominated by non-thermal emission. One possible explanation is that the photospheric emission is suppressed in a magnetized outflow \citep{2009ApJ...700L..65Z,2011ApJ...726...90Z}. We examine this possibility following this physical prescription, as subsequently applied to GRB~230307A, GRB~221009A, and GRB~161117A \citep{2024MNRAS.529L..67D,2023ApJ...947L..11Y,2026ApJ...998...24C}. Following \citet{2009ApJ...700L..65Z}, we define the magnetization at the photosphere as $\sigma_{\rm ph}\equiv L_{\rm P}/L_{\rm b}$, where $L_{\rm P}$ and $L_{\rm b}$ are the Poynting-flux and baryonic luminosities at the photosphere, respectively. This quantity is distinct from the central-engine magnetization $\sigma_{0}$ discussed for EI.

For each EII interval, we obtained the isotropic gamma-ray luminosity $L_{\gamma}$ from the joint GBM+BAT Band fit and conservatively set the wind luminosity to $L_{\rm w}=L_{\gamma}$. Assuming that the Poynting flux is not dissipated below the photosphere, the luminosity available to the baryonic photosphere is $L_{\rm b}=L_{\rm w}/(1+\sigma_{\rm ph})$. We adopted $r_{0}=10^{9}$ cm and the maximum-temperature fireball case, in which the photospheric radius is close to the saturation radius, $r_{\rm ph}\simeq r_{\rm s}=r_{0}\eta_{*}$, where $\eta_{*}=[L_{\rm b}\sigma_{\rm T}/(8\pi m_{\rm p}c^{3}r_{0})]^{1/4}$. This choice fixes the hottest pseudo-thermal component allowed by the adopted nondissipative baryonic-fireball prescription; the blackbody temperature is therefore not a free parameter and is not selected by visual tangency. Defining the unmagnetized reference temperature as $T_{0}^{\rm obs}=[L_{\rm w}/(4\pi r_{0}^{2}\sigma_{\rm SB})]^{1/4}(1+z)^{-1}$, for each trial $\sigma_{\rm ph}$ we calculated the temperature as $T_{\rm BB}^{\rm obs}(\sigma_{\rm ph})=T_{0}^{\rm obs}(1+\sigma_{\rm ph})^{-1/4}$ and the bolometric flux as $F_{\rm BB}(\sigma_{\rm ph})=L_{\rm w}/[4\pi d_{L}^{2}(1+\sigma_{\rm ph})]$. The pseudo-thermal SED was then written as
\begin{equation}
(\nu F_{\nu})_{\rm BB}(E)=A_{\rm BB}
\frac{E^{4}}{\exp[E/(kT_{\rm BB}^{\rm obs})]-1},
\end{equation}
where $A_{\rm BB}$ was determined numerically from $\int(\nu F_{\nu})_{\rm BB}\,d\ln E=F_{\rm BB}$. Thus both the temperature and normalization were recalculated self-consistently for every trial $\sigma_{\rm ph}$.

We used the joint \textit{Fermi}/GBM+\textit{Swift}/BAT Band fits as the reference non-thermal spectra in EII. Both the Band and pseudo-thermal spectra were evaluated as $\nu F_{\nu}=E^{2}N(E)$. For each trial $\sigma_{\rm ph}$, we recalculated the temperature, normalization, and Planck spectrum. We define $\sigma_{\min}$ as the minimum value of $\sigma_{\rm ph}$ for which the pseudo-thermal spectrum does not exceed the Band spectrum at any energy. The apparent tangency is therefore the numerical boundary of this condition rather than a temperature selected to force tangency.

A blackbody lying below the Band curve may still produce a statistically detectable spectral feature. The present comparison only determines whether the predicted pseudo-thermal spectrum exceeds the observed total spectrum. We therefore interpret $\sigma_{\min}$ as a conservative, model-dependent minimum constraint rather than a formal statistical limit.

\begin{figure}[htbp]
\centering
\begin{minipage}[t]{\textwidth}
\centering
\includegraphics[width=0.6\textwidth]{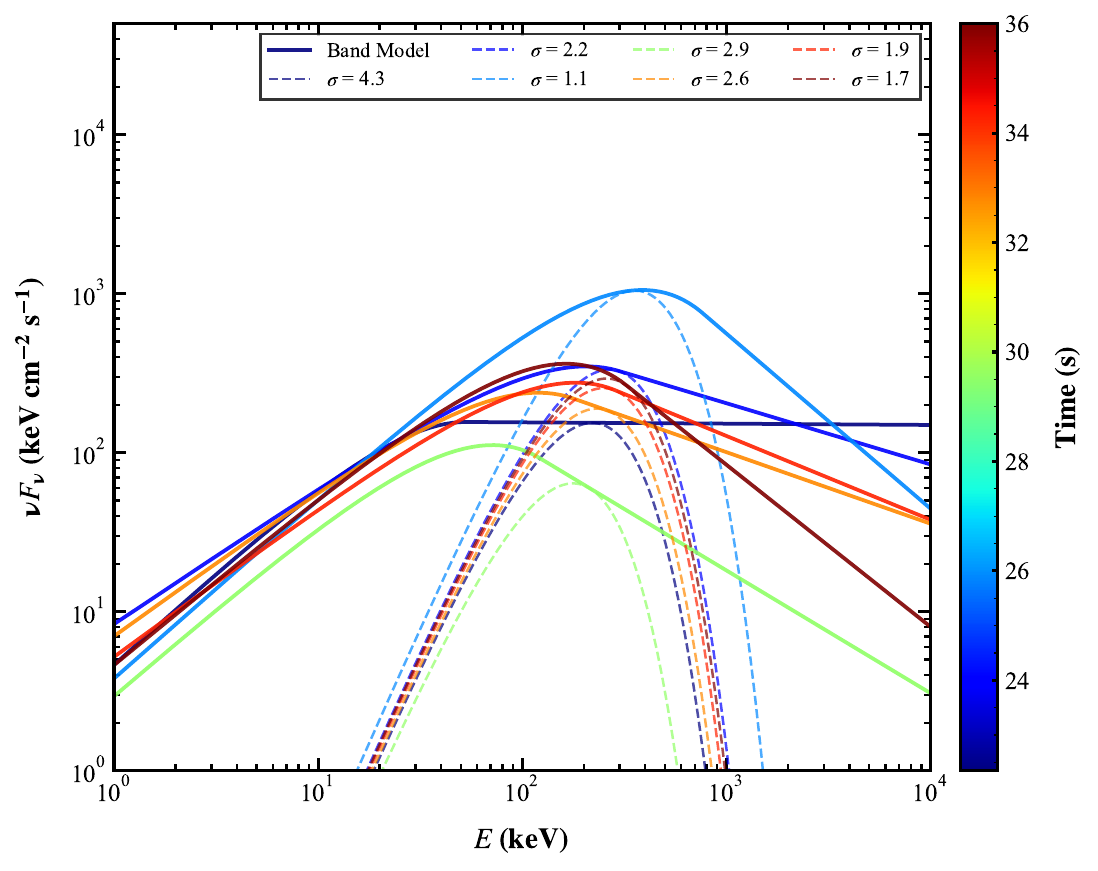}
\end{minipage}
\caption{Envelope-based constraints on the photospheric magnetization $\sigma_{\rm ph}$ for the EII spectra of GRB 250920C. Solid curves show the Band spectra from the joint \textit{Fermi}/GBM+\textit{Swift}/BAT fits. Dashed curves show the boundary pseudo-thermal spectra corresponding to $\sigma_{\min}$, the minimum value of $\sigma_{\rm ph}$ for which the blackbody curve nowhere exceeds the Band curve. These curves define a conservative, model-dependent envelope condition rather than formal statistical limits. \label{fig:sigma_constraint}}
\end{figure}

Figure~\ref{fig:sigma_constraint} shows the constraints for the different EII intervals. We obtain $\sigma_{\min}\sim1.1$--$4.3$. Since $\sigma_{\rm ph}=L_{\rm P}/L_{\rm b}$, these values correspond to minimum Poynting-flux fractions $f_{\rm P}=\sigma_{\rm ph}/(1+\sigma_{\rm ph})\simeq0.52$--$0.81$ and suggest that EII may be Poynting-flux dominated. However, $\sigma_{\min}$ is a model-dependent envelope constraint rather than a formal statistical limit. It is also a photospheric constraint and should not be compared directly with the central-engine magnetization $\sigma_{0}$ inferred for EI.

\section{Discussion and Summary}
 \label{sec6}
GRB 250920C was jointly observed by \textit{Fermi}/GBM and \textit{Swift}/BAT. Its prompt emission shows two distinct emission episodes, EI and EII, separated by a significant quiescent interval. In this work, we performed a systematic temporal and spectral analysis of the two episodes using Bayesian MCMC inference. The main analysis results are summarized as follows:

\begin{enumerate}
\item GRB 250920C exhibits two distinct prompt-emission episodes, EI and EII, separated by a significant quiescent interval. Its overall temporal properties are broadly consistent with those of long GRBs.

\item EI shows clear evidence for an additional thermal component. The BB+PL model is preferred in most bright time bins, and the thermal component remains significant in the joint \textit{Fermi}/GBM+\textit{Swift}/BAT fits.

\item The thermal-component parameters in EI show systematic evolution. The blackbody temperature generally decreases with time, the blackbody flux follows the pulse structure, and the thermal flux fraction remains high, indicating an important contribution from photospheric emission to the early prompt emission.

\item The fireball-parameter estimates indicate that EI has a photospheric Lorentz factor of the order of a few hundred and an initial radius of $r_{0}\sim10^{9}$--$10^{10}$ cm. The inferred $1+\sigma_{0}$ is of order unity and $\eta\gg1$, supporting a low-magnetization, thermally dominated baryonic fireball outflow.

\item EII is mainly dominated by non-thermal emission. Its low-energy photon indices do not significantly exceed the synchrotron line of death, and the spectra are well described by the fast-cooling synchrotron model in a decaying magnetic field.

\item The diagnostic gives $\sigma_{\min}\sim1.1$--$4.3$. Since all values exceed unity, they suggest that EII may be Poynting-flux dominated.
\end{enumerate}

These results suggest that the two emission episodes of GRB 250920C may be governed by different dominant radiation mechanisms. In EI, the significant thermal component, the high thermal flux fraction, and the fireball-parameter estimates all support a photospheric-like thermal-emission interpretation. The low magnetization and high dimensionless entropy indicate that the outflow in this phase is mainly driven by thermal energy, and EI can therefore be naturally interpreted as a baryonic fireball-dominated photospheric emission phase.

In contrast, EII shows more prominent non-thermal spectral properties. The low-energy photon index $\alpha$ does not significantly exceed the synchrotron line of death and is broadly consistent with the expectation of fast-cooling synchrotron radiation in a decaying magnetic field. The MDFSYN model provides a good description of the time-resolved spectra in EII, while the strong $F$--$E_{\rm p}$ correlation indicates a clear intensity-tracking behavior. This suggests that the spectral evolution in EII is mainly controlled by the instantaneous physical conditions in the emitting region.

The magnetization diagnostic gives $\sigma_{\min}\sim1.1$--$4.3$. Since $\sigma_{\min}>1$ in every interval, EII may be Poynting-flux dominated. We therefore tentatively interpret GRB~250920C as transitioning from a fireball-dominated EI phase to a possibly Poynting-flux-dominated EII phase.

The inferred evolution of GRB~250920C has observational precedents. The clearest example is GRB~160625B, whose prompt emission comprises three isolated episodes separated by long quiescent intervals. Its short first episode contains a prominent photospheric component, whereas the subsequent bright episode has a featureless non-thermal spectrum and was interpreted as arising from a Poynting-flux-dominated outflow \citep{2017ApJ...849...71L,2018NatAs...2...69Z}. In a broader study of 43 clear multipulse \textit{Fermi} GRBs, \citet{2019ApJS..242...16L} identified 9 events (about 21\%) that may evolve from an early thermal-like pulse to later non-thermal pulses. Detailed time-resolved analysis further showed that three of four suitable events contain an early thermal-like precursor followed by emission compatible with optically thin synchrotron radiation. Related spectral sequences have also been reported in GRB~161117A \citep{2026ApJ...998...24C} and GRB~240619A \citep{2025ApJ...993..147C}. The latter was interpreted in a compact-merger context and therefore provides only an observational comparison. Despite differences in progenitor and temporal structure, these events show that distinct prompt-emission episodes can probe different physical states of a GRB jet.

Within a collapsar interpretation, a hyperaccreting black hole offers a physically motivated route for such an evolution \citep{1993ApJ...405..273W,1999ApJ...524..262M}. At an early high accretion rate, $\nu\bar{\nu}$ annihilation above the disk can efficiently launch a thermally dominated baryonic fireball and produce a photospheric component such as that observed in EI \citep{1999ApJ...518..356P,2002ApJ...579..706D,2013ApJ...762..102L,2007ApJ...657..383C,2015ApJS..218...12L,2024ApJ...975..225L}. \citet{2017ApJ...849...47L} showed that an initially slowly rotating black hole may first produce a neutrino-annihilation-powered fireball and, after being spun up by accretion, subsequently launch a more powerful Poynting-flux-dominated jet through the Blandford--Znajek mechanism. As the neutrino-annihilation power weakens and dynamically important magnetic flux develops near the black hole, the Blandford--Znajek process can become increasingly important \citep{1977MNRAS.179..433B,2013ApJ...765..125L}. This sequence naturally connects the thermal-rich EI outflow with the more strongly magnetized outflow inferred for EII.

The quiescent interval adds an episodic dimension to this picture. In prompt-emission models, such intervals can arise either from a genuine interruption of the central engine or from a continuously operating but strongly modulated outflow. In the latter case, successive ejecta may have an unfavorable Lorentz-factor distribution and fail to dissipate efficiently in the observable gamma-ray band \citep{2001MNRAS.324.1147R}. Both situations would appear observationally as a temporary decrease of the gamma-ray luminosity below the detection threshold, so the quiescent interval does not by itself specify the state of the central engine.

A possible connection with the evolving jet composition is provided by magnetic regulation of the accretion flow. Magnetic flux accumulated around the black hole can temporarily impede inner accretion, allowing matter to build up outside the magnetic barrier until accretion and energy extraction resume \citep{2006MNRAS.370L..61P}. Likewise, transitions into and out of a magnetically arrested disk can produce active episodes separated by quiescent intervals \citep{2016MNRAS.461.1045L}. Combined with the central-engine evolution discussed by \citet{2017ApJ...849...47L}, this suggests a self-consistent sequence in which EI is powered mainly by a neutrino-annihilation fireball, the gamma-ray output then decreases while the inner accretion flow and magnetic-flux configuration evolve, and EII is launched after magnetic extraction has become more important. This scenario is not unique, because outflow modulation without complete engine dormancy remains viable. The evidence for changing jet composition therefore rests primarily on the contrasting spectral components and inferred outflow properties of EI and EII, while the quiescent interval indicates that the evolution occurred episodically.

In summary, GRB 250920C shows a transition from a thermal-rich, fireball-compatible EI phase to a non-thermal, synchrotron-compatible EII phase. Since the $\sigma_{\min}$ values exceed unity, EII may be Poynting-flux dominated. These results therefore suggest a possible transition from a fireball-dominated jet to a Poynting-flux-dominated jet.

\begingroup
\def\IncludeMainTables{}
\ifdefined\IncludeMainTables
\begin{landscape}
\begingroup
\tiny
\renewcommand{\arraystretch}{0.92}
\setlength{\tabcolsep}{1.0pt}
\begin{longtable}{llcccccccccccc}
\caption{Time-resolved Band, CPL, mBB, and BB+PL spectral-fitting results
using \textit{Fermi}/GBM-only data.\label{tab:fermi_empirical}}\\
\hline\hline
Time & Model & $\alpha$ & $\beta$ & $E_{\rm p}$ & $m$ & $kT_{\min}$
& $kT_{\max}/kT$ & $\alpha_{\rm PL}$ & $F$ & BIC & PG-stat/dof
& RSS/dof & PPC\\
(s) & & & & (keV) & & (keV) & (keV) &
& $(10^{-6}\,\mathrm{erg\,cm^{-2}\,s^{-1}})$ & & & &\\
\hline
\endfirsthead

\caption[]{Time-resolved empirical-model results using
\textit{Fermi}/GBM-only data (continued).}\\
\hline\hline
Time & Model & $\alpha$ & $\beta$ & $E_{\rm p}$ & $m$ & $kT_{\min}$
& $kT_{\max}/kT$ & $\alpha_{\rm PL}$ & $F$ & BIC & PG-stat/dof
& RSS/dof & PPC\\
(s) & & & & (keV) & & (keV) & (keV) &
& $(10^{-6}\,\mathrm{erg\,cm^{-2}\,s^{-1}})$ & & & &\\
\hline
\endhead

\hline
\multicolumn{14}{r}{\textit{Continued on next page}}\\
\endfoot

\hline
\endlastfoot

$0.001\!\sim\!0.194$ & Band & $-0.63_{-0.16}^{+0.26}$ & $-2.65_{-0.43}^{+0.32}$ & $171.66_{-25.79}^{+32.18}$ & $\cdots$ & $\cdots$ & $\cdots$ & $\cdots$ & $1.345_{-0.122}^{+0.145}$ & $595.99$ & $571.25/481$ & $295.95/481$ & $0.594$\\
 & CPL & $-0.58_{-0.18}^{+0.22}$ & $\cdots$ & $177.61_{-23.01}^{+29.59}$ & $\cdots$ & $\cdots$ & $\cdots$ & $\cdots$ & $1.228_{-0.125}^{+0.130}$ & $589.21$ & $570.66/482$ & $292.95/482$ & $0.542$\\
 & mBB & $\cdots$ & $\cdots$ & $\cdots$ & $0.29_{-0.22}^{+0.29}$ & $0.75_{-0.05}^{+3.08}$ & $72.32_{-10.92}^{+14.39}$ & $\cdots$ & $1.167_{-0.104}^{+0.118}$ & $633.88$ & $609.15/481$ & $308.35/481$ & $0.518$\\
 & BB+PL & $\cdots$ & $\cdots$ & $\cdots$ & $\cdots$ & $\cdots$ & $37.59_{-5.03}^{+3.46}$ & $-1.72_{-0.27}^{+0.16}$ & $1.244_{-0.127}^{+0.131}$ & $434.54$ & $409.81/481$ & $303.39/481$ & $0.532$\\
\hline
$0.194\!\sim\!0.355$ & Band & $-0.25_{-0.16}^{+0.15}$ & $-2.72_{-0.43}^{+0.20}$ & $161.47_{-13.38}^{+16.97}$ & $\cdots$ & $\cdots$ & $\cdots$ & $\cdots$ & $3.262_{-0.186}^{+0.197}$ & $575.36$ & $550.62/481$ & $299.58/481$ & $0.677$\\
 & CPL & $-0.30_{-0.13}^{+0.16}$ & $\cdots$ & $172.38_{-12.32}^{+15.18}$ & $\cdots$ & $\cdots$ & $\cdots$ & $\cdots$ & $2.988_{-0.175}^{+0.186}$ & $565.65$ & $547.10/482$ & $314.68/482$ & $0.573$\\
 & mBB & $\cdots$ & $\cdots$ & $\cdots$ & $0.62_{-0.23}^{+0.16}$ & $0.84_{-0.17}^{+2.76}$ & $66.09_{-4.28}^{+9.29}$ & $\cdots$ & $2.910_{-0.164}^{+0.160}$ & $602.02$ & $577.29/481$ & $318.38/481$ & $0.545$\\
 & BB+PL & $\cdots$ & $\cdots$ & $\cdots$ & $\cdots$ & $\cdots$ & $35.73_{-2.45}^{+2.04}$ & $-1.67_{-0.12}^{+0.10}$ & $3.010_{-0.166}^{+0.174}$ & $411.67$ & $386.93/481$ & $322.21/481$ & $0.546$\\
\hline
$0.355\!\sim\!0.570$ & Band & $-0.12_{-0.13}^{+0.13}$ & $-2.70_{-0.32}^{+0.13}$ & $143.10_{-8.57}^{+10.78}$ & $\cdots$ & $\cdots$ & $\cdots$ & $\cdots$ & $4.740_{-0.196}^{+0.201}$ & $911.44$ & $886.70/481$ & $384.26/481$ & $0.293$\\
 & CPL & $-0.26_{-0.10}^{+0.11}$ & $\cdots$ & $162.53_{-8.51}^{+8.71}$ & $\cdots$ & $\cdots$ & $\cdots$ & $\cdots$ & $4.352_{-0.165}^{+0.175}$ & $902.41$ & $883.85/482$ & $400.02/482$ & $0.134$\\
 & mBB & $\cdots$ & $\cdots$ & $\cdots$ & $0.59_{-0.57}^{+0.02}$ & $0.92_{-0.79}^{+6.77}$ & $63.93_{-1.19}^{+15.50}$ & $\cdots$ & $4.296_{-0.174}^{+0.175}$ & $936.58$ & $911.85/481$ & $397.97/481$ & $0.107$\\
 & BB+PL & $\cdots$ & $\cdots$ & $\cdots$ & $\cdots$ & $\cdots$ & $32.15_{-1.43}^{+1.24}$ & $-1.60_{-0.04}^{+0.07}$ & $4.521_{-0.161}^{+0.167}$ & $761.56$ & $736.82/481$ & $390.43/481$ & $0.138$\\
\hline
$0.570\!\sim\!1.053$ & Band & $-0.13_{-0.12}^{+0.18}$ & $-3.03_{-0.27}^{+0.20}$ & $89.07_{-4.92}^{+5.40}$ & $\cdots$ & $\cdots$ & $\cdots$ & $\cdots$ & $1.924_{-0.081}^{+0.086}$ & $1565.77$ & $1541.03/481$ & $344.90/481$ & $0.573$\\
 & CPL & $-0.27_{-0.09}^{+0.13}$ & $\cdots$ & $98.15_{-4.48}^{+3.72}$ & $\cdots$ & $\cdots$ & $\cdots$ & $\cdots$ & $1.755_{-0.060}^{+0.062}$ & $1545.66$ & $1527.10/482$ & $363.64/482$ & $0.395$\\
 & mBB & $\cdots$ & $\cdots$ & $\cdots$ & $-0.21_{-0.50}^{+0.33}$ & $6.52_{-1.27}^{+1.57}$ & $49.07_{-5.22}^{+11.13}$ & $\cdots$ & $1.760_{-0.060}^{+0.069}$ & $1584.81$ & $1560.07/481$ & $363.03/481$ & $0.505$\\
 & BB+PL & $\cdots$ & $\cdots$ & $\cdots$ & $\cdots$ & $\cdots$ & $20.52_{-0.79}^{+0.90}$ & $-1.69_{-0.08}^{+0.06}$ & $2.005_{-0.083}^{+0.085}$ & $1410.51$ & $1385.78/481$ & $346.78/481$ & $0.393$\\
\hline
$1.053\!\sim\!1.136$ & Band & $-0.41_{-0.16}^{+0.18}$ & $-2.63_{-0.38}^{+0.22}$ & $154.04_{-17.23}^{+18.76}$ & $\cdots$ & $\cdots$ & $\cdots$ & $\cdots$ & $4.590_{-0.303}^{+0.317}$ & $209.27$ & $184.54/481$ & $284.28/481$ & $0.910$\\
 & CPL & $-0.53_{-0.11}^{+0.19}$ & $\cdots$ & $173.73_{-18.06}^{+18.48}$ & $\cdots$ & $\cdots$ & $\cdots$ & $\cdots$ & $4.232_{-0.274}^{+0.321}$ & $200.84$ & $182.29/482$ & $288.03/482$ & $0.818$\\
 & mBB & $\cdots$ & $\cdots$ & $\cdots$ & $0.44_{-0.65}^{+0.05}$ & $0.80_{-0.20}^{+6.52}$ & $65.53_{-2.43}^{+27.94}$ & $\cdots$ & $4.109_{-0.271}^{+0.318}$ & $243.44$ & $218.71/481$ & $303.51/481$ & $0.782$\\
 & BB+PL & $\cdots$ & $\cdots$ & $\cdots$ & $\cdots$ & $\cdots$ & $31.73_{-2.30}^{+2.33}$ & $-1.60_{-0.10}^{+0.07}$ & $4.351_{-0.256}^{+0.256}$ & $61.24$ & $36.50/481$ & $286.88/481$ & $0.841$\\
\hline
$1.136\!\sim\!1.389$ & Band & $-0.24_{-0.15}^{+0.17}$ & $-2.87_{-0.31}^{+0.19}$ & $89.00_{-6.10}^{+6.41}$ & $\cdots$ & $\cdots$ & $\cdots$ & $\cdots$ & $2.179_{-0.119}^{+0.120}$ & $957.48$ & $932.74/481$ & $362.45/481$ & $0.191$\\
 & CPL & $-0.37_{-0.12}^{+0.16}$ & $\cdots$ & $97.31_{-5.06}^{+5.48}$ & $\cdots$ & $\cdots$ & $\cdots$ & $\cdots$ & $1.945_{-0.089}^{+0.093}$ & $944.14$ & $925.59/482$ & $381.27/482$ & $0.071$\\
 & mBB & $\cdots$ & $\cdots$ & $\cdots$ & $0.51_{-0.87}^{+0.02}$ & $0.71_{-1.04}^{+5.92}$ & $39.64_{-0.95}^{+13.03}$ & $\cdots$ & $1.920_{-0.087}^{+0.096}$ & $972.86$ & $948.12/481$ & $386.02/481$ & $0.073$\\
 & BB+PL & $\cdots$ & $\cdots$ & $\cdots$ & $\cdots$ & $\cdots$ & $20.17_{-1.10}^{+1.01}$ & $-1.69_{-0.08}^{+0.08}$ & $2.268_{-0.120}^{+0.121}$ & $798.71$ & $773.98/481$ & $368.15/481$ & $0.154$\\
\hline
$1.389\!\sim\!1.614$ & Band & $-0.56_{-0.16}^{+0.24}$ & $-2.77_{-0.34}^{+0.20}$ & $77.97_{-8.23}^{+7.51}$ & $\cdots$ & $\cdots$ & $\cdots$ & $\cdots$ & $1.312_{-0.093}^{+0.105}$ & $737.42$ & $712.69/481$ & $293.47/481$ & $0.724$\\
 & CPL & $-0.65_{-0.15}^{+0.23}$ & $\cdots$ & $86.37_{-8.28}^{+6.97}$ & $\cdots$ & $\cdots$ & $\cdots$ & $\cdots$ & $1.162_{-0.074}^{+0.083}$ & $726.52$ & $707.97/482$ & $306.67/482$ & $0.594$\\
 & mBB & $\cdots$ & $\cdots$ & $\cdots$ & $0.32_{-1.23}^{+0.05}$ & $0.70_{-0.90}^{+5.70}$ & $34.87_{-0.05}^{+30.36}$ & $\cdots$ & $1.139_{-0.073}^{+0.091}$ & $765.36$ & $740.62/481$ & $307.97/481$ & $0.596$\\
 & BB+PL & $\cdots$ & $\cdots$ & $\cdots$ & $\cdots$ & $\cdots$ & $17.69_{-1.73}^{+1.19}$ & $-1.81_{-0.12}^{+0.11}$ & $1.348_{-0.115}^{+0.115}$ & $593.23$ & $568.50/481$ & $300.23/481$ & $0.617$\\
\hline
$1.614\!\sim\!2.158$ & Band & $-0.75_{-0.11}^{+0.37}$ & $-2.83_{-0.37}^{+0.20}$ & $45.67_{-6.06}^{+3.34}$ & $\cdots$ & $\cdots$ & $\cdots$ & $\cdots$ & $0.453_{-0.036}^{+0.045}$ & $1521.90$ & $1497.16/481$ & $368.61/481$ & $0.373$\\
 & CPL & $-0.76_{-0.14}^{+0.34}$ & $\cdots$ & $47.93_{-4.47}^{+4.35}$ & $\cdots$ & $\cdots$ & $\cdots$ & $\cdots$ & $0.395_{-0.030}^{+0.032}$ & $1510.95$ & $1492.39/482$ & $373.70/482$ & $0.269$\\
 & mBB & $\cdots$ & $\cdots$ & $\cdots$ & $0.27_{-1.73}^{+0.03}$ & $0.88_{-0.43}^{+4.01}$ & $20.67_{-1.11}^{+32.47}$ & $\cdots$ & $0.385_{-0.031}^{+0.033}$ & $1544.78$ & $1520.05/481$ & $373.69/481$ & $0.257$\\
 & BB+PL & $\cdots$ & $\cdots$ & $\cdots$ & $\cdots$ & $\cdots$ & $10.65_{-0.97}^{+1.30}$ & $-0.94_{-1.29}^{+0.85}$ & $0.462_{-0.059}^{+0.060}$ & $1317.89$ & $1293.16/481$ & $377.10/481$ & $0.299$\\
\hline
$2.158\!\sim\!3.458$ & Band & $-0.19_{-0.07}^{+0.11}$ & $-3.53_{-0.25}^{+0.30}$ & $108.68_{-4.28}^{+3.44}$ & $\cdots$ & $\cdots$ & $\cdots$ & $\cdots$ & $1.591_{-0.045}^{+0.048}$ & $2767.44$ & $2742.70/481$ & $358.80/481$ & $0.936$\\
 & CPL & $-0.22_{-0.07}^{+0.09}$ & $\cdots$ & $111.65_{-3.48}^{+3.17}$ & $\cdots$ & $\cdots$ & $\cdots$ & $\cdots$ & $1.503_{-0.038}^{+0.039}$ & $2733.23$ & $2714.68/482$ & $358.64/482$ & $0.933$\\
 & mBB & $\cdots$ & $\cdots$ & $\cdots$ & $0.56_{-0.18}^{+0.08}$ & $0.66_{-0.20}^{+3.41}$ & $45.59_{-1.55}^{+3.05}$ & $\cdots$ & $1.482_{-0.035}^{+0.036}$ & $2760.70$ & $2735.97/481$ & $359.92/481$ & $0.906$\\
 & BB+PL & $\cdots$ & $\cdots$ & $\cdots$ & $\cdots$ & $\cdots$ & $25.18_{-0.70}^{+0.80}$ & $-1.78_{-0.07}^{+0.05}$ & $1.636_{-0.050}^{+0.052}$ & $2613.70$ & $2588.97/481$ & $383.60/481$ & $0.722$\\
\hline
$3.458\!\sim\!4.035$ & Band & $-0.39_{-0.14}^{+0.19}$ & $-3.01_{-0.38}^{+0.19}$ & $91.48_{-7.02}^{+5.72}$ & $\cdots$ & $\cdots$ & $\cdots$ & $\cdots$ & $0.929_{-0.051}^{+0.062}$ & $1647.04$ & $1622.30/481$ & $358.42/481$ & $0.622$\\
 & CPL & $-0.41_{-0.12}^{+0.20}$ & $\cdots$ & $94.31_{-6.61}^{+5.87}$ & $\cdots$ & $\cdots$ & $\cdots$ & $\cdots$ & $0.843_{-0.043}^{+0.046}$ & $1632.06$ & $1613.51/482$ & $359.97/482$ & $0.516$\\
 & mBB & $\cdots$ & $\cdots$ & $\cdots$ & $0.48_{-0.34}^{+0.15}$ & $0.82_{-0.01}^{+3.00}$ & $37.81_{-2.31}^{+5.96}$ & $\cdots$ & $0.821_{-0.040}^{+0.040}$ & $1664.03$ & $1639.30/481$ & $359.92/481$ & $0.497$\\
 & BB+PL & $\cdots$ & $\cdots$ & $\cdots$ & $\cdots$ & $\cdots$ & $21.52_{-1.41}^{+1.19}$ & $-1.85_{-0.14}^{+0.12}$ & $0.949_{-0.071}^{+0.070}$ & $1486.44$ & $1461.71/481$ & $366.00/481$ & $0.512$\\
\hline
$4.035\!\sim\!4.755$ & Band & $-0.53_{-0.10}^{+0.37}$ & $-3.08_{-0.39}^{+0.21}$ & $52.37_{-5.51}^{+3.28}$ & $\cdots$ & $\cdots$ & $\cdots$ & $\cdots$ & $0.391_{-0.030}^{+0.034}$ & $1895.06$ & $1870.33/481$ & $408.01/481$ & $0.042$\\
 & CPL & $-0.44_{-0.15}^{+0.27}$ & $\cdots$ & $52.21_{-3.18}^{+4.33}$ & $\cdots$ & $\cdots$ & $\cdots$ & $\cdots$ & $0.348_{-0.023}^{+0.024}$ & $1880.85$ & $1862.30/482$ & $411.42/482$ & $0.031$\\
 & mBB & $\cdots$ & $\cdots$ & $\cdots$ & $0.78_{-0.64}^{+0.45}$ & $0.85_{-0.10}^{+3.25}$ & $20.13_{-2.05}^{+3.17}$ & $\cdots$ & $0.331_{-0.022}^{+0.024}$ & $1903.46$ & $1878.72/481$ & $409.69/481$ & $0.024$\\
 & BB+PL & $\cdots$ & $\cdots$ & $\cdots$ & $\cdots$ & $\cdots$ & $12.46_{-0.55}^{+1.09}$ & $-4.86_{-1.34}^{+2.96}$ & $0.340_{-0.032}^{+0.040}$ & $1661.23$ & $1636.50/481$ & $407.82/481$ & $0.119$\\
\hline
$4.755\!\sim\!5.333$ & Band & $-0.43_{-0.12}^{+0.29}$ & $-3.43_{-0.30}^{+0.25}$ & $44.89_{-2.99}^{+2.44}$ & $\cdots$ & $\cdots$ & $\cdots$ & $\cdots$ & $0.548_{-0.031}^{+0.034}$ & $1579.54$ & $1554.81/481$ & $327.02/481$ & $0.927$\\
 & CPL & $-0.38_{-0.15}^{+0.27}$ & $\cdots$ & $45.80_{-2.55}^{+2.59}$ & $\cdots$ & $\cdots$ & $\cdots$ & $\cdots$ & $0.506_{-0.026}^{+0.029}$ & $1552.14$ & $1533.59/482$ & $327.17/482$ & $0.903$\\
 & mBB & $\cdots$ & $\cdots$ & $\cdots$ & $0.71_{-0.53}^{+0.37}$ & $0.84_{-0.13}^{+2.74}$ & $17.89_{-1.39}^{+2.35}$ & $\cdots$ & $0.488_{-0.026}^{+0.027}$ & $1566.19$ & $1541.46/481$ & $328.26/481$ & $0.907$\\
 & BB+PL & $\cdots$ & $\cdots$ & $\cdots$ & $\cdots$ & $\cdots$ & $11.16_{-0.23}^{+1.06}$ & $-3.96_{-0.58}^{+1.80}$ & $0.507_{-0.039}^{+0.042}$ & $1333.98$ & $1309.24/481$ & $331.92/481$ & $0.836$\\
\hline
$5.333\!\sim\!5.802$ & Band & $-0.87_{-0.15}^{+0.45}$ & $-2.84_{-0.40}^{+0.17}$ & $34.55_{-4.72}^{+5.18}$ & $\cdots$ & $\cdots$ & $\cdots$ & $\cdots$ & $0.300_{-0.030}^{+0.036}$ & $1338.33$ & $1313.60/481$ & $357.04/481$ & $0.216$\\
 & CPL & $-0.88_{-0.12}^{+0.40}$ & $\cdots$ & $37.50_{-4.72}^{+4.88}$ & $\cdots$ & $\cdots$ & $\cdots$ & $\cdots$ & $0.264_{-0.026}^{+0.027}$ & $1325.98$ & $1307.43/482$ & $362.03/482$ & $0.178$\\
 & mBB & $\cdots$ & $\cdots$ & $\cdots$ & $0.49_{-1.91}^{+0.31}$ & $0.85_{-0.02}^{+3.68}$ & $14.91_{-1.12}^{+19.98}$ & $\cdots$ & $0.247_{-0.026}^{+0.028}$ & $1358.19$ & $1333.45/481$ & $362.31/481$ & $0.156$\\
 & BB+PL & $\cdots$ & $\cdots$ & $\cdots$ & $\cdots$ & $\cdots$ & $8.70_{-0.84}^{+0.78}$ & $-3.90_{-0.26}^{+2.40}$ & $0.224_{-0.027}^{+0.038}$ & $1119.69$ & $1094.95/481$ & $367.80/481$ & $0.196$\\
\hline
$5.802\!\sim\!6.999$ & Band & $-0.75_{-0.13}^{+0.46}$ & $-2.60_{-0.41}^{+0.26}$ & $25.56_{-5.55}^{+7.58}$ & $\cdots$ & $\cdots$ & $\cdots$ & $\cdots$ & $0.085_{-0.017}^{+0.020}$ & $2436.43$ & $2411.69/481$ & $390.82/481$ & $0.699$\\
 & CPL & $-0.80_{-0.09}^{+0.52}$ & $\cdots$ & $25.61_{-4.52}^{+8.83}$ & $\cdots$ & $\cdots$ & $\cdots$ & $\cdots$ & $0.063_{-0.011}^{+0.012}$ & $2430.47$ & $2411.92/482$ & $391.81/482$ & $0.664$\\
 & mBB & $\cdots$ & $\cdots$ & $\cdots$ & $0.79_{-2.29}^{+0.17}$ & $0.83_{-0.11}^{+2.87}$ & $9.93_{-0.34}^{+131.17}$ & $\cdots$ & $0.068_{-0.017}^{+0.022}$ & $2452.86$ & $2428.13/481$ & $389.08/481$ & $0.679$\\
 & BB+PL & $\cdots$ & $\cdots$ & $\cdots$ & $\cdots$ & $\cdots$ & $6.17_{-0.93}^{+1.68}$ & $-1.96_{-1.32}^{+1.01}$ & $0.052_{-0.010}^{+0.027}$ & $2216.55$ & $2191.82/481$ & $387.79/481$ & $0.640$\\
\hline
$22.000\!\sim\!22.666$ & Band & $-1.30_{-0.06}^{+0.33}$ & $-2.66_{-0.33}^{+0.57}$ & $142.43_{-64.02}^{+14.28}$ & $\cdots$ & $\cdots$ & $\cdots$ & $\cdots$ & $1.849_{-0.086}^{+0.085}$ & $1825.43$ & $1800.70/481$ & $359.53/481$ & $0.565$\\
 & CPL & $-1.31_{-0.07}^{+0.06}$ & $\cdots$ & $148.87_{-11.89}^{+19.83}$ & $\cdots$ & $\cdots$ & $\cdots$ & $\cdots$ & $1.761_{-0.080}^{+0.088}$ & $1818.09$ & $1799.54/482$ & $373.91/482$ & $0.496$\\
 & mBB & $\cdots$ & $\cdots$ & $\cdots$ & $-0.66_{-0.11}^{+0.05}$ & $3.36_{-0.42}^{+0.52}$ & $117.08_{-10.08}^{+36.87}$ & $\cdots$ & $1.740_{-0.084}^{+0.087}$ & $1877.87$ & $1853.14/481$ & $362.50/481$ & $0.604$\\
 & BB+PL & $\cdots$ & $\cdots$ & $\cdots$ & $\cdots$ & $\cdots$ & $3.00_{-0.01}^{+0.07}$ & $-1.69_{-0.02}^{+0.01}$ & $2.232_{-0.062}^{+0.053}$ & $1896.68$ & $1871.95/481$ & $417.66/481$ & $0.020$\\
\hline
$22.666\!\sim\!22.947$ & Band & $-0.97_{-0.14}^{+0.24}$ & $-2.74_{-0.38}^{+0.28}$ & $94.52_{-14.99}^{+11.82}$ & $\cdots$ & $\cdots$ & $\cdots$ & $\cdots$ & $1.038_{-0.085}^{+0.093}$ & $787.61$ & $762.88/481$ & $297.48/481$ & $0.972$\\
 & CPL & $-0.96_{-0.14}^{+0.21}$ & $\cdots$ & $96.80_{-10.79}^{+11.77}$ & $\cdots$ & $\cdots$ & $\cdots$ & $\cdots$ & $0.929_{-0.070}^{+0.075}$ & $779.21$ & $760.66/482$ & $292.58/482$ & $0.961$\\
 & mBB & $\cdots$ & $\cdots$ & $\cdots$ & $-0.16_{-0.38}^{+0.06}$ & $0.84_{-0.06}^{+2.85}$ & $48.51_{-3.94}^{+19.05}$ & $\cdots$ & $0.903_{-0.066}^{+0.074}$ & $826.29$ & $801.56/481$ & $301.05/481$ & $0.961$\\
 & BB+PL & $\cdots$ & $\cdots$ & $\cdots$ & $\cdots$ & $\cdots$ & $3.00_{-0.01}^{+0.10}$ & $-1.72_{-0.04}^{+0.05}$ & $1.347_{-0.098}^{+0.095}$ & $846.82$ & $822.08/481$ & $326.15/481$ & $0.649$\\
\hline
$22.947\!\sim\!23.965$ & Band & $-0.95_{-0.11}^{+0.34}$ & $-2.82_{-0.33}^{+0.24}$ & $58.99_{-8.46}^{+5.31}$ & $\cdots$ & $\cdots$ & $\cdots$ & $\cdots$ & $0.438_{-0.032}^{+0.034}$ & $2173.82$ & $2149.09/481$ & $325.26/481$ & $0.927$\\
 & CPL & $-0.99_{-0.11}^{+0.25}$ & $\cdots$ & $63.42_{-6.26}^{+5.42}$ & $\cdots$ & $\cdots$ & $\cdots$ & $\cdots$ & $0.393_{-0.026}^{+0.026}$ & $2162.75$ & $2144.20/482$ & $327.07/482$ & $0.885$\\
 & mBB & $\cdots$ & $\cdots$ & $\cdots$ & $-0.28_{-0.41}^{+0.11}$ & $0.64_{-0.10}^{+2.69}$ & $34.51_{-3.37}^{+12.59}$ & $\cdots$ & $0.387_{-0.024}^{+0.027}$ & $2205.17$ & $2180.43/481$ & $327.12/481$ & $0.885$\\
 & BB+PL & $\cdots$ & $\cdots$ & $\cdots$ & $\cdots$ & $\cdots$ & $3.03_{-0.02}^{+0.02}$ & $-1.66_{-0.01}^{+0.02}$ & $0.908_{-0.007}^{+0.011}$ & $2207.75$ & $2183.01/481$ & $414.71/481$ & $0.178$\\
\hline
$23.965\!\sim\!24.158$ & Band & $-1.06_{-0.24}^{+0.23}$ & $-2.37_{-0.51}^{+0.19}$ & $106.78_{-20.90}^{+48.33}$ & $\cdots$ & $\cdots$ & $\cdots$ & $\cdots$ & $1.130_{-0.111}^{+0.122}$ & $543.65$ & $518.91/481$ & $315.40/481$ & $0.291$\\
 & CPL & $-1.20_{-0.16}^{+0.22}$ & $\cdots$ & $130.74_{-22.73}^{+51.15}$ & $\cdots$ & $\cdots$ & $\cdots$ & $\cdots$ & $1.041_{-0.121}^{+0.157}$ & $539.81$ & $521.26/482$ & $326.42/482$ & $0.246$\\
 & mBB & $\cdots$ & $\cdots$ & $\cdots$ & $-0.33_{-0.30}^{+0.08}$ & $0.60_{-0.04}^{+2.65}$ & $67.93_{-6.50}^{+56.87}$ & $\cdots$ & $1.009_{-0.108}^{+0.134}$ & $591.35$ & $566.62/481$ & $333.61/481$ & $0.223$\\
 & BB+PL & $\cdots$ & $\cdots$ & $\cdots$ & $\cdots$ & $\cdots$ & $3.01_{-0.01}^{+0.13}$ & $-1.67_{-0.04}^{+0.07}$ & $1.381_{-0.118}^{+0.129}$ & $578.93$ & $554.19/481$ & $330.40/481$ & $0.120$\\
\hline
$24.158\!\sim\!25.007$ & Band & $-1.05_{-0.06}^{+0.05}$ & $-2.44_{-0.38}^{+0.15}$ & $282.64_{-25.20}^{+37.35}$ & $\cdots$ & $\cdots$ & $\cdots$ & $\cdots$ & $3.362_{-0.103}^{+0.099}$ & $2277.18$ & $2252.45/481$ & $441.55/481$ & $0.121$\\
 & CPL & $-1.09_{-0.04}^{+0.05}$ & $\cdots$ & $316.91_{-26.77}^{+31.62}$ & $\cdots$ & $\cdots$ & $\cdots$ & $\cdots$ & $3.383_{-0.114}^{+0.123}$ & $2275.40$ & $2256.85/482$ & $446.91/482$ & $0.075$\\
 & mBB & $\cdots$ & $\cdots$ & $\cdots$ & $-0.30_{-0.06}^{+0.07}$ & $3.30_{-0.89}^{+0.53}$ & $169.78_{-18.00}^{+28.30}$ & $\cdots$ & $3.340_{-0.131}^{+0.136}$ & $2344.77$ & $2320.03/481$ & $447.24/481$ & $0.088$\\
 & BB+PL & $\cdots$ & $\cdots$ & $\cdots$ & $\cdots$ & $\cdots$ & $3.00_{-0.00}^{+0.03}$ & $-1.53_{-0.01}^{+0.02}$ & $3.461_{-0.072}^{+0.068}$ & $2466.57$ & $2441.83/481$ & $637.58/481$ & $0.000$\\
\hline
$25.007\!\sim\!25.342$ & Band & $-1.00_{-0.28}^{+0.19}$ & $-2.09_{-0.50}^{+0.02}$ & $96.34_{-12.71}^{+71.23}$ & $\cdots$ & $\cdots$ & $\cdots$ & $\cdots$ & $1.381_{-0.099}^{+0.096}$ & $1036.81$ & $1012.07/481$ & $342.39/481$ & $0.370$\\
 & CPL & $-1.30_{-0.14}^{+0.12}$ & $\cdots$ & $179.88_{-30.28}^{+80.94}$ & $\cdots$ & $\cdots$ & $\cdots$ & $\cdots$ & $1.363_{-0.129}^{+0.147}$ & $1035.75$ & $1017.19/482$ & $352.06/482$ & $0.290$\\
 & mBB & $\cdots$ & $\cdots$ & $\cdots$ & $-0.60_{-0.20}^{+0.04}$ & $3.03_{-0.43}^{+1.36}$ & $138.34_{-10.80}^{+204.62}$ & $\cdots$ & $1.400_{-0.135}^{+0.145}$ & $1100.04$ & $1075.30/481$ & $350.31/481$ & $0.364$\\
 & BB+PL & $\cdots$ & $\cdots$ & $\cdots$ & $\cdots$ & $\cdots$ & $3.01_{-0.06}^{+0.10}$ & $-1.63_{-0.18}^{+0.14}$ & $1.359_{-0.036}^{+0.076}$ & $1108.14$ & $1083.40/481$ & $400.45/481$ & $0.016$\\
\hline
$25.342\!\sim\!25.753$ & Band & $-0.73_{-0.06}^{+0.06}$ & $-2.91_{-0.38}^{+0.25}$ & $404.73_{-32.75}^{+37.64}$ & $\cdots$ & $\cdots$ & $\cdots$ & $\cdots$ & $5.721_{-0.197}^{+0.200}$ & $1472.99$ & $1448.25/481$ & $390.43/481$ & $0.245$\\
 & CPL & $-0.74_{-0.05}^{+0.06}$ & $\cdots$ & $413.33_{-30.45}^{+35.69}$ & $\cdots$ & $\cdots$ & $\cdots$ & $\cdots$ & $5.786_{-0.193}^{+0.195}$ & $1460.80$ & $1442.25/482$ & $387.37/482$ & $0.235$\\
 & mBB & $\cdots$ & $\cdots$ & $\cdots$ & $0.13_{-0.06}^{+0.04}$ & $0.78_{-0.20}^{+1.72}$ & $174.56_{-10.01}^{+18.74}$ & $\cdots$ & $5.737_{-0.210}^{+0.224}$ & $1517.16$ & $1492.43/481$ & $390.95/481$ & $0.260$\\
 & BB+PL & $\cdots$ & $\cdots$ & $\cdots$ & $\cdots$ & $\cdots$ & $3.03_{-0.01}^{+0.07}$ & $-1.41_{-0.01}^{+0.01}$ & $5.012_{-0.102}^{+0.072}$ & $1776.28$ & $1751.55/481$ & $629.47/481$ & $0.000$\\
\hline
$25.753\!\sim\!26.042$ & Band & $-0.78_{-0.12}^{+0.10}$ & $-2.41_{-0.43}^{+0.20}$ & $228.44_{-27.33}^{+47.30}$ & $\cdots$ & $\cdots$ & $\cdots$ & $\cdots$ & $3.100_{-0.159}^{+0.153}$ & $985.49$ & $960.75/481$ & $351.56/481$ & $0.587$\\
 & CPL & $-0.86_{-0.10}^{+0.10}$ & $\cdots$ & $270.81_{-30.95}^{+41.25}$ & $\cdots$ & $\cdots$ & $\cdots$ & $\cdots$ & $3.052_{-0.190}^{+0.205}$ & $982.70$ & $964.15/482$ & $359.13/482$ & $0.454$\\
 & mBB & $\cdots$ & $\cdots$ & $\cdots$ & $-0.21_{-0.17}^{+0.14}$ & $4.76_{-1.60}^{+1.16}$ & $141.14_{-18.59}^{+54.73}$ & $\cdots$ & $3.053_{-0.207}^{+0.241}$ & $1052.60$ & $1027.87/481$ & $355.39/481$ & $0.512$\\
 & BB+PL & $\cdots$ & $\cdots$ & $\cdots$ & $\cdots$ & $\cdots$ & $3.00_{-0.01}^{+0.06}$ & $-1.50_{-0.02}^{+0.02}$ & $3.214_{-0.119}^{+0.118}$ & $1039.08$ & $1014.34/481$ & $419.66/481$ & $0.000$\\
\hline
$26.042\!\sim\!26.568$ & Band & $-0.42_{-0.05}^{+0.06}$ & $-3.03_{-0.41}^{+0.18}$ & $329.24_{-17.60}^{+19.38}$ & $\cdots$ & $\cdots$ & $\cdots$ & $\cdots$ & $6.617_{-0.180}^{+0.180}$ & $1818.53$ & $1793.79/481$ & $435.10/481$ & $0.099$\\
 & CPL & $-0.43_{-0.05}^{+0.05}$ & $\cdots$ & $342.13_{-15.48}^{+16.41}$ & $\cdots$ & $\cdots$ & $\cdots$ & $\cdots$ & $6.657_{-0.182}^{+0.197}$ & $1802.23$ & $1783.68/482$ & $433.70/482$ & $0.082$\\
 & mBB & $\cdots$ & $\cdots$ & $\cdots$ & $0.41_{-0.10}^{+0.02}$ & $0.59_{-0.37}^{+3.89}$ & $139.12_{-4.27}^{+12.24}$ & $\cdots$ & $6.593_{-0.192}^{+0.193}$ & $1855.28$ & $1830.54/481$ & $443.10/481$ & $0.087$\\
 & BB+PL & $\cdots$ & $\cdots$ & $\cdots$ & $\cdots$ & $\cdots$ & $3.00_{-0.00}^{+0.02}$ & $-1.38_{-0.01}^{+0.01}$ & $5.800_{-0.109}^{+0.106}$ & $2598.96$ & $2574.23/481$ & $1061.87/481$ & $0.000$\\
\hline
$26.568\!\sim\!26.876$ & Band & $-1.14_{-0.11}^{+0.10}$ & $-2.52_{-0.47}^{+0.25}$ & $204.34_{-35.04}^{+45.40}$ & $\cdots$ & $\cdots$ & $\cdots$ & $\cdots$ & $2.097_{-0.124}^{+0.135}$ & $992.92$ & $968.18/481$ & $342.34/481$ & $0.590$\\
 & CPL & $-1.16_{-0.09}^{+0.10}$ & $\cdots$ & $216.80_{-28.89}^{+43.46}$ & $\cdots$ & $\cdots$ & $\cdots$ & $\cdots$ & $2.039_{-0.141}^{+0.152}$ & $987.54$ & $968.98/482$ & $349.06/482$ & $0.556$\\
 & mBB & $\cdots$ & $\cdots$ & $\cdots$ & $-0.39_{-0.14}^{+0.09}$ & $2.81_{-1.45}^{+1.02}$ & $126.39_{-19.25}^{+37.63}$ & $\cdots$ & $1.995_{-0.143}^{+0.140}$ & $1050.74$ & $1026.01/481$ & $341.68/481$ & $0.581$\\
 & BB+PL & $\cdots$ & $\cdots$ & $\cdots$ & $\cdots$ & $\cdots$ & $3.00_{-0.01}^{+0.09}$ & $-1.58_{-0.03}^{+0.03}$ & $2.380_{-0.100}^{+0.107}$ & $998.94$ & $974.20/481$ & $357.49/481$ & $0.121$\\
\hline
$26.876\!\sim\!27.355$ & Band & $-0.96_{-0.20}^{+0.24}$ & $-2.37_{-0.54}^{+0.19}$ & $120.87_{-23.54}^{+44.81}$ & $\cdots$ & $\cdots$ & $\cdots$ & $\cdots$ & $0.773_{-0.069}^{+0.080}$ & $1338.44$ & $1313.70/481$ & $358.01/481$ & $0.440$\\
 & CPL & $-1.03_{-0.18}^{+0.18}$ & $\cdots$ & $141.55_{-20.70}^{+41.18}$ & $\cdots$ & $\cdots$ & $\cdots$ & $\cdots$ & $0.707_{-0.073}^{+0.082}$ & $1334.60$ & $1316.04/482$ & $370.27/482$ & $0.398$\\
 & mBB & $\cdots$ & $\cdots$ & $\cdots$ & $-0.25_{-0.29}^{+0.08}$ & $0.90_{-0.10}^{+2.75}$ & $74.72_{-8.31}^{+48.41}$ & $\cdots$ & $0.693_{-0.066}^{+0.087}$ & $1383.51$ & $1358.78/481$ & $371.25/481$ & $0.361$\\
 & BB+PL & $\cdots$ & $\cdots$ & $\cdots$ & $\cdots$ & $\cdots$ & $3.02_{-0.01}^{+0.00}$ & $-1.63_{-0.01}^{+0.04}$ & $1.005_{-0.024}^{+0.031}$ & $1368.20$ & $1343.46/481$ & $397.72/481$ & $0.071$\\
\hline
$27.355\!\sim\!28.510$ & Band & $-0.81_{-0.24}^{+0.40}$ & $-2.23_{-0.37}^{+0.06}$ & $44.66_{-5.62}^{+11.77}$ & $\cdots$ & $\cdots$ & $\cdots$ & $\cdots$ & $0.299_{-0.033}^{+0.038}$ & $2384.55$ & $2359.81/481$ & $404.04/481$ & $0.186$\\
 & CPL & $-1.29_{-0.15}^{+0.36}$ & $\cdots$ & $73.51_{-15.20}^{+19.43}$ & $\cdots$ & $\cdots$ & $\cdots$ & $\cdots$ & $0.260_{-0.031}^{+0.038}$ & $2385.38$ & $2366.83/482$ & $412.83/482$ & $0.124$\\
 & mBB & $\cdots$ & $\cdots$ & $\cdots$ & $-0.91_{-0.29}^{+0.17}$ & $3.17_{-1.10}^{+1.17}$ & $80.60_{-15.63}^{+186.97}$ & $\cdots$ & $0.280_{-0.037}^{+0.044}$ & $2433.24$ & $2408.50/481$ & $404.34/481$ & $0.187$\\
 & BB+PL & $\cdots$ & $\cdots$ & $\cdots$ & $\cdots$ & $\cdots$ & $3.01_{-0.00}^{+0.00}$ & $-1.54_{-0.00}^{+0.00}$ & $1.597_{-0.001}^{+0.002}$ & $3243.73$ & $3219.00/481$ & $1282.38/481$ & $0.000$\\
\hline
$28.510\!\sim\!28.725$ & Band & $-0.82_{-0.16}^{+0.27}$ & $-2.67_{-0.46}^{+0.23}$ & $95.80_{-13.89}^{+16.91}$ & $\cdots$ & $\cdots$ & $\cdots$ & $\cdots$ & $0.962_{-0.083}^{+0.102}$ & $591.61$ & $566.87/481$ & $315.19/481$ & $0.736$\\
 & CPL & $-0.80_{-0.17}^{+0.24}$ & $\cdots$ & $101.08_{-12.11}^{+14.17}$ & $\cdots$ & $\cdots$ & $\cdots$ & $\cdots$ & $0.857_{-0.075}^{+0.081}$ & $583.39$ & $564.84/482$ & $319.05/482$ & $0.699$\\
 & mBB & $\cdots$ & $\cdots$ & $\cdots$ & $0.06_{-0.43}^{+0.13}$ & $0.61_{-0.17}^{+3.41}$ & $44.75_{-4.06}^{+16.63}$ & $\cdots$ & $0.829_{-0.064}^{+0.083}$ & $627.31$ & $602.57/481$ & $314.82/481$ & $0.679$\\
 & BB+PL & $\cdots$ & $\cdots$ & $\cdots$ & $\cdots$ & $\cdots$ & $3.01_{-0.01}^{+0.18}$ & $-1.66_{-0.06}^{+0.07}$ & $1.291_{-0.189}^{+0.094}$ & $596.95$ & $572.21/481$ & $340.67/481$ & $0.146$\\
\hline
$28.725\!\sim\!31.563$ & Band & $-0.75_{-0.09}^{+0.20}$ & $-3.09_{-0.34}^{+0.20}$ & $61.42_{-4.34}^{+3.19}$ & $\cdots$ & $\cdots$ & $\cdots$ & $\cdots$ & $0.437_{-0.020}^{+0.022}$ & $3710.81$ & $3686.07/481$ & $453.70/481$ & $0.492$\\
 & CPL & $-0.79_{-0.11}^{+0.13}$ & $\cdots$ & $64.36_{-3.43}^{+3.61}$ & $\cdots$ & $\cdots$ & $\cdots$ & $\cdots$ & $0.405_{-0.016}^{+0.016}$ & $3692.96$ & $3674.41/482$ & $450.74/482$ & $0.486$\\
 & mBB & $\cdots$ & $\cdots$ & $\cdots$ & $-0.47_{-0.23}^{+0.24}$ & $3.29_{-0.78}^{+0.73}$ & $38.88_{-5.26}^{+5.65}$ & $\cdots$ & $0.397_{-0.016}^{+0.016}$ & $3738.74$ & $3714.00/481$ & $449.01/481$ & $0.537$\\
 & BB+PL & $\cdots$ & $\cdots$ & $\cdots$ & $\cdots$ & $\cdots$ & $3.00_{-0.00}^{+0.01}$ & $-1.68_{-0.03}^{+0.02}$ & $0.889_{-0.009}^{+0.008}$ & $3958.86$ & $3934.12/481$ & $798.26/481$ & $0.000$\\
\hline
$31.563\!\sim\!31.788$ & Band & $-0.71_{-0.14}^{+0.22}$ & $-2.66_{-0.41}^{+0.24}$ & $110.15_{-12.53}^{+16.00}$ & $\cdots$ & $\cdots$ & $\cdots$ & $\cdots$ & $1.520_{-0.109}^{+0.125}$ & $719.45$ & $694.72/481$ & $346.64/481$ & $0.088$\\
 & CPL & $-0.73_{-0.14}^{+0.19}$ & $\cdots$ & $118.15_{-11.50}^{+13.89}$ & $\cdots$ & $\cdots$ & $\cdots$ & $\cdots$ & $1.356_{-0.093}^{+0.103}$ & $712.45$ & $693.89/482$ & $363.91/482$ & $0.058$\\
 & mBB & $\cdots$ & $\cdots$ & $\cdots$ & $0.10_{-0.26}^{+0.15}$ & $0.75_{-0.09}^{+2.76}$ & $51.39_{-4.33}^{+12.03}$ & $\cdots$ & $1.314_{-0.086}^{+0.094}$ & $757.07$ & $732.33/481$ & $353.79/481$ & $0.071$\\
 & BB+PL & $\cdots$ & $\cdots$ & $\cdots$ & $\cdots$ & $\cdots$ & $3.00_{-0.01}^{+0.09}$ & $-1.64_{-0.02}^{+0.06}$ & $1.964_{-0.120}^{+0.120}$ & $757.83$ & $733.09/481$ & $404.48/481$ & $0.001$\\
\hline
$31.788\!\sim\!32.616$ & Band & $-0.77_{-0.10}^{+0.26}$ & $-2.91_{-0.31}^{+0.22}$ & $61.43_{-5.94}^{+4.14}$ & $\cdots$ & $\cdots$ & $\cdots$ & $\cdots$ & $0.671_{-0.042}^{+0.044}$ & $2016.72$ & $1991.98/481$ & $418.93/481$ & $0.070$\\
 & CPL & $-0.82_{-0.12}^{+0.20}$ & $\cdots$ & $65.54_{-4.76}^{+4.35}$ & $\cdots$ & $\cdots$ & $\cdots$ & $\cdots$ & $0.604_{-0.030}^{+0.032}$ & $2005.11$ & $1986.55/482$ & $425.19/482$ & $0.048$\\
 & mBB & $\cdots$ & $\cdots$ & $\cdots$ & $-0.05_{-0.55}^{+0.05}$ & $0.75_{-0.22}^{+3.09}$ & $31.72_{-1.93}^{+10.21}$ & $\cdots$ & $0.589_{-0.031}^{+0.033}$ & $2046.40$ & $2021.66/481$ & $427.80/481$ & $0.038$\\
 & BB+PL & $\cdots$ & $\cdots$ & $\cdots$ & $\cdots$ & $\cdots$ & $3.01_{-0.00}^{+0.05}$ & $-1.77_{-0.02}^{+0.00}$ & $1.135_{-0.015}^{+0.013}$ & $2051.48$ & $2026.74/481$ & $523.63/481$ & $0.000$\\
\hline
$32.616\!\sim\!33.224$ & Band & $-0.78_{-0.10}^{+0.12}$ & $-2.55_{-0.34}^{+0.20}$ & $150.96_{-15.05}^{+18.22}$ & $\cdots$ & $\cdots$ & $\cdots$ & $\cdots$ & $2.050_{-0.088}^{+0.093}$ & $1730.23$ & $1705.49/481$ & $389.02/481$ & $0.450$\\
 & CPL & $-0.88_{-0.08}^{+0.10}$ & $\cdots$ & $175.31_{-16.25}^{+15.20}$ & $\cdots$ & $\cdots$ & $\cdots$ & $\cdots$ & $1.929_{-0.090}^{+0.102}$ & $1726.91$ & $1708.36/482$ & $396.14/482$ & $0.372$\\
 & mBB & $\cdots$ & $\cdots$ & $\cdots$ & $-0.04_{-0.40}^{+0.00}$ & $0.78_{-0.51}^{+4.52}$ & $79.88_{-0.53}^{+36.03}$ & $\cdots$ & $1.891_{-0.095}^{+0.105}$ & $1783.38$ & $1758.65/481$ & $403.83/481$ & $0.284$\\
 & BB+PL & $\cdots$ & $\cdots$ & $\cdots$ & $\cdots$ & $\cdots$ & $3.01_{-0.00}^{+0.03}$ & $-1.57_{-0.02}^{+0.01}$ & $2.381_{-0.063}^{+0.077}$ & $1855.48$ & $1830.75/481$ & $545.30/481$ & $0.000$\\
\hline
$33.224\!\sim\!33.478$ & Band & $-1.09_{-0.10}^{+0.38}$ & $-2.86_{-0.40}^{+0.26}$ & $62.94_{-11.55}^{+6.83}$ & $\cdots$ & $\cdots$ & $\cdots$ & $\cdots$ & $0.706_{-0.061}^{+0.073}$ & $664.67$ & $639.93/481$ & $266.56/481$ & $0.952$\\
 & CPL & $-1.02_{-0.17}^{+0.26}$ & $\cdots$ & $64.23_{-7.61}^{+8.63}$ & $\cdots$ & $\cdots$ & $\cdots$ & $\cdots$ & $0.639_{-0.052}^{+0.057}$ & $653.14$ & $634.59/482$ & $273.01/482$ & $0.953$\\
 & mBB & $\cdots$ & $\cdots$ & $\cdots$ & $-0.22_{-0.52}^{+0.06}$ & $0.72_{-0.04}^{+2.62}$ & $33.56_{-1.84}^{+16.01}$ & $\cdots$ & $0.631_{-0.053}^{+0.057}$ & $696.64$ & $671.91/481$ & $280.46/481$ & $0.949$\\
 & BB+PL & $\cdots$ & $\cdots$ & $\cdots$ & $\cdots$ & $\cdots$ & $3.00_{-0.02}^{+0.16}$ & $-1.90_{-0.00}^{+0.12}$ & $0.950_{-0.055}^{+0.072}$ & $673.56$ & $648.82/481$ & $289.45/481$ & $0.713$\\
\hline
$33.478\!\sim\!34.140$ & Band & $-0.82_{-0.20}^{+0.39}$ & $-2.62_{-0.40}^{+0.14}$ & $35.61_{-4.32}^{+6.59}$ & $\cdots$ & $\cdots$ & $\cdots$ & $\cdots$ & $0.315_{-0.031}^{+0.036}$ & $1620.76$ & $1596.02/481$ & $331.36/481$ & $0.721$\\
 & CPL & $-1.06_{-0.12}^{+0.36}$ & $\cdots$ & $44.21_{-6.07}^{+5.53}$ & $\cdots$ & $\cdots$ & $\cdots$ & $\cdots$ & $0.277_{-0.026}^{+0.027}$ & $1610.61$ & $1592.06/482$ & $332.83/482$ & $0.633$\\
 & mBB & $\cdots$ & $\cdots$ & $\cdots$ & $-1.04_{-0.53}^{+0.29}$ & $3.32_{-0.64}^{+0.89}$ & $38.40_{-8.66}^{+75.47}$ & $\cdots$ & $0.277_{-0.025}^{+0.033}$ & $1658.15$ & $1633.41/481$ & $331.10/481$ & $0.732$\\
 & BB+PL & $\cdots$ & $\cdots$ & $\cdots$ & $\cdots$ & $\cdots$ & $3.00_{-0.00}^{+0.04}$ & $-1.69_{-0.02}^{+0.06}$ & $0.899_{-0.015}^{+0.019}$ & $1634.36$ & $1609.63/481$ & $397.76/481$ & $0.238$\\
\hline
$34.140\!\sim\!34.711$ & Band & $-0.70_{-0.08}^{+0.10}$ & $-2.54_{-0.38}^{+0.18}$ & $198.27_{-18.55}^{+21.87}$ & $\cdots$ & $\cdots$ & $\cdots$ & $\cdots$ & $2.716_{-0.105}^{+0.113}$ & $1650.00$ & $1625.27/481$ & $347.48/481$ & $0.948$\\
 & CPL & $-0.75_{-0.08}^{+0.08}$ & $\cdots$ & $221.75_{-15.93}^{+20.03}$ & $\cdots$ & $\cdots$ & $\cdots$ & $\cdots$ & $2.619_{-0.120}^{+0.123}$ & $1645.85$ & $1627.30/482$ & $354.97/482$ & $0.896$\\
 & mBB & $\cdots$ & $\cdots$ & $\cdots$ & $0.09_{-0.16}^{+0.02}$ & $0.75_{-0.12}^{+2.94}$ & $96.54_{-4.38}^{+16.65}$ & $\cdots$ & $2.534_{-0.109}^{+0.111}$ & $1702.19$ & $1677.46/481$ & $360.44/481$ & $0.880$\\
 & BB+PL & $\cdots$ & $\cdots$ & $\cdots$ & $\cdots$ & $\cdots$ & $3.00_{-0.00}^{+0.03}$ & $-1.51_{-0.01}^{+0.02}$ & $2.996_{-0.093}^{+0.085}$ & $1832.30$ & $1807.56/481$ & $535.88/481$ & $0.000$\\
\hline
$34.711\!\sim\!35.211$ & Band & $-0.95_{-0.14}^{+0.22}$ & $-2.92_{-0.40}^{+0.23}$ & $80.99_{-8.73}^{+8.45}$ & $\cdots$ & $\cdots$ & $\cdots$ & $\cdots$ & $0.743_{-0.051}^{+0.057}$ & $1373.90$ & $1349.16/481$ & $373.83/481$ & $0.563$\\
 & CPL & $-0.93_{-0.14}^{+0.18}$ & $\cdots$ & $83.07_{-7.51}^{+9.45}$ & $\cdots$ & $\cdots$ & $\cdots$ & $\cdots$ & $0.680_{-0.046}^{+0.050}$ & $1361.20$ & $1342.65/482$ & $369.39/482$ & $0.554$\\
 & mBB & $\cdots$ & $\cdots$ & $\cdots$ & $-0.17_{-0.19}^{+0.14}$ & $0.69_{-0.14}^{+1.48}$ & $41.94_{-3.94}^{+8.23}$ & $\cdots$ & $0.661_{-0.041}^{+0.044}$ & $1404.18$ & $1379.44/481$ & $370.46/481$ & $0.535$\\
 & BB+PL & $\cdots$ & $\cdots$ & $\cdots$ & $\cdots$ & $\cdots$ & $3.00_{-0.01}^{+0.07}$ & $-1.77_{-0.02}^{+0.07}$ & $1.001_{-0.062}^{+0.066}$ & $1391.42$ & $1366.68/481$ & $400.12/481$ & $0.049$\\
\hline
$35.211\!\sim\!35.891$ & Band & $-1.28_{-0.03}^{+0.52}$ & $-2.74_{-0.39}^{+0.28}$ & $43.10_{-10.19}^{+8.23}$ & $\cdots$ & $\cdots$ & $\cdots$ & $\cdots$ & $0.276_{-0.029}^{+0.034}$ & $1661.09$ & $1636.36/481$ & $326.64/481$ & $0.497$\\
 & CPL & $-1.25_{-0.16}^{+0.34}$ & $\cdots$ & $48.80_{-8.13}^{+9.29}$ & $\cdots$ & $\cdots$ & $\cdots$ & $\cdots$ & $0.251_{-0.027}^{+0.030}$ & $1650.58$ & $1632.03/482$ & $325.88/482$ & $0.480$\\
 & mBB & $\cdots$ & $\cdots$ & $\cdots$ & $-0.65_{-0.32}^{+0.17}$ & $0.83_{-0.10}^{+1.43}$ & $36.40_{-5.10}^{+28.33}$ & $\cdots$ & $0.257_{-0.029}^{+0.032}$ & $1688.23$ & $1663.49/481$ & $324.67/481$ & $0.507$\\
 & BB+PL & $\cdots$ & $\cdots$ & $\cdots$ & $\cdots$ & $\cdots$ & $3.08_{-0.00}^{+0.01}$ & $-1.78_{-0.00}^{+0.02}$ & $0.894_{-0.007}^{+0.016}$ & $1754.04$ & $1729.30/481$ & $480.82/481$ & $0.000$\\
\hline
$35.891\!\sim\!36.091$ & Band & $-0.95_{-0.15}^{+0.22}$ & $-2.72_{-0.40}^{+0.30}$ & $113.98_{-21.00}^{+17.82}$ & $\cdots$ & $\cdots$ & $\cdots$ & $\cdots$ & $1.306_{-0.109}^{+0.116}$ & $544.82$ & $520.08/481$ & $283.20/481$ & $0.886$\\
 & CPL & $-0.96_{-0.18}^{+0.16}$ & $\cdots$ & $120.89_{-15.29}^{+22.02}$ & $\cdots$ & $\cdots$ & $\cdots$ & $\cdots$ & $1.210_{-0.107}^{+0.124}$ & $536.22$ & $517.67/482$ & $276.74/482$ & $0.857$\\
 & mBB & $\cdots$ & $\cdots$ & $\cdots$ & $-0.40_{-0.31}^{+0.21}$ & $3.65_{-1.81}^{+1.21}$ & $66.63_{-7.68}^{+38.24}$ & $\cdots$ & $1.194_{-0.108}^{+0.115}$ & $594.66$ & $569.92/481$ & $281.84/481$ & $0.876$\\
 & BB+PL & $\cdots$ & $\cdots$ & $\cdots$ & $\cdots$ & $\cdots$ & $3.01_{-0.00}^{+0.11}$ & $-1.69_{-0.04}^{+0.05}$ & $1.616_{-0.108}^{+0.126}$ & $565.78$ & $541.04/481$ & $298.81/481$ & $0.237$\\
\hline
$36.091\!\sim\!36.367$ & Band & $-0.71_{-0.10}^{+0.14}$ & $-2.87_{-0.34}^{+0.25}$ & $135.60_{-14.03}^{+9.69}$ & $\cdots$ & $\cdots$ & $\cdots$ & $\cdots$ & $2.516_{-0.126}^{+0.134}$ & $921.96$ & $897.22/481$ & $328.95/481$ & $0.632$\\
 & CPL & $-0.73_{-0.11}^{+0.11}$ & $\cdots$ & $140.68_{-9.41}^{+12.51}$ & $\cdots$ & $\cdots$ & $\cdots$ & $\cdots$ & $2.352_{-0.119}^{+0.122}$ & $910.79$ & $892.23/482$ & $331.92/482$ & $0.532$\\
 & mBB & $\cdots$ & $\cdots$ & $\cdots$ & $0.04_{-0.10}^{+0.11}$ & $0.80_{-0.25}^{+1.45}$ & $64.96_{-5.37}^{+7.12}$ & $\cdots$ & $2.286_{-0.101}^{+0.102}$ & $958.35$ & $933.62/481$ & $347.91/481$ & $0.512$\\
 & BB+PL & $\cdots$ & $\cdots$ & $\cdots$ & $\cdots$ & $\cdots$ & $3.00_{-0.01}^{+0.04}$ & $-1.62_{-0.02}^{+0.01}$ & $3.078_{-0.058}^{+0.057}$ & $990.38$ & $965.64/481$ & $433.57/481$ & $0.000$\\
\hline
$36.367\!\sim\!36.685$ & Band & $-0.57_{-0.08}^{+0.13}$ & $-2.54_{-0.35}^{+0.23}$ & $207.75_{-26.40}^{+18.26}$ & $\cdots$ & $\cdots$ & $\cdots$ & $\cdots$ & $4.851_{-0.180}^{+0.187}$ & $1177.66$ & $1152.92/481$ & $384.47/481$ & $0.390$\\
 & CPL & $-0.65_{-0.07}^{+0.08}$ & $\cdots$ & $234.84_{-16.38}^{+16.45}$ & $\cdots$ & $\cdots$ & $\cdots$ & $\cdots$ & $4.682_{-0.195}^{+0.194}$ & $1173.75$ & $1155.19/482$ & $393.35/482$ & $0.237$\\
 & mBB & $\cdots$ & $\cdots$ & $\cdots$ & $-0.06_{-0.20}^{+0.19}$ & $5.42_{-2.29}^{+1.66}$ & $117.89_{-14.80}^{+23.12}$ & $\cdots$ & $4.665_{-0.200}^{+0.238}$ & $1242.83$ & $1218.09/481$ & $383.87/481$ & $0.191$\\
 & BB+PL & $\cdots$ & $\cdots$ & $\cdots$ & $\cdots$ & $\cdots$ & $3.00_{-0.00}^{+0.04}$ & $-1.49_{-0.01}^{+0.01}$ & $5.135_{-0.137}^{+0.118}$ & $1420.57$ & $1395.83/481$ & $607.24/481$ & $0.000$\\
\hline
$36.685\!\sim\!36.943$ & Band & $-0.63_{-0.13}^{+0.17}$ & $-2.81_{-0.39}^{+0.20}$ & $109.36_{-8.59}^{+11.89}$ & $\cdots$ & $\cdots$ & $\cdots$ & $\cdots$ & $1.928_{-0.115}^{+0.126}$ & $846.75$ & $822.01/481$ & $357.91/481$ & $0.342$\\
 & CPL & $-0.68_{-0.11}^{+0.15}$ & $\cdots$ & $119.37_{-10.17}^{+8.70}$ & $\cdots$ & $\cdots$ & $\cdots$ & $\cdots$ & $1.763_{-0.093}^{+0.100}$ & $835.80$ & $817.25/482$ & $351.42/482$ & $0.288$\\
 & mBB & $\cdots$ & $\cdots$ & $\cdots$ & $0.13_{-0.29}^{+0.08}$ & $0.75_{-0.18}^{+3.50}$ & $51.88_{-3.14}^{+10.98}$ & $\cdots$ & $1.712_{-0.084}^{+0.099}$ & $879.39$ & $854.66/481$ & $359.29/481$ & $0.265$\\
 & BB+PL & $\cdots$ & $\cdots$ & $\cdots$ & $\cdots$ & $\cdots$ & $3.01_{-0.00}^{+0.07}$ & $-1.64_{-0.03}^{+0.03}$ & $2.527_{-0.121}^{+0.112}$ & $871.66$ & $846.93/481$ & $423.23/481$ & $0.000$\\
\hline
$36.943\!\sim\!37.304$ & Band & $-0.93_{-0.16}^{+0.36}$ & $-2.65_{-0.42}^{+0.24}$ & $71.63_{-11.93}^{+10.54}$ & $\cdots$ & $\cdots$ & $\cdots$ & $\cdots$ & $0.608_{-0.060}^{+0.066}$ & $998.79$ & $974.06/481$ & $316.84/481$ & $0.657$\\
 & CPL & $-0.90_{-0.19}^{+0.28}$ & $\cdots$ & $74.38_{-8.42}^{+12.02}$ & $\cdots$ & $\cdots$ & $\cdots$ & $\cdots$ & $0.524_{-0.047}^{+0.053}$ & $992.04$ & $973.49/482$ & $324.99/482$ & $0.537$\\
 & mBB & $\cdots$ & $\cdots$ & $\cdots$ & $-0.05_{-0.44}^{+0.12}$ & $0.93_{-0.25}^{+2.13}$ & $34.71_{-2.63}^{+15.16}$ & $\cdots$ & $0.508_{-0.042}^{+0.048}$ & $1033.76$ & $1009.03/481$ & $328.31/481$ & $0.509$\\
 & BB+PL & $\cdots$ & $\cdots$ & $\cdots$ & $\cdots$ & $\cdots$ & $3.10_{-0.02}^{+0.09}$ & $-1.60_{-0.02}^{+0.04}$ & $1.001_{-0.037}^{+0.055}$ & $1047.39$ & $1022.66/481$ & $322.87/481$ & $0.518$\\
\hline
$37.304\!\sim\!39.515$ & Band & $-0.97_{-0.12}^{+0.50}$ & $-2.35_{-0.29}^{+0.13}$ & $32.65_{-6.70}^{+5.70}$ & $\cdots$ & $\cdots$ & $\cdots$ & $\cdots$ & $0.181_{-0.021}^{+0.024}$ & $3294.77$ & $3270.03/481$ & $447.04/481$ & $0.349$\\
 & CPL & $-1.32_{-0.16}^{+0.39}$ & $\cdots$ & $46.77_{-8.56}^{+12.34}$ & $\cdots$ & $\cdots$ & $\cdots$ & $\cdots$ & $0.150_{-0.018}^{+0.023}$ & $3297.00$ & $3278.44/482$ & $450.94/482$ & $0.256$\\
 & mBB & $\cdots$ & $\cdots$ & $\cdots$ & $-1.03_{-0.24}^{+0.17}$ & $2.42_{-1.38}^{+0.73}$ & $62.49_{-9.28}^{+180.70}$ & $\cdots$ & $0.170_{-0.022}^{+0.026}$ & $3334.11$ & $3309.37/481$ & $444.60/481$ & $0.285$\\
 & BB+PL & $\cdots$ & $\cdots$ & $\cdots$ & $\cdots$ & $\cdots$ & $3.02_{-0.00}^{+0.01}$ & $-1.52_{-0.00}^{+0.01}$ & $1.002_{-0.003}^{+0.004}$ & $3973.97$ & $3949.23/481$ & $1201.65/481$ & $0.000$\\
\hline
$39.515\!\sim\!40.576$ & Band & $-0.76_{-0.22}^{+0.20}$ & $-2.30_{-0.37}^{+0.13}$ & $87.48_{-12.16}^{+23.08}$ & $\cdots$ & $\cdots$ & $\cdots$ & $\cdots$ & $0.841_{-0.057}^{+0.055}$ & $2272.45$ & $2247.72/481$ & $361.52/481$ & $0.872$\\
 & CPL & $-1.01_{-0.11}^{+0.12}$ & $\cdots$ & $120.67_{-11.48}^{+14.95}$ & $\cdots$ & $\cdots$ & $\cdots$ & $\cdots$ & $0.748_{-0.045}^{+0.048}$ & $2268.41$ & $2249.86/482$ & $370.34/482$ & $0.790$\\
 & mBB & $\cdots$ & $\cdots$ & $\cdots$ & $-0.57_{-0.26}^{+0.13}$ & $4.08_{-0.78}^{+1.16}$ & $85.81_{-11.82}^{+52.56}$ & $\cdots$ & $0.764_{-0.057}^{+0.067}$ & $2327.46$ & $2302.72/481$ & $359.78/481$ & $0.845$\\
 & BB+PL & $\cdots$ & $\cdots$ & $\cdots$ & $\cdots$ & $\cdots$ & $3.09_{-0.01}^{+0.06}$ & $-1.63_{-0.02}^{+0.04}$ & $1.055_{-0.019}^{+0.026}$ & $2315.40$ & $2290.66/481$ & $478.47/481$ & $0.011$\\
\hline
$40.576\!\sim\!42.998$ & Band & $-0.92_{-0.01}^{+0.51}$ & $-2.72_{-0.33}^{+0.24}$ & $24.05_{-4.20}^{+7.00}$ & $\cdots$ & $\cdots$ & $\cdots$ & $\cdots$ & $0.086_{-0.012}^{+0.015}$ & $3435.12$ & $3410.38/481$ & $485.33/481$ & $0.149$\\
 & CPL & $-0.94_{-0.08}^{+0.50}$ & $\cdots$ & $28.75_{-5.38}^{+6.31}$ & $\cdots$ & $\cdots$ & $\cdots$ & $\cdots$ & $0.071_{-0.009}^{+0.010}$ & $3426.70$ & $3408.15/482$ & $486.23/482$ & $0.116$\\
 & mBB & $\cdots$ & $\cdots$ & $\cdots$ & $-0.59_{-0.96}^{+0.09}$ & $1.02_{-0.03}^{+1.97}$ & $19.08_{-1.76}^{+108.61}$ & $\cdots$ & $0.080_{-0.013}^{+0.015}$ & $3454.06$ & $3429.32/481$ & $487.01/481$ & $0.145$\\
 & BB+PL & $\cdots$ & $\cdots$ & $\cdots$ & $\cdots$ & $\cdots$ & $3.00_{-0.00}^{+0.00}$ & $-1.53_{-0.01}^{+0.00}$ & $0.966_{-0.007}^{+0.004}$ & $4409.45$ & $4384.71/481$ & $1544.97/481$ & $0.000$\\
\hline

\end{longtable}
\noindent\parbox{0.96\linewidth}{\scriptsize
Note. All fitted intervals, including those with $S<10$, are listed. The
analysis in the main text focuses primarily on intervals with $S>10$. The columns $\alpha$, $m$, and
$\alpha_{\rm PL}$ denote, respectively, the low-energy photon index of Band/CPL,
the mBB temperature-distribution index, and the photon index of the PL component
in BB+PL. In the $kT_{\max}/kT$ column, the mBB and BB+PL rows give $kT_{\max}$
and $kT$, respectively. Parameter values and uncertainties are the reported
central values and 68\% credible errors. PGSTAT is $-2\ln L_{\max}$ summed
over all active channels and GBM detectors. RSS is the sum of the squared
\texttt{ThreeML} significance residuals; RSS/dof is only an auxiliary measure
of residual amplitude and is not a reduced chi-square. PPC is the posterior
predictive-check probability defined in Section~4; non-extreme values indicate
posterior-predictive compatibility, whereas values close to 0 or 1 indicate
possible model--data tension.}
\endgroup
\end{landscape}

\begin{landscape}
\begingroup
\tiny
\renewcommand{\arraystretch}{0.92}
\setlength{\tabcolsep}{1.0pt}
\begin{longtable}{llcccccccccccc}
\caption{Time-resolved Band, CPL, mBB, and BB+PL spectral-fitting results
using joint \textit{Fermi}/GBM+\textit{Swift}/BAT data.
\label{tab:joint_empirical}}\\
\hline\hline
Time & Model & $\alpha$ & $\beta$ & $E_{\rm p}$ & $m$ & $kT_{\min}$
& $kT_{\max}/kT$ & $\alpha_{\rm PL}$ & $F$ & BIC & PG-stat/dof
& RSS/dof & PPC\\
(s) & & & & (keV) & & (keV) & (keV) &
& $(10^{-6}\,\mathrm{erg\,cm^{-2}\,s^{-1}})$ & & & &\\
\hline
\endfirsthead

\caption[]{Time-resolved empirical-model results using joint
\textit{Fermi}/GBM+\textit{Swift}/BAT data (continued).}\\
\hline\hline
Time & Model & $\alpha$ & $\beta$ & $E_{\rm p}$ & $m$ & $kT_{\min}$
& $kT_{\max}/kT$ & $\alpha_{\rm PL}$ & $F$ & BIC & PG-stat/dof
& RSS/dof & PPC\\
(s) & & & & (keV) & & (keV) & (keV) &
& $(10^{-6}\,\mathrm{erg\,cm^{-2}\,s^{-1}})$ & & & &\\
\hline
\endhead

\hline
\multicolumn{14}{r}{\textit{Continued on next page}}\\
\endfoot

\hline
\endlastfoot

$0.000\!\sim\!1.400$ & Band & $-0.32_{-0.06}^{+0.09}$ & $-2.70_{-0.18}^{+0.15}$ & $118.82_{-6.85}^{+4.90}$ & $\cdots$ & $\cdots$ & $\cdots$ & $\cdots$ & $2.518_{-0.072}^{+0.071}$ & $3114.33$ & $3089.13/541$ & $543.09/541$ & $0.038$\\
 & CPL & $-0.47_{-0.05}^{+0.06}$ & $\cdots$ & $135.42_{-4.53}^{+3.41}$ & $\cdots$ & $\cdots$ & $\cdots$ & $\cdots$ & $2.274_{-0.049}^{+0.049}$ & $3112.48$ & $3093.58/542$ & $564.37/542$ & $0.004$\\
 & mBB & $\cdots$ & $\cdots$ & $\cdots$ & $-0.26_{-0.18}^{+0.09}$ & $6.55_{-0.49}^{+0.88}$ & $74.80_{-4.29}^{+9.41}$ & $\cdots$ & $2.313_{-0.056}^{+0.061}$ & $3170.53$ & $3145.33/541$ & $548.36/541$ & $0.018$\\
 & BB+PL & $\cdots$ & $\cdots$ & $\cdots$ & $\cdots$ & $\cdots$ & $25.38_{-0.62}^{+0.75}$ & $-1.62_{-0.03}^{+0.02}$ & $2.566_{-0.052}^{+0.055}$ & $3008.50$ & $2983.29/541$ & $620.34/541$ & $0.112$\\
\hline
$1.400\!\sim\!3.800$ & Band & $-0.45_{-0.07}^{+0.07}$ & $-3.27_{-0.31}^{+0.22}$ & $101.41_{-4.00}^{+3.62}$ & $\cdots$ & $\cdots$ & $\cdots$ & $\cdots$ & $1.157_{-0.031}^{+0.035}$ & $3747.75$ & $3722.55/541$ & $523.17/541$ & $0.268$\\
 & CPL & $-0.49_{-0.06}^{+0.07}$ & $\cdots$ & $105.06_{-3.23}^{+3.21}$ & $\cdots$ & $\cdots$ & $\cdots$ & $\cdots$ & $1.086_{-0.025}^{+0.026}$ & $3721.27$ & $3702.37/542$ & $522.28/542$ & $0.266$\\
 & mBB & $\cdots$ & $\cdots$ & $\cdots$ & $0.02_{-0.11}^{+0.15}$ & $4.00_{-1.12}^{+0.63}$ & $50.32_{-3.42}^{+3.15}$ & $\cdots$ & $1.068_{-0.025}^{+0.026}$ & $3778.40$ & $3753.20/541$ & $519.88/541$ & $0.270$\\
 & BB+PL & $\cdots$ & $\cdots$ & $\cdots$ & $\cdots$ & $\cdots$ & $22.64_{-0.68}^{+0.63}$ & $-1.79_{-0.04}^{+0.04}$ & $1.248_{-0.036}^{+0.035}$ & $3599.28$ & $3574.08/541$ & $578.19/541$ & $0.218$\\
\hline
$3.800\!\sim\!7.000$ & Band & $-0.63_{-0.11}^{+0.25}$ & $-2.98_{-0.35}^{+0.16}$ & $45.98_{-2.73}^{+2.82}$ & $\cdots$ & $\cdots$ & $\cdots$ & $\cdots$ & $0.281_{-0.015}^{+0.017}$ & $4067.58$ & $4042.37/541$ & $581.62/541$ & $0.034$\\
 & CPL & $-0.71_{-0.12}^{+0.21}$ & $\cdots$ & $48.51_{-2.42}^{+2.45}$ & $\cdots$ & $\cdots$ & $\cdots$ & $\cdots$ & $0.251_{-0.010}^{+0.011}$ & $4051.36$ & $4032.46/542$ & $581.50/542$ & $0.022$\\
 & mBB & $\cdots$ & $\cdots$ & $\cdots$ & $-0.44_{-0.48}^{+0.36}$ & $3.20_{-1.44}^{+0.94}$ & $26.60_{-3.16}^{+8.33}$ & $\cdots$ & $0.245_{-0.011}^{+0.012}$ & $4101.55$ & $4076.34/541$ & $577.71/541$ & $0.018$\\
 & BB+PL & $\cdots$ & $\cdots$ & $\cdots$ & $\cdots$ & $\cdots$ & $11.95_{-0.92}^{+0.61}$ & $-2.06_{-0.07}^{+0.13}$ & $0.314_{-0.022}^{+0.023}$ & $3881.05$ & $3855.85/541$ & $590.68/541$ & $0.210$\\
\hline
$21.500\!\sim\!23.200$ & Band & $-0.82_{-0.55}^{+0.10}$ & $-2.01_{-0.62}^{+0.02}$ & $52.68_{-1.20}^{+65.90}$ & $\cdots$ & $\cdots$ & $\cdots$ & $\cdots$ & $0.989_{-0.051}^{+0.038}$ & $3112.29$ & $3087.08/541$ & $502.74/541$ & $0.274$\\
 & CPL & $-1.40_{-0.06}^{+0.06}$ & $\cdots$ & $128.14_{-12.60}^{+17.75}$ & $\cdots$ & $\cdots$ & $\cdots$ & $\cdots$ & $0.910_{-0.043}^{+0.049}$ & $3105.09$ & $3086.19/542$ & $504.57/542$ & $0.196$\\
 & mBB & $\cdots$ & $\cdots$ & $\cdots$ & $-0.75_{-0.10}^{+0.04}$ & $3.20_{-0.28}^{+0.46}$ & $118.58_{-14.07}^{+52.22}$ & $\cdots$ & $0.908_{-0.051}^{+0.058}$ & $3169.72$ & $3144.52/541$ & $491.67/541$ & $0.366$\\
 & BB+PL & $\cdots$ & $\cdots$ & $\cdots$ & $\cdots$ & $\cdots$ & $1.47_{-0.37}^{+0.27}$ & $-1.78_{-0.03}^{+0.02}$ & $1.184_{-0.035}^{+0.037}$ & $3134.58$ & $3109.37/541$ & $547.87/541$ & $0.062$\\
\hline
$23.200\!\sim\!24.800$ & Band & $-1.14_{-0.06}^{+0.07}$ & $-2.39_{-0.41}^{+0.16}$ & $205.17_{-22.19}^{+30.55}$ & $\cdots$ & $\cdots$ & $\cdots$ & $\cdots$ & $1.666_{-0.060}^{+0.057}$ & $3061.52$ & $3036.32/541$ & $458.08/541$ & $0.584$\\
 & CPL & $-1.17_{-0.05}^{+0.05}$ & $\cdots$ & $230.04_{-18.23}^{+26.90}$ & $\cdots$ & $\cdots$ & $\cdots$ & $\cdots$ & $1.632_{-0.068}^{+0.067}$ & $3060.07$ & $3041.17/542$ & $465.51/542$ & $0.496$\\
 & mBB & $\cdots$ & $\cdots$ & $\cdots$ & $-0.37_{-0.06}^{+0.07}$ & $2.68_{-1.53}^{+0.62}$ & $130.93_{-15.84}^{+18.10}$ & $\cdots$ & $1.583_{-0.065}^{+0.072}$ & $3140.88$ & $3115.67/541$ & $468.92/541$ & $0.508$\\
 & BB+PL & $\cdots$ & $\cdots$ & $\cdots$ & $\cdots$ & $\cdots$ & $1.02_{-0.07}^{+0.69}$ & $-1.61_{-0.01}^{+0.01}$ & $1.892_{-0.044}^{+0.045}$ & $3173.74$ & $3148.54/541$ & $561.54/541$ & $0.006$\\
\hline
$24.800\!\sim\!27.000$ & Band & $-0.87_{-0.03}^{+0.03}$ & $-3.10_{-0.43}^{+0.17}$ & $386.56_{-16.54}^{+20.28}$ & $\cdots$ & $\cdots$ & $\cdots$ & $\cdots$ & $3.925_{-0.078}^{+0.074}$ & $3856.67$ & $3831.47/541$ & $660.96/541$ & $0.100$\\
 & CPL & $-0.87_{-0.03}^{+0.03}$ & $\cdots$ & $393.34_{-17.47}^{+19.45}$ & $\cdots$ & $\cdots$ & $\cdots$ & $\cdots$ & $3.952_{-0.077}^{+0.081}$ & $3837.22$ & $3818.31/542$ & $656.24/542$ & $0.080$\\
 & mBB & $\cdots$ & $\cdots$ & $\cdots$ & $-0.02_{-0.02}^{+0.03}$ & $2.55_{-1.68}^{+0.23}$ & $175.34_{-9.19}^{+7.62}$ & $\cdots$ & $3.906_{-0.086}^{+0.082}$ & $3910.49$ & $3885.29/541$ & $651.02/541$ & $0.000$\\
 & BB+PL & $\cdots$ & $\cdots$ & $\cdots$ & $\cdots$ & $\cdots$ & $1.06_{-0.02}^{+0.48}$ & $-1.45_{-0.01}^{+0.01}$ & $3.604_{-0.042}^{+0.040}$ & $4744.24$ & $4719.04/541$ & $1600.16/541$ & $0.080$\\
\hline
$27.000\!\sim\!32.000$ & Band & $-0.89_{-0.11}^{+0.11}$ & $-2.77_{-0.37}^{+0.16}$ & $71.46_{-4.64}^{+5.47}$ & $\cdots$ & $\cdots$ & $\cdots$ & $\cdots$ & $0.477_{-0.021}^{+0.022}$ & $4676.54$ & $4651.34/541$ & $597.68/541$ & $0.026$\\
 & CPL & $-1.01_{-0.07}^{+0.09}$ & $\cdots$ & $78.42_{-3.93}^{+3.87}$ & $\cdots$ & $\cdots$ & $\cdots$ & $\cdots$ & $0.435_{-0.013}^{+0.015}$ & $4667.03$ & $4648.13/542$ & $596.58/542$ & $0.030$\\
 & mBB & $\cdots$ & $\cdots$ & $\cdots$ & $-0.52_{-0.17}^{+0.08}$ & $3.15_{-0.41}^{+0.65}$ & $49.94_{-3.55}^{+10.00}$ & $\cdots$ & $0.427_{-0.014}^{+0.016}$ & $4728.44$ & $4703.24/541$ & $594.43/541$ & $0.026$\\
 & BB+PL & $\cdots$ & $\cdots$ & $\cdots$ & $\cdots$ & $\cdots$ & $1.06_{-0.01}^{+0.48}$ & $-1.81_{-0.02}^{+0.01}$ & $0.687_{-0.018}^{+0.018}$ & $4733.44$ & $4708.24/541$ & $794.47/541$ & $0.040$\\
\hline
$32.000\!\sim\!33.500$ & Band & $-1.07_{-0.10}^{+0.08}$ & $-2.45_{-0.44}^{+0.13}$ & $119.51_{-11.38}^{+18.77}$ & $\cdots$ & $\cdots$ & $\cdots$ & $\cdots$ & $1.156_{-0.050}^{+0.053}$ & $2991.13$ & $2965.92/541$ & $534.63/541$ & $0.008$\\
 & CPL & $-1.16_{-0.06}^{+0.07}$ & $\cdots$ & $139.91_{-11.15}^{+12.02}$ & $\cdots$ & $\cdots$ & $\cdots$ & $\cdots$ & $1.083_{-0.047}^{+0.048}$ & $2988.37$ & $2969.47/542$ & $544.10/542$ & $0.004$\\
 & mBB & $\cdots$ & $\cdots$ & $\cdots$ & $-0.46_{-0.09}^{+0.09}$ & $2.90_{-0.75}^{+0.59}$ & $87.03_{-10.00}^{+14.09}$ & $\cdots$ & $1.052_{-0.045}^{+0.047}$ & $3062.24$ & $3037.03/541$ & $546.41/541$ & $0.002$\\
 & BB+PL & $\cdots$ & $\cdots$ & $\cdots$ & $\cdots$ & $\cdots$ & $1.43_{-0.33}^{+0.28}$ & $-1.70_{-0.02}^{+0.01}$ & $1.477_{-0.039}^{+0.040}$ & $2998.27$ & $2973.07/541$ & $643.51/541$ & $0.020$\\
\hline
$33.500\!\sim\!35.000$ & Band & $-1.06_{-0.08}^{+0.06}$ & $-2.53_{-0.40}^{+0.21}$ & $182.23_{-14.86}^{+24.49}$ & $\cdots$ & $\cdots$ & $\cdots$ & $\cdots$ & $1.274_{-0.056}^{+0.057}$ & $2959.10$ & $2933.89/541$ & $497.61/541$ & $0.154$\\
 & CPL & $-1.10_{-0.06}^{+0.06}$ & $\cdots$ & $201.23_{-17.30}^{+22.55}$ & $\cdots$ & $\cdots$ & $\cdots$ & $\cdots$ & $1.229_{-0.057}^{+0.063}$ & $2956.14$ & $2937.24/542$ & $502.39/542$ & $0.104$\\
 & mBB & $\cdots$ & $\cdots$ & $\cdots$ & $-0.27_{-0.05}^{+0.05}$ & $2.08_{-1.28}^{+0.43}$ & $99.95_{-8.41}^{+14.21}$ & $\cdots$ & $1.167_{-0.053}^{+0.062}$ & $3024.96$ & $2999.76/541$ & $502.00/541$ & $0.092$\\
 & BB+PL & $\cdots$ & $\cdots$ & $\cdots$ & $\cdots$ & $\cdots$ & $1.44_{-0.35}^{+0.24}$ & $-1.61_{-0.02}^{+0.02}$ & $1.530_{-0.041}^{+0.040}$ & $2984.13$ & $2958.93/541$ & $593.82/541$ & $0.002$\\
\hline
$35.000\!\sim\!37.000$ & Band & $-0.94_{-0.05}^{+0.05}$ & $-3.02_{-0.34}^{+0.24}$ & $163.77_{-8.80}^{+10.10}$ & $\cdots$ & $\cdots$ & $\cdots$ & $\cdots$ & $1.530_{-0.044}^{+0.048}$ & $3510.38$ & $3485.18/541$ & $604.89/541$ & $0.070$\\
 & CPL & $-0.95_{-0.04}^{+0.05}$ & $\cdots$ & $169.86_{-8.95}^{+9.18}$ & $\cdots$ & $\cdots$ & $\cdots$ & $\cdots$ & $1.477_{-0.044}^{+0.049}$ & $3493.62$ & $3474.72/542$ & $605.11/542$ & $0.010$\\
 & mBB & $\cdots$ & $\cdots$ & $\cdots$ & $-0.15_{-0.05}^{+0.03}$ & $0.59_{-0.06}^{+1.46}$ & $82.95_{-4.01}^{+5.83}$ & $\cdots$ & $1.427_{-0.042}^{+0.043}$ & $3558.29$ & $3533.09/541$ & $605.04/541$ & $0.000$\\
 & BB+PL & $\cdots$ & $\cdots$ & $\cdots$ & $\cdots$ & $\cdots$ & $1.18_{-0.10}^{+0.43}$ & $-1.61_{-0.01}^{+0.01}$ & $1.958_{-0.036}^{+0.034}$ & $3732.11$ & $3706.91/541$ & $933.17/541$ & $0.002$\\
\hline
$37.000\!\sim\!41.000$ & Band & $-1.24_{-0.20}^{+0.24}$ & $-2.12_{-0.53}^{+0.05}$ & $62.20_{-11.90}^{+30.53}$ & $\cdots$ & $\cdots$ & $\cdots$ & $\cdots$ & $0.366_{-0.025}^{+0.024}$ & $4278.96$ & $4253.75/541$ & $560.17/541$ & $0.172$\\
 & CPL & $-1.48_{-0.11}^{+0.07}$ & $\cdots$ & $101.49_{-10.72}^{+27.50}$ & $\cdots$ & $\cdots$ & $\cdots$ & $\cdots$ & $0.339_{-0.024}^{+0.029}$ & $4280.66$ & $4261.76/542$ & $559.80/542$ & $0.152$\\
 & mBB & $\cdots$ & $\cdots$ & $\cdots$ & $-0.76_{-0.09}^{+0.07}$ & $2.35_{-0.59}^{+0.51}$ & $94.23_{-12.73}^{+55.52}$ & $\cdots$ & $0.336_{-0.026}^{+0.030}$ & $4330.65$ & $4305.45/541$ & $555.97/541$ & $0.196$\\
 & BB+PL & $\cdots$ & $\cdots$ & $\cdots$ & $\cdots$ & $\cdots$ & $1.14_{-0.04}^{+0.62}$ & $-1.84_{-0.03}^{+0.03}$ & $0.457_{-0.018}^{+0.019}$ & $4356.76$ & $4331.55/541$ & $582.79/541$ & $0.106$\\
\hline

\end{longtable}
\noindent\parbox{0.96\linewidth}{\scriptsize
Note. Parameter definitions are the same as in Table~\ref{tab:fermi_empirical}.
For the joint fits, the column labeled PG-stat reports the total
$-2\ln L_{\max}$ from the Gaussian BAT and Poisson--Gaussian GBM likelihood
contributions and is therefore not a pure PGSTAT. The label is retained for
consistent table formatting. All available joint-fit intervals are listed.}
\endgroup
\end{landscape}

\begin{landscape}
\begingroup
\tiny
\renewcommand{\arraystretch}{0.9}
\setlength{\tabcolsep}{0.8pt}
\begin{longtable}{lccccccccccccc}
\caption{Time-resolved MDFSYN spectral-fitting results for EII using
\textit{Fermi}/GBM-only data.\label{tab:eii_mdfsyn}}\\
\hline\hline
Time & $\hat{t}$ & $a$ & $\log Q_0$ & $\log\Gamma$ & $\log R_0$
& $\log\gamma_{\rm inj}$ & $\log B_0$ & $p$ & $F$ & BIC
& PG-stat/dof & RSS/dof & PPC\\
(s) & (s) & & $(s^{-1})$ & & (cm) & & (G) & &
$(10^{-6}\,\mathrm{erg\,cm^{-2}\,s^{-1}})$ & & & &\\
\hline
\endfirsthead
\caption[]{Time-resolved MDFSYN spectral-fitting results for EII (continued).}\\
\hline\hline
Time & $\hat{t}$ & $a$ & $\log Q_0$ & $\log\Gamma$ & $\log R_0$
& $\log\gamma_{\rm inj}$ & $\log B_0$ & $p$ & $F$ & BIC
& PG-stat/dof & RSS/dof & PPC\\
(s) & (s) & & $(s^{-1})$ & & (cm) & & (G) & &
$(10^{-6}\,\mathrm{erg\,cm^{-2}\,s^{-1}})$ & & & &\\
\hline
\endhead
\hline
\multicolumn{14}{r}{\textit{Continued on next page}}\\
\endfoot
\hline
\endlastfoot
$22.000\!\sim\!22.666$ & $1.95_{-0.40}^{+2.23}$ & $1.02_{-0.28}^{+0.28}$ & $58.87_{-2.97}^{+0.17}$ & $2.85_{-0.64}^{+0.05}$ & $15.48_{-1.05}^{+0.08}$ & $4.93_{-0.11}^{+0.15}$ & $1.57_{-0.30}^{+0.56}$ & $3.28_{-0.66}^{+0.08}$ & $1.907_{-0.072}^{+0.071}$ & $1815.74$ & $1766.27/477$ & $359.87/477$ & $0.632$\\
$22.666\!\sim\!22.947$ & $1.13_{-0.18}^{+2.94}$ & $1.09_{-0.33}^{+0.22}$ & $64.09_{-6.59}^{+1.70}$ & $2.61_{-0.36}^{+0.22}$ & $14.53_{-0.21}^{+0.96}$ & $5.24_{-0.42}^{+0.05}$ & $1.36_{-0.07}^{+0.77}$ & $3.94_{-1.00}^{+0.17}$ & $1.134_{-0.078}^{+0.089}$ & $786.72$ & $737.24/477$ & $290.28/477$ & $0.960$\\
$22.947\!\sim\!23.965$ & $3.98_{-2.57}^{+0.09}$ & $1.16_{-0.42}^{+0.15}$ & $65.13_{-7.39}^{+1.88}$ & $2.00_{-0.16}^{+0.75}$ & $14.20_{-0.26}^{+1.35}$ & $5.28_{-0.67}^{+0.24}$ & $1.48_{-0.18}^{+0.68}$ & $4.09_{-0.94}^{+0.06}$ & $0.478_{-0.032}^{+0.034}$ & $2173.15$ & $2123.68/477$ & $328.66/477$ & $0.892$\\
$23.965\!\sim\!24.158$ & $2.81_{-1.64}^{+1.17}$ & $0.85_{-0.25}^{+0.45}$ & $59.82_{-4.30}^{+0.05}$ & $2.45_{-0.26}^{+0.31}$ & $14.30_{-0.26}^{+1.39}$ & $5.16_{-0.64}^{+0.14}$ & $1.62_{-0.16}^{+1.07}$ & $3.26_{-0.64}^{+0.16}$ & $1.214_{-0.111}^{+0.124}$ & $540.88$ & $491.40/477$ & $322.14/477$ & $0.274$\\
$24.158\!\sim\!25.007$ & $2.09_{-0.59}^{+1.94}$ & $1.26_{-0.36}^{+0.14}$ & $63.07_{-4.83}^{+1.11}$ & $2.16_{-0.07}^{+0.64}$ & $14.26_{-0.01}^{+0.97}$ & $5.48_{-0.21}^{+0.04}$ & $1.73_{-0.42}^{+0.29}$ & $3.58_{-0.69}^{+0.09}$ & $3.446_{-0.088}^{+0.094}$ & $2271.56$ & $2222.09/477$ & $437.33/477$ & $0.160$\\
$25.007\!\sim\!25.342$ & $2.92_{-1.73}^{+1.09}$ & $0.90_{-0.21}^{+0.43}$ & $57.79_{-2.07}^{+2.16}$ & $2.76_{-0.56}^{+0.01}$ & $15.16_{-0.80}^{+0.33}$ & $5.10_{-0.29}^{+0.12}$ & $1.33_{-0.07}^{+0.96}$ & $2.98_{-0.40}^{+0.33}$ & $1.440_{-0.095}^{+0.104}$ & $1031.14$ & $981.66/477$ & $343.79/477$ & $0.378$\\
$25.342\!\sim\!25.753$ & $0.70_{-0.48}^{+3.11}$ & $1.62_{-0.44}^{+0.01}$ & $64.98_{-1.15}^{+0.62}$ & $2.11_{-0.05}^{+0.69}$ & $14.09_{-0.06}^{+0.93}$ & $5.79_{-0.05}^{+0.12}$ & $1.13_{-0.04}^{+0.82}$ & $3.64_{-0.08}^{+0.23}$ & $5.812_{-0.158}^{+0.163}$ & $1472.15$ & $1422.68/477$ & $385.44/477$ & $0.190$\\
$25.753\!\sim\!26.042$ & $2.01_{-0.64}^{+1.95}$ & $1.11_{-0.15}^{+0.27}$ & $64.99_{-5.74}^{+1.47}$ & $2.64_{-0.22}^{+0.24}$ & $14.63_{-0.33}^{+0.59}$ & $5.65_{-0.26}^{+0.02}$ & $1.30_{-0.11}^{+0.70}$ & $3.89_{-0.84}^{+0.19}$ & $3.297_{-0.141}^{+0.154}$ & $981.70$ & $932.22/477$ & $351.88/477$ & $0.590$\\
$26.042\!\sim\!26.568$ & $1.05_{-0.22}^{+0.40}$ & $1.19_{-0.08}^{+0.09}$ & $65.95_{-0.40}^{+0.01}$ & $2.99_{-0.25}^{+0.02}$ & $15.04_{-0.53}^{+0.02}$ & $5.75_{-0.03}^{+0.12}$ & $1.02_{-0.01}^{+0.18}$ & $4.00_{-0.15}^{+0.01}$ & $7.008_{-0.147}^{+0.149}$ & $1865.87$ & $1816.39/477$ & $481.69/477$ & $0.170$\\
$26.568\!\sim\!26.876$ & $0.72_{-0.67}^{+3.35}$ & $0.78_{-0.02}^{+0.63}$ & $60.00_{-3.91}^{+0.26}$ & $2.65_{-0.40}^{+0.17}$ & $14.12_{-0.30}^{+1.39}$ & $4.98_{-0.00}^{+0.30}$ & $2.01_{-0.72}^{+0.33}$ & $3.41_{-0.77}^{+0.06}$ & $2.190_{-0.117}^{+0.125}$ & $989.27$ & $939.80/477$ & $339.07/477$ & $0.568$\\
$26.876\!\sim\!27.355$ & $1.14_{-0.27}^{+3.08}$ & $0.98_{-0.26}^{+0.41}$ & $59.01_{-3.15}^{+1.77}$ & $2.82_{-0.60}^{+0.03}$ & $14.89_{-0.47}^{+0.67}$ & $5.05_{-0.18}^{+0.17}$ & $1.51_{-0.22}^{+0.86}$ & $3.25_{-0.58}^{+0.27}$ & $0.841_{-0.070}^{+0.078}$ & $1335.93$ & $1286.45/477$ & $355.77/477$ & $0.460$\\
$27.355\!\sim\!28.510$ & $3.11_{-1.99}^{+1.02}$ & $1.05_{-0.48}^{+0.26}$ & $58.45_{-3.13}^{+0.48}$ & $2.63_{-0.50}^{+0.05}$ & $14.36_{-0.21}^{+1.34}$ & $4.84_{-0.54}^{+0.03}$ & $2.46_{-1.03}^{+0.09}$ & $3.27_{-0.58}^{+0.12}$ & $0.315_{-0.029}^{+0.032}$ & $2380.01$ & $2330.54/477$ & $407.06/477$ & $0.192$\\
$28.510\!\sim\!28.725$ & $4.09_{-2.88}^{+0.09}$ & $0.92_{-0.20}^{+0.43}$ & $63.02_{-6.42}^{+1.49}$ & $2.54_{-0.28}^{+0.29}$ & $14.29_{-0.14}^{+1.12}$ & $5.37_{-0.54}^{+0.17}$ & $1.53_{-0.26}^{+0.67}$ & $3.73_{-0.92}^{+0.12}$ & $1.083_{-0.090}^{+0.094}$ & $590.27$ & $540.79/477$ & $312.74/477$ & $0.740$\\
$28.725\!\sim\!31.563$ & $1.81_{-0.42}^{+2.37}$ & $1.57_{-0.60}^{+0.10}$ & $65.19_{-3.09}^{+0.89}$ & $2.01_{-0.07}^{+0.86}$ & $14.61_{-0.10}^{+0.82}$ & $5.15_{-0.28}^{+0.03}$ & $1.27_{-0.04}^{+0.66}$ & $4.16_{-0.28}^{+0.01}$ & $0.490_{-0.018}^{+0.019}$ & $3715.93$ & $3666.46/477$ & $469.76/477$ & $0.356$\\
$31.563\!\sim\!31.788$ & $1.15_{-0.42}^{+3.13}$ & $0.99_{-0.18}^{+0.34}$ & $64.03_{-5.48}^{+0.78}$ & $2.89_{-0.53}^{+0.05}$ & $14.84_{-0.56}^{+0.35}$ & $5.30_{-0.21}^{+0.14}$ & $1.25_{-0.04}^{+0.87}$ & $3.97_{-0.93}^{+0.20}$ & $1.708_{-0.105}^{+0.119}$ & $718.76$ & $669.29/477$ & $345.60/477$ & $0.292$\\
$31.788\!\sim\!32.616$ & $0.75_{-0.42}^{+3.11}$ & $1.42_{-0.55}^{+0.03}$ & $64.37_{-4.21}^{+1.08}$ & $2.05_{-0.09}^{+0.74}$ & $14.41_{-0.02}^{+1.06}$ & $5.02_{-0.22}^{+0.05}$ & $1.32_{-0.05}^{+0.62}$ & $4.04_{-0.52}^{+0.08}$ & $0.748_{-0.036}^{+0.040}$ & $2017.53$ & $1968.06/477$ & $428.58/477$ & $0.278$\\
$32.616\!\sim\!33.224$ & $1.08_{-0.81}^{+3.28}$ & $0.68_{-0.16}^{+0.61}$ & $63.79_{-1.92}^{+0.53}$ & $2.92_{-0.50}^{+0.04}$ & $14.03_{-0.15}^{+0.97}$ & $5.38_{-0.03}^{+0.21}$ & $1.11_{-0.11}^{+0.77}$ & $3.87_{-0.37}^{+0.00}$ & $2.211_{-0.084}^{+0.084}$ & $1729.46$ & $1679.98/477$ & $388.21/477$ & $0.488$\\
$33.224\!\sim\!33.478$ & $3.58_{-2.39}^{+0.30}$ & $1.11_{-0.48}^{+0.19}$ & $62.66_{-5.93}^{+1.61}$ & $2.24_{-0.09}^{+0.53}$ & $14.38_{-0.20}^{+1.30}$ & $5.03_{-0.70}^{+0.19}$ & $1.90_{-0.48}^{+0.67}$ & $3.80_{-0.87}^{+0.09}$ & $0.770_{-0.062}^{+0.068}$ & $660.21$ & $610.73/477$ & $262.53/477$ & $0.936$\\
$33.478\!\sim\!34.140$ & $1.63_{-0.58}^{+2.33}$ & $1.18_{-0.63}^{+0.09}$ & $61.93_{-5.02}^{+1.21}$ & $2.22_{-0.08}^{+0.48}$ & $14.28_{-0.31}^{+1.44}$ & $4.76_{-0.54}^{+0.06}$ & $2.08_{-0.67}^{+0.41}$ & $3.83_{-0.79}^{+0.06}$ & $0.346_{-0.031}^{+0.033}$ & $1614.77$ & $1565.30/477$ & $331.79/477$ & $0.756$\\
$34.140\!\sim\!34.711$ & $1.25_{-0.19}^{+2.34}$ & $0.79_{-0.08}^{+0.46}$ & $64.77_{-2.09}^{+0.03}$ & $2.92_{-0.36}^{+0.01}$ & $14.12_{-0.06}^{+0.73}$ & $5.61_{-0.11}^{+0.09}$ & $1.05_{-0.09}^{+0.59}$ & $3.89_{-0.34}^{+0.02}$ & $2.950_{-0.098}^{+0.098}$ & $1650.47$ & $1601.00/477$ & $351.81/477$ & $0.952$\\
$34.711\!\sim\!35.211$ & $4.81_{-3.60}^{+0.72}$ & $0.77_{-0.02}^{+0.54}$ & $64.38_{-6.11}^{+1.61}$ & $2.36_{-0.17}^{+0.40}$ & $14.19_{-0.12}^{+1.25}$ & $5.25_{-0.54}^{+0.15}$ & $1.44_{-0.09}^{+0.95}$ & $4.06_{-0.89}^{+0.18}$ & $0.821_{-0.052}^{+0.058}$ & $1374.68$ & $1325.21/477$ & $372.87/477$ & $0.516$\\
$35.211\!\sim\!35.891$ & $3.04_{-2.03}^{+0.85}$ & $1.02_{-0.48}^{+0.22}$ & $58.52_{-2.16}^{+1.85}$ & $2.41_{-0.28}^{+0.26}$ & $15.54_{-0.96}^{+0.17}$ & $4.21_{-0.03}^{+0.43}$ & $2.43_{-0.92}^{+0.07}$ & $3.48_{-0.53}^{+0.29}$ & $0.293_{-0.030}^{+0.034}$ & $1651.05$ & $1601.58/477$ & $321.52/477$ & $0.554$\\
$35.891\!\sim\!36.091$ & $1.48_{-0.08}^{+2.56}$ & $1.27_{-0.46}^{+0.11}$ & $63.66_{-5.86}^{+0.64}$ & $2.02_{-0.24}^{+0.80}$ & $14.09_{-0.24}^{+1.29}$ & $5.28_{-0.36}^{+0.04}$ & $1.57_{-0.30}^{+0.52}$ & $3.75_{-0.75}^{+0.03}$ & $1.416_{-0.101}^{+0.115}$ & $542.74$ & $493.27/477$ & $278.49/477$ & $0.876$\\
$36.091\!\sim\!36.367$ & $0.43_{-0.93}^{+3.98}$ & $1.49_{-0.66}^{+0.15}$ & $64.33_{-2.72}^{+0.20}$ & $2.09_{-0.19}^{+0.74}$ & $14.09_{-0.09}^{+0.88}$ & $5.30_{-0.07}^{+0.24}$ & $1.26_{-0.05}^{+0.62}$ & $3.78_{-0.28}^{+0.10}$ & $2.795_{-0.121}^{+0.130}$ & $920.74$ & $871.26/477$ & $330.76/477$ & $0.538$\\
$36.367\!\sim\!36.685$ & $0.88_{-0.27}^{+2.49}$ & $1.00_{-0.00}^{+0.27}$ & $65.95_{-1.38}^{+0.13}$ & $2.67_{-0.06}^{+0.25}$ & $14.10_{-0.19}^{+0.97}$ & $5.67_{-0.04}^{+0.13}$ & $1.16_{-0.07}^{+0.18}$ & $3.94_{-0.19}^{+0.02}$ & $5.253_{-0.168}^{+0.173}$ & $1176.02$ & $1126.54/477$ & $378.64/477$ & $0.326$\\
$36.685\!\sim\!36.943$ & $0.28_{-0.53}^{+3.48}$ & $0.83_{-0.02}^{+0.47}$ & $63.99_{-2.40}^{+0.63}$ & $2.84_{-0.55}^{+0.02}$ & $14.22_{-0.04}^{+0.70}$ & $5.01_{-0.12}^{+0.44}$ & $1.47_{-0.28}^{+0.36}$ & $4.07_{-0.55}^{+0.10}$ & $2.172_{-0.107}^{+0.110}$ & $845.49$ & $796.01/477$ & $354.80/477$ & $0.244$\\
$36.943\!\sim\!37.304$ & $3.85_{-2.48}^{+0.26}$ & $1.09_{-0.43}^{+0.20}$ & $63.37_{-7.45}^{+2.15}$ & $2.19_{-0.00}^{+0.60}$ & $14.24_{-0.23}^{+1.31}$ & $5.26_{-0.72}^{+0.21}$ & $1.65_{-0.34}^{+0.80}$ & $3.81_{-1.06}^{+0.16}$ & $0.673_{-0.057}^{+0.065}$ & $998.00$ & $948.52/477$ & $316.56/477$ & $0.628$\\
$37.304\!\sim\!39.515$ & $4.29_{-3.07}^{+0.21}$ & $1.02_{-0.49}^{+0.19}$ & $57.87_{-2.85}^{+0.81}$ & $2.94_{-0.79}^{+0.25}$ & $15.34_{-0.83}^{+0.34}$ & $4.69_{-0.51}^{+0.09}$ & $1.88_{-0.40}^{+0.65}$ & $3.32_{-0.61}^{+0.11}$ & $0.189_{-0.019}^{+0.020}$ & $3287.00$ & $3237.53/477$ & $446.78/477$ & $0.314$\\
$39.515\!\sim\!40.576$ & $1.29_{-0.28}^{+2.74}$ & $1.01_{-0.15}^{+0.33}$ & $64.12_{-6.35}^{+1.57}$ & $2.37_{-0.05}^{+0.47}$ & $14.09_{-0.27}^{+1.22}$ & $5.37_{-0.38}^{+0.01}$ & $1.33_{-0.13}^{+0.57}$ & $3.85_{-0.87}^{+0.13}$ & $0.888_{-0.043}^{+0.049}$ & $2271.07$ & $2221.60/477$ & $362.94/477$ & $0.876$\\
$40.576\!\sim\!42.998$ & $1.31_{-0.13}^{+2.64}$ & $1.03_{-0.36}^{+0.36}$ & $57.79_{-1.86}^{+2.12}$ & $2.16_{-0.02}^{+0.56}$ & $14.80_{-0.34}^{+0.82}$ & $4.34_{-0.22}^{+0.16}$ & $1.93_{-0.44}^{+0.63}$ & $3.25_{-0.29}^{+0.50}$ & $0.095_{-0.012}^{+0.014}$ & $3421.28$ & $3371.81/477$ & $487.56/477$ & $0.124$\\

\end{longtable}
\noindent\parbox{0.96\linewidth}{\scriptsize
Note. All fitted EII intervals, including those with $S<10$, are listed, while
the analysis in the main text focuses primarily on intervals with $S>10$.
Parameter values and uncertainties are the reported central values and 68\%
credible errors. PGSTAT is $-2\ln L_{\max}$ summed over all active channels
and GBM detectors. RSS/dof is an auxiliary significance-residual measure, not a
reduced chi-square; PPC is the posterior predictive-check probability defined
in Section~4, with values close to 0 or 1 indicating possible model--data tension.}
\endgroup
\end{landscape}

\begin{landscape}
\begingroup
\tiny
\renewcommand{\arraystretch}{0.9}
\setlength{\tabcolsep}{0.8pt}
\begin{longtable}{lccccccccccccc}
\caption{Time-resolved MDFSYN spectral-fitting results for EII using joint
\textit{Fermi}/GBM+\textit{Swift}/BAT data.\label{tab:joint_mdfsyn}}\\
\hline\hline
Time & $\hat{t}$ & $a$ & $\log Q_0$ & $\log\Gamma$ & $\log R_0$
& $\log\gamma_{\rm inj}$ & $\log B_0$ & $p$ & $F$ & BIC
& PG-stat/dof & RSS/dof & PPC\\
(s) & (s) & & $(s^{-1})$ & & (cm) & & (G) & &
$(10^{-6}\,\mathrm{erg\,cm^{-2}\,s^{-1}})$ & & & &\\
\hline
\endfirsthead
\caption[]{Joint \textit{Fermi}/GBM+\textit{Swift}/BAT MDFSYN results
(continued).}\\
\hline\hline
Time & $\hat{t}$ & $a$ & $\log Q_0$ & $\log\Gamma$ & $\log R_0$
& $\log\gamma_{\rm inj}$ & $\log B_0$ & $p$ & $F$ & BIC
& PG-stat/dof & RSS/dof & PPC\\
(s) & (s) & & $(s^{-1})$ & & (cm) & & (G) & &
$(10^{-6}\,\mathrm{erg\,cm^{-2}\,s^{-1}})$ & & & &\\
\hline
\endhead
\hline
\multicolumn{14}{r}{\textit{Continued on next page}}\\
\endfoot
\hline
\endlastfoot
$21.500\!\sim\!23.200$ & $1.47_{-0.15}^{+2.72}$ & $1.25_{-0.51}^{+0.13}$ & $57.88_{-2.47}^{+0.43}$ & $2.03_{-0.14}^{+0.71}$ & $14.34_{-0.14}^{+1.27}$ & $5.05_{-0.31}^{+0.04}$ & $1.32_{-0.06}^{+0.84}$ & $2.79_{-0.20}^{+0.32}$ & $0.995_{-0.034}^{+0.037}$ & $3091.94$ & $3041.53/537$ & $499.38/537$ & $0.316$\\
$23.200\!\sim\!24.800$ & $0.64_{-1.08}^{+3.54}$ & $1.04_{-0.22}^{+0.35}$ & $58.44_{-1.68}^{+1.40}$ & $2.14_{-0.12}^{+0.64}$ & $14.41_{-0.09}^{+0.91}$ & $5.14_{-0.06}^{+0.18}$ & $1.30_{-0.03}^{+0.97}$ & $2.91_{-0.16}^{+0.34}$ & $1.722_{-0.048}^{+0.053}$ & $3048.56$ & $2998.15/537$ & $455.13/537$ & $0.662$\\
$24.800\!\sim\!27.000$ & $0.91_{-0.00}^{+2.20}$ & $1.66_{-0.39}^{+0.05}$ & $66.00_{-1.33}^{+0.32}$ & $2.16_{-0.11}^{+0.56}$ & $14.12_{-0.04}^{+0.51}$ & $5.76_{-0.11}^{+0.10}$ & $1.46_{-0.13}^{+0.46}$ & $3.87_{-0.19}^{+0.05}$ & $3.965_{-0.063}^{+0.061}$ & $3835.89$ & $3785.48/537$ & $677.82/537$ & $0.210$\\
$27.000\!\sim\!32.000$ & $1.82_{-0.22}^{+2.41}$ & $1.09_{-0.24}^{+0.10}$ & $64.17_{-2.48}^{+0.72}$ & $2.06_{-0.14}^{+0.82}$ & $14.38_{-0.16}^{+0.88}$ & $5.20_{-0.19}^{+0.01}$ & $1.21_{-0.15}^{+0.61}$ & $3.98_{-0.25}^{+0.02}$ & $0.527_{-0.014}^{+0.015}$ & $4662.83$ & $4612.43/537$ & $604.91/537$ & $0.108$\\
$32.000\!\sim\!33.500$ & $3.57_{-1.92}^{+0.72}$ & $1.02_{-0.16}^{+0.26}$ & $62.35_{-2.70}^{+0.05}$ & $2.78_{-0.46}^{+0.01}$ & $15.03_{-0.73}^{+0.26}$ & $5.23_{-0.17}^{+0.09}$ & $1.60_{-0.34}^{+0.35}$ & $3.78_{-0.52}^{+0.09}$ & $1.216_{-0.038}^{+0.041}$ & $2984.13$ & $2933.72/537$ & $535.62/537$ & $0.022$\\
$33.500\!\sim\!35.000$ & $3.90_{-2.57}^{+0.31}$ & $1.23_{-0.32}^{+0.22}$ & $60.64_{-3.04}^{+0.18}$ & $2.86_{-0.66}^{+0.05}$ & $15.13_{-0.78}^{+0.23}$ & $5.24_{-0.16}^{+0.13}$ & $2.20_{-0.89}^{+0.04}$ & $3.52_{-0.62}^{+0.08}$ & $1.355_{-0.046}^{+0.047}$ & $2953.23$ & $2902.83/537$ & $498.93/537$ & $0.170$\\
$35.000\!\sim\!37.000$ & $3.99_{-1.56}^{+0.48}$ & $0.96_{-0.01}^{+0.30}$ & $64.01_{-2.39}^{+0.73}$ & $2.67_{-0.14}^{+0.23}$ & $14.15_{-0.00}^{+0.66}$ & $5.50_{-0.24}^{+0.04}$ & $1.84_{-0.21}^{+0.68}$ & $3.86_{-0.28}^{+0.06}$ & $1.676_{-0.037}^{+0.041}$ & $3507.05$ & $3456.65/537$ & $618.64/537$ & $0.120$\\
$37.000\!\sim\!41.000$ & $3.10_{-2.02}^{+0.93}$ & $1.01_{-0.46}^{+0.21}$ & $56.92_{-1.85}^{+1.09}$ & $2.29_{-0.16}^{+0.38}$ & $14.59_{-0.09}^{+1.15}$ & $4.90_{-0.60}^{+0.12}$ & $1.66_{-0.19}^{+1.01}$ & $2.84_{-0.21}^{+0.37}$ & $0.373_{-0.016}^{+0.018}$ & $4261.93$ & $4211.52/537$ & $558.51/537$ & $0.196$\\

\end{longtable}
\noindent\parbox{0.96\linewidth}{\scriptsize
Note. All available joint-fit intervals are listed. Parameter values and
uncertainties are the reported central values and 68\% credible errors. The
column labeled PG-stat reports the total joint GBM+BAT likelihood statistic
$-2\ln L_{\max}$ and is not a pure PGSTAT; the label is retained for consistent
table formatting. RSS/dof and PPC have the same definitions as in
Table~\ref{tab:eii_mdfsyn}.}
\endgroup
\end{landscape}
\fi

\endgroup

\normalem
\begin{acknowledgements}
This work is supported by the National Natural Science Foundation of China (NSFC 12233006). Dr. Shan Chang acknowledges support from the National Natural Science Foundation of China 12103046 and the Xingdian Talent Support Plan - Youth Project. This work is also supported by the Postdoctoral Fellowship Program of CPSF under Grant Number GZC20252090. We acknowledge the use of the public data from the Fermi data archives.

\end{acknowledgements}


\bibliographystyle{raa}
\bibliography{ms2026-0344.bib}

\end{document}